\documentclass[reprint,floatfix,showpacs]{revtex4-2}
\usepackage{graphicx}
\usepackage{dcolumn}
\usepackage{amsmath}
\usepackage{amssymb}
\usepackage{color}

\begin{document}

\title{Using Covariant Lyapunov Vectors to Quantify High Dimensional Chaos with a Conservation Law}
\author{J. Barbish}
\affiliation{Department of Mechanical Engineering, Virginia Tech, Blacksburg, Virginia 24061}
\author{M. R. Paul}
\email{Corresponding author: mrp@vt.edu}
\affiliation{Department of Mechanical Engineering, Virginia Tech, Blacksburg, Virginia 24061}

\date{\today}

\begin{abstract}
We explore the high dimensional chaos of a one-dimensional lattice of diffusively coupled tent maps using the covariant Lyapunov vectors (CLVs). We investigate the connection between the dynamics of the maps in the physical space and the dynamics of the covariant Lyapunov vectors and covariant Lyapunov exponents that describe the direction and growth (or decay) of small perturbations in the tangent space. We explore the tangent space splitting into physical and transient modes and find that the splitting persists for all of the conditions we explore. In general, the leading CLVs are highly localized in space and the CLVs become more delocalized with increasing Lyapunov index. We consider the dynamics with a conservation law whose strength is controlled by a parameter that can be continuously varied. Our results indicate that a conservation law delocalizes the spatial variation of the CLVs.  We find that when a conservation law is present, the leading CLVs are entangled with fewer of their neighboring CLVs than in the absence of a conservation law.
\end{abstract}

\maketitle

\section{Introduction}
The complex dynamics of high-dimensional systems are at the center of many important problems~\cite{cross:1993}. Examples include weather prediction and the complex dynamics of fluid turbulence. The dimension of these systems is large, in part, due to the large number of degrees of freedom that contribute to the dynamics. Many powerful approaches are available when the dimension of the dynamics is small~\cite{guckenheimer:1983,abarbanel:1996}, however few of these generalize in a straightforward way to very large systems which are of intense current interest~\cite{cross:2009}.

Although powerful new approaches have emerged with exciting potential to describe high dimensional dynamics, such as Koopman mode decomposition~\cite{rowley:2009,mezic:2013}, dynamic mode decomposition~\cite{schmid:2010,tu:2014}, machine learning approaches~\cite{brunton:2016,pathak:2018}, and the use of exact coherent structures~\cite{waleffe:1997,kerswell:2005,kawahara:2012}, an exact and rigorous representation of high dimensional chaos remains an open challenge.  The use of covariant Lyapunov vectors (CLVs), an approach rooted in a fundamental dynamical systems description~\cite{eckmann:1985}, has been shown to shed new physical insights into high dimensional systems which has yielded a deeper understanding of chaotic dynamics~\cite{pikovsky:2016}.

We use the CLVs to probe the high-dimensional dynamics of large one-dimensional coupled map lattices (CMLs)~\cite{kaneko:1993} for a wide range of conditions.  By using CMLs we are able to numerically iterate large lattices for very long times while computing the entire spectrum CLVs. We use the spectrum of CLVs to investigate fundamental features of the dynamics.

In particular, we explore the dynamics for two different values of the control parameter for an individual map which yield significantly different chaotic dynamics.  We also investigate the influence of a conservation law on the CLVs and the dynamics. Such a broad study for large systems over long times remains computationally prohibitive using the governing partial differential equations of most laboratory scale systems as the Navier-Stokes equations of fluid dynamics. The fundamental insights gained by studying CMLs are useful in guiding the development of theoretical ideas and numerical approaches to build our physical understanding of the high-dimensional chaotic dynamics of large systems.

\section{Approach}
\label{section:approach}
\subsection{Diffusively Coupled Maps with a Conservation Law}

We are interested in studying the dynamics of large spatially extended systems where the chaos is generated locally and the dynamics are coupled spatially.  In particular, we explore the influence of nearest neighbor diffusive coupling as well as a global coupling due to a conservation law.

\begin{figure}[h!]
\begin{center}
\includegraphics[width=2in]{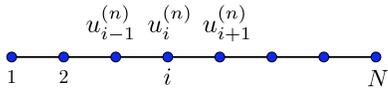}
\end{center}
\caption{A one-dimensional lattice of coupled maps with lattice index $i$, time step $n$, total number of maps $N$, and periodic boundary conditions. The dynamics of the lattice are given by iterating Eq.~(\ref{eq:cml-full}).}
\label{fig:lattice1dschematic}
\end{figure}

We explore the dynamics of a large one-dimensional lattice of coupled maps for a range of conditions. A schematic of the lattice indicating our variable conventions is shown in Fig.~\ref{fig:lattice1dschematic}.  The most complex situation we will explore is a lattice with diffusive coupling and a conservation law and we will use this situation to describe the details of our approach.

The dynamics of the lattice is given by
\begin{multline}
u_i^{(n+1)} = f(u_i^{(n)}) + \frac{\epsilon}{2} \left(f(u_{i+1}^{(n)}) - 2 f(u_i^{(n)}) + f(u_{i-1}^{(n)}) \right)  \\
+ \beta \left( \frac{c_0}{N} - \langle g(u_1^{(n)}, \ldots, u_N^{(n)}) \rangle \right)
\label{eq:cml-full}
\end{multline}
where $u_i^{(n+1)}$ is the continuous state of the map located at site $i$ at discrete time $n\!+\!1$. The lattice has a total of $N$ sites and the index $i$ specifies an individual lattice site where $i\!=\!1, 2, \ldots N$. We use periodic boundary conditions such that $u_i^{(n)} \!=\! u_{i+N}^{(n)}$ for all $n$. We are interested in the long time dynamics of the lattice and in a typical simulation we use $n_f \!\gtrsim\! 10^6$ where $n_f$ is the total number of discrete time steps.

The function $f(u)$ gives the state of an isolated map at the next time step, $u^{(n+1)}\!=\!f(u^{(n)})$. In our study, an identical tent map is placed at every lattice site where 
\begin{equation}
f(u^{(n)}) = 1 - a |u^{(n)}|
\label{eq:tent-map}
\end{equation}
and $a\!>\!0$ is a constant. The Lyapunov exponent $\lambda$ for a single tent-map is $\lambda = \ln a$.

The second term on the right hand side of Eq.~(\ref{eq:cml-full}) is the  nearest neighbor diffusive coupling with a strength of $\epsilon$. In this case, $\epsilon$ has the physical interpretation of determining the magnitude of a diffusion coefficient that multiplies a finite difference operator for a second-order spatial derivative with a step size of unity.

An additive conservation law is included with the final term of Eq.~(\ref{eq:cml-full}) which ensures that
\begin{equation}
\sum_{i=1}^N u_i^{(n)} \!=\! c_0
\label{eq:c0}
\end{equation}
at every $n$ where $c_0$ is a constant determined by the initial conditions. The choice of $c_0$ has a significant impact upon the dynamics. The conservation law is a form of global coupling involving the states of all of the lattice sites.

The strength of the conservation law is determined by the constant $\beta$ where $0\!\le\!\beta\!\le\!1$. The term $\langle g(u_1^{(n)}, \ldots, u_N^{(n)}) \rangle$ is the mean value of the state of the maps over all of the lattice sites at time $n+1$ in the \emph{absence} of the global coupling ($\beta\!=\!0$). For the symmetric diffusive coupling that we consider, all that remains in this term is the sum of the states of the individual maps if evolved forward by one time step. This yields
\begin{multline}
u_i^{(n+1)} = f(u_i^{(n)}) + \frac{\epsilon}{2} \left(f(u_{i+1}^{(n)}) - 2 f(u_i^{(n)}) + f(u_{i-1}^{(n)}) \right)  \\
+ \frac{\beta }{N}\left( c_0 - \sum_{i=1}^{N} f(u_i^{(n)}) \right).
\label{eq:cml}
\end{multline}
The presence of the conservation law in Eq.~(\ref{eq:cml}) can be validated by setting $\beta=1$ and summing over all of the lattice sites while using Eq.~(\ref{eq:c0}) and recognizing that the diffusive term vanishes when periodic boundary conditions are used. In Eq.~(\ref{eq:cml-full}) the conservation law is applied synchronously~\cite{oono:1987} as opposed to sequentially~\cite{kaneko:1989}.

When $\beta\!=\!0$, the conservation law is not imposed and the result is a lattice of diffusively coupled maps. In this case, our formulation is similar to the approach used by Takeuchi \emph{et al.}~\cite{takeuchi:2011}.  When $\beta\!=\!1$, the conservation law is enforced and the sum of all of the states of the maps across the lattice remain at the constant value $c_0$.  When $0 \!<\! \beta \!<\! 1$, the conservation law is partially imposed, we will refer to this as a broken conservation law. In this light, the constant $\beta$ provides a single parameter that we can use to study the lattice dynamics as a function of the strength of the conservation law.

\subsection{The Covariant Lyapunov Vectors}
We use the dynamic algorithm proposed by Ginelli \emph{et al.}~\cite{ginelli:2007} to compute the spectrum of CLVs. In the following, we present only the necessary details regarding their computation for the lattices we explore, see Refs.~\cite{wolfe:2007,kuptsov:2012,pikovsky:2016} for an in-depth discussion of the computational aspects of determining the CLVs.

There are essentially three steps in the calculation. We begin by listing these steps and then by providing more detail about their implementation for our calculations. (i) A long forward-time calculation of the lattice dynamics is conducted. (ii) The forward-time calculation is then continued while also computing the dynamics of the perturbations which provide a description of the tangent space. During this time, periodic reduced $\mathbf{QR}$ decompositions of the perturbations are computed and stored for use in the next step. (iii) A set of vectors are chosen which are evolved backwards in time to compute the combination matrix that is used to construct the CLVs from the stored $\mathbf{Q}$ and $\mathbf{R}$ matrices.

The implementation of this algorithm for our computations can be described as follows. Starting from an initial condition $u_i^{(0)}$, Eq.~(\ref{eq:cml}) is iterated for a long time $n_f$ to ensure initial transients have decayed. The goal is to evolve the lattice dynamics long enough such that the nonlinear trajectory in the $N$ dimensional state space is now on the attractor. In a typical simulation we use $n_f \!\gtrsim\! 10^6$.

At this point, $N_\lambda$ linearized equations are simultaneously evolved with the nonlinear lattice. $N_\lambda$ is the number of Lyapunov vectors and exponents that will be calculated where $N_\lambda$ is chosen such that $1 \!\le\! N_\lambda \!\le\! N$.  The linearized equations quantify the growth or decay of small perturbations and collectively they describe the tangent space. The evolution of the $k$th perturbation vector $\delta \vec{u}_k$ is given by
\begin{equation}
    \delta \vec{u}_k^{(n+1)} = \mathbf{J}^{(n)} \delta \vec{u}_k^{(n)}
\end{equation}
where $k = 1, \ldots, N_\lambda$ is the Lyapunov vector index and each $\delta \vec{u}_k$ has $N$ elements. $\mathbf{J}$ is the $N \times N$ Jacobian matrix
\begin{equation}
     \mathbf{J}^{(n)} = \left(\frac{\partial \vec{f}}{\partial \vec{u}}\, \right)^{(n)}
\end{equation}
that is evaluated along the nonlinear trajectory where ${\vec{f}} \!=\! \vec{f}(\vec{u}^{(n)})$. As the evolution proceeds forward in time, the perturbation vectors are periodically reorthonormalized using a reduced $\mathbf{QR}$ decomposition where the $\mathbf{Q}$ and $\mathbf{R}$ matrices are stored for later use in the algorithm. For the calculation of $N_\lambda$ CLVs, the columns of the $N \!\times\! N_\lambda$ matrix $\mathbf{Q}$ contain the orthonormalized vectors $\vec{q}_k^{\,(n)}$ where $k=1, \ldots, N_\lambda$ and each $\vec{q}_k$ has $N$ elements. These orthonormalized vectors $\vec{q}_k$ are often referred to as the Gram-Schmidt vectors or the backward Lyapunov vectors.  The upper triangular matrix $\mathbf{R}$ is a $N_\lambda \!\times\! N_\lambda$ matrix containing the expansion and contraction factors for the $\vec{q}_k^{\,(n)}$. The diagonal elements of $\mathbf{R}$ are directly related to the Gram-Schmidt Lyapunov exponents and the off-diagonal elements are essential in determining the directions of the CLVs. 

Next, $N_\lambda$ linearly independent vectors (often, random vectors are used) of length $k$, where $k=1,\ldots,N_\lambda$,  are chosen which are iterated backwards in time using the stored $\mathbf{Q}$ and $\mathbf{R}$ matrices in order to compute the $N_\lambda \times N_\lambda$ upper triangular combination matrix $\mathbf{C}$. The matrix $\mathbf{C}$ is computed during the backwards evolution as
\begin{equation}
    C_{j,i}^{(n-1)} = \left(R_{j,i}^{(n)}\right)^{-1} C_{j,i}^{(n)}
\end{equation}
which requires the inverse of $\mathbf{R}$.

The CLVs, $\vec{v}_k^{\,(n)}$, are then computed as the linear combination of the $\vec{q}_k^{\,(n)}$ where the coefficients used in the linear combination are given by the columns of $\mathbf{C}$. Specifically, column $k$ of $\mathbf{C}$ provides the $k$ coefficients that are used when forming the linear combination of the $\vec{q}_j^{\,(n)}$ where $j \!=\! 1,\ldots,k$. This can be represented as
\begin{equation}
    \vec{v}_k^{\,(n)} = \sum_{j=1}^k C_{j,k}^{(n)} \vec{q}_j^{\,(n)}
\end{equation}
where each $\vec{v}_k^{\,(n)}$ has $N$ elements.

In our study, we have found it to be more useful to use the backward integration process to compute the CLVs, $\vec{v}_k^{\,(n)}$, at many different time steps rather than to integrate a particular CLV forward or backward in time using the Jacobian. The computed CLVs can be used to visualize their spatiotemporal dynamics for further analysis. The CLVs can also be integrated forward in time using the Jacobian to provide values of the instantaneous covariant Lyapunov exponents. The long-time average of the instantaneous covariant Lyapunov exponents yields the spectrum of Lyapunov exponents.

\section{Results and Discussion}

The value of the control parameter $a$ in Eq.~(\ref{eq:tent-map}) determines the dynamics of a single tent-map.
We have focused our investigation on the two values $a\!=\!1.1$ and $a\!=\!1.6$. For these two values of $a$, the dynamics of a single map is chaotic while the details of the dynamics are quite different. We use this to explore how this is reflected in the CLVs and in the powerful diagnostics that can be calculated using the CLVs.

\begin{figure}[h!]
\begin{center}
\includegraphics[width=1.7in]{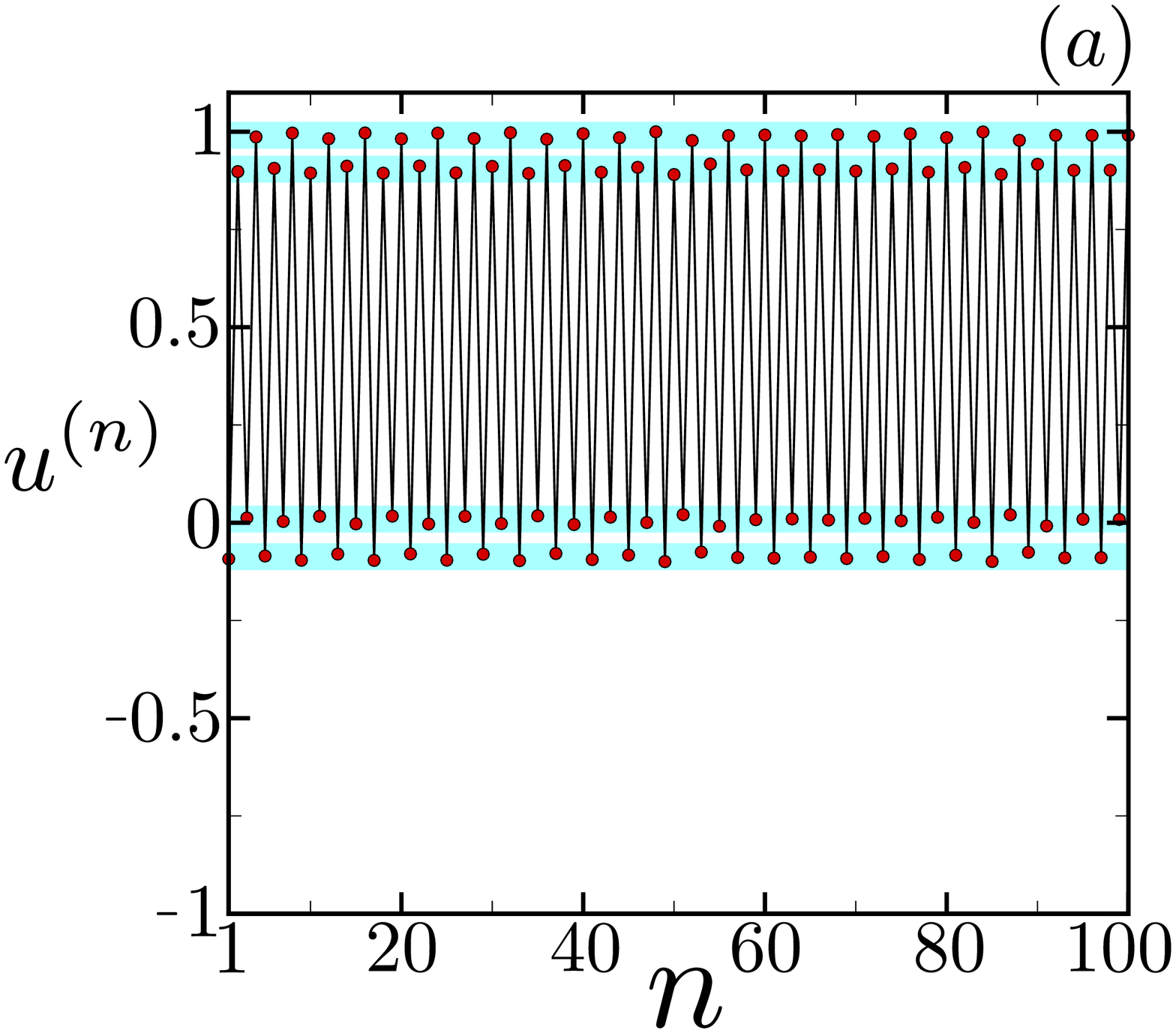} \hspace{-0.2cm}
\includegraphics[width=1.7in]{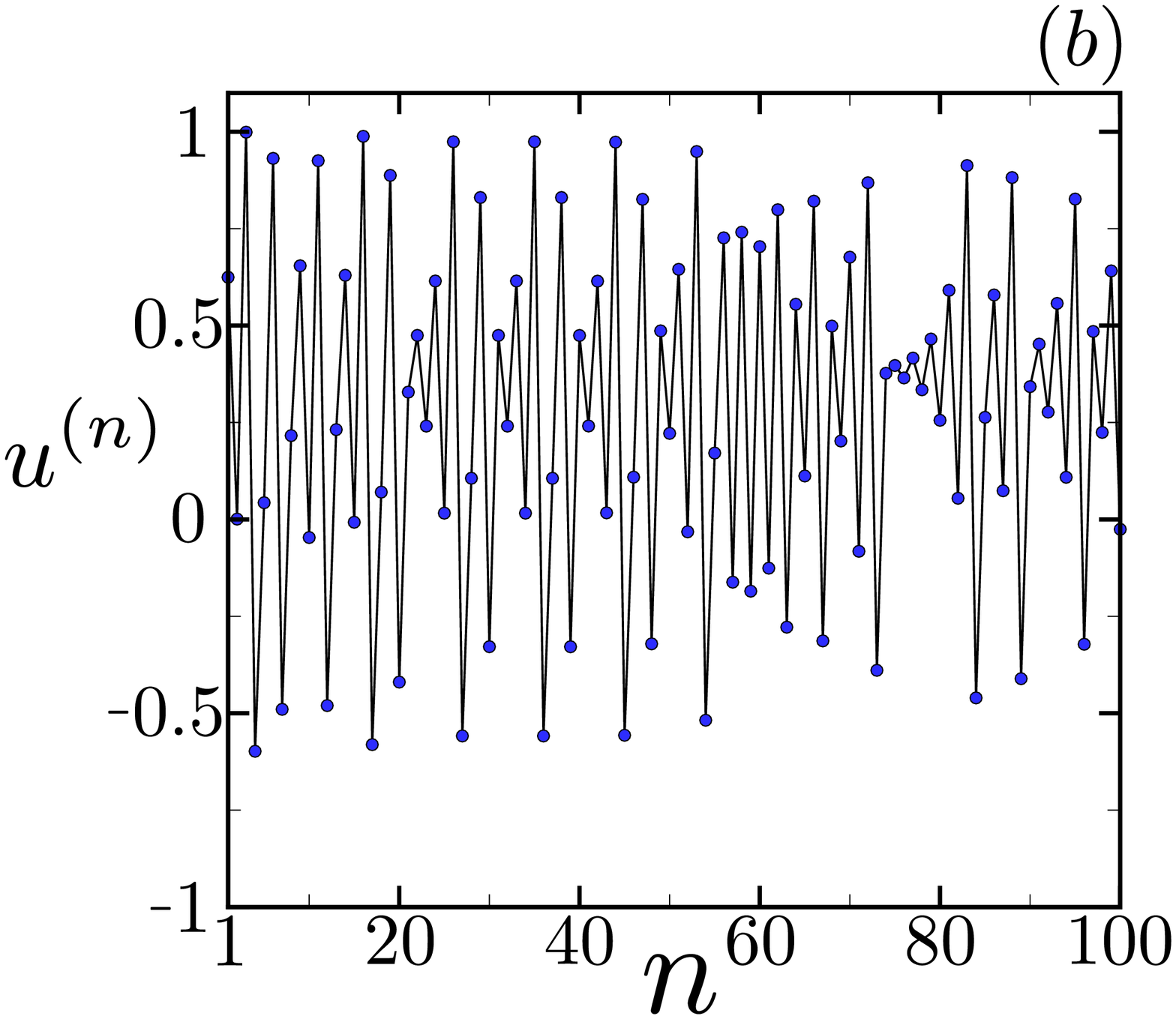}
\end{center}
\caption{The chaotic dynamics of a single tent-map for (a)~four-band chaos ($a\!=\!1.1$, $\lambda \!=\! 0.095$) and (b)~homogeneous chaos ($a\!=\!1.6$, $\lambda \!=\! 0.47$) where $u^{(0)}\!=\!0.2$, $n_f\!=\!10^4$, and the last 100 values are shown. Solid lines are included between the points as a guide. (a)~Four-band chaos, the map proceeds from one band to the next in a sequential order as indicated by the shaded regions. The chaos is due to the variation of the values within each band. (b)~Homogeneous chaos, the chaotic dynamics are not limited to a band structure and the map values are more homogeneously distributed.}
\label{fig:tent-map-single}
\end{figure}

For $a\!=\!1.1$, a single tent map exhibits what is called four-band chaos with a positive Lyapunov exponent $\lambda \!=\! 0.095$. The dynamics are illustrated in Fig.~\ref{fig:tent-map-single}(a) which shows how the tent-map proceeds through the four bands in a repeating sequence. However, the values within each band vary to yield the chaotic dynamics. Figure~\ref{fig:tent-map-single}(a) shows the variation of $u^{(n)}$ for 100 time steps after initial transients have decayed. The lines are included as a guide and the four different bands are indicated by the shaded regions. 

Figure~\ref{fig:tent-map-single}(b) shows the chaotic dynamics for $a\!=\!1.6$. In this case, $\lambda \!=\! 0.47$ which is nearly 5 times larger than the Lyapunov exponent for the four-band chaos we explore. For $a\!=\!1.6$, the four-band structure is no longer present and the values of the state of the map are more homogeneously distributed. We will refer to these dynamics as homogeneous chaos.

We next discuss our investigation of one dimensional lattices of tent-maps with periodic boundary conditions for a range of conditions. In all cases we have used diffusive coupling with a strength of $\epsilon\!=\!0.65$. In addition, the initial conditions for all simulations are random where the value of $c_0$ in Eq.~(\ref{eq:c0}) has been set to $c_0\!=\!1.2$.  Therefore, for all of the $N\!=\!512$ lattices that we discuss, each lattice was initiated from the identical initial condition for all of the figures shown. When we present results for multiple initial conditions these have also been implemented consistently when comparing different cases for the same system size. 

Unless noted otherwise, the following describes the details of our numerical approach. Each lattice was first iterated for times $n_f \!\ge\! 10^6$. The lattices and the perturbation vectors were then iterated forward in time for $10^4$ time units where reduced $\mathbf{QR}$ decompositions were computed every time step. The lattices were then iterated backward in time for $4 \!\times\! 10^3$ time units to compute the CLVs. We have verified that our results do not change significantly using simulations with much longer time integrations.  

The organization of the remainder of the paper is as follows. We first discuss our results for a diffusively coupled lattice without a conservation law in \S\ref{section:without-conservation-law}. This is followed by \S\ref{section:with-conservation-law} where we include a conservation law. In \S\ref{section:broken-conservation-law} we explore the dynamics when the conservation law is partially imposed. Lastly, we present our conclusions in \S\ref{section:conclusion}.

\subsection{Chaotic Dynamics without a Conservation Law}
\label{section:without-conservation-law}

\begin{figure}[h!]
\begin{center}
\includegraphics[width=1.6in]{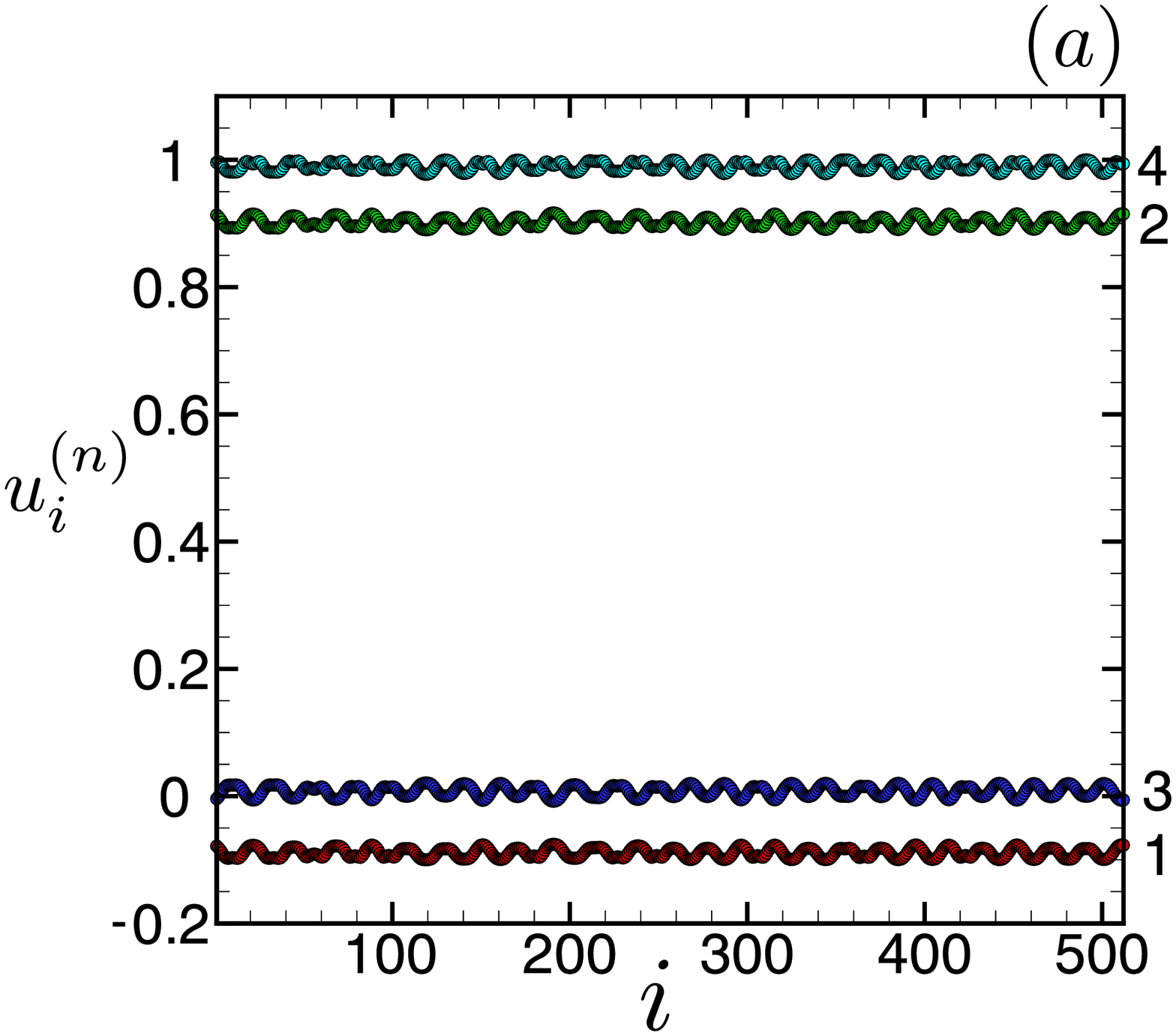} \hspace{-0.25cm}
\includegraphics[width=1.82in]{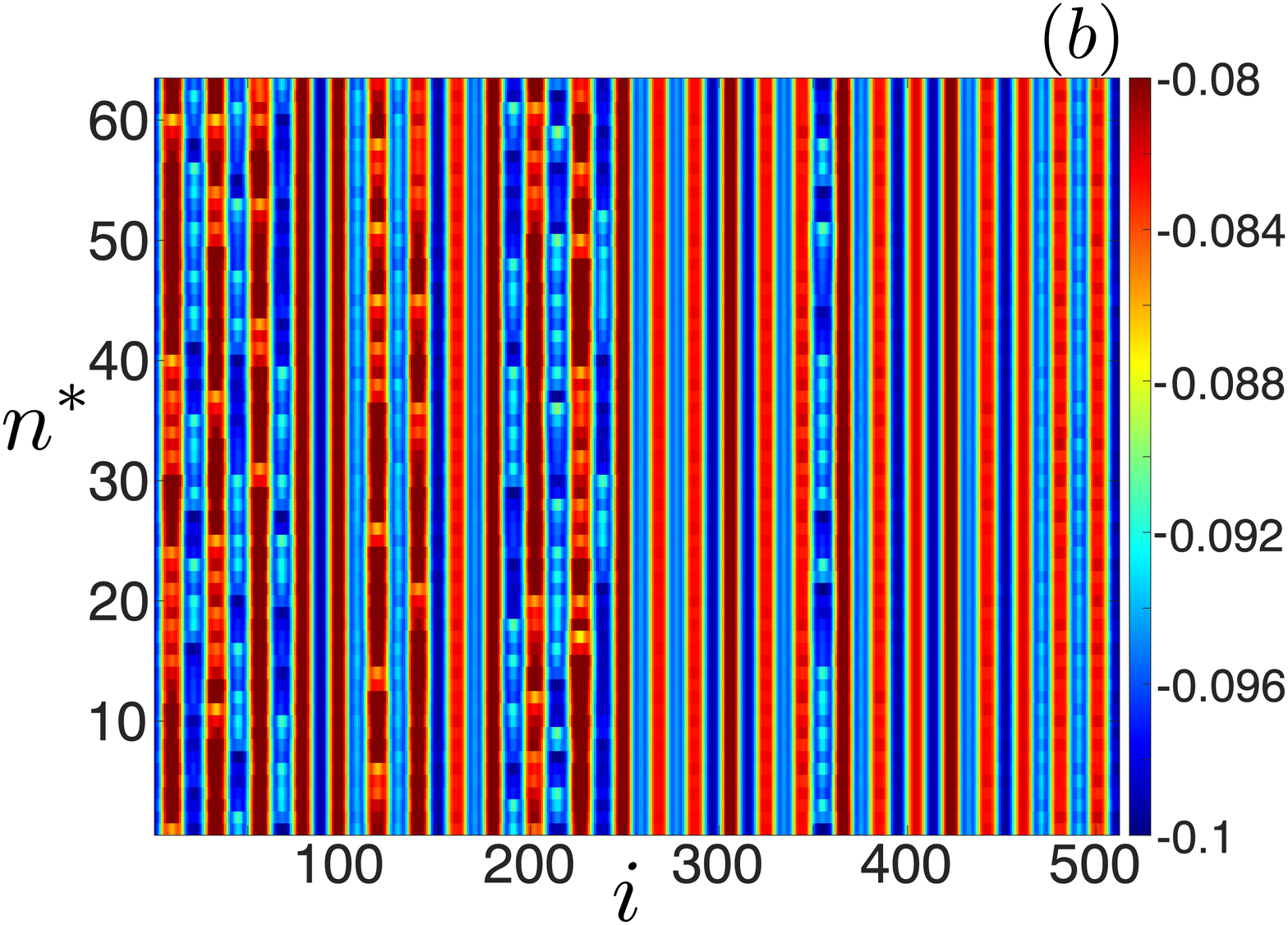}
\end{center}
\caption{The dynamics of diffusively coupled maps in the four-band chaos regime without a conservation law ($N \!=\!512$, $\epsilon\!=\!0.65$, $\beta\!=\!0$, $a\!=\!1.1$). (a)~The entire lattice exhibits four band chaos. The values of $u_i^{(n)}$ are shown for four consecutive time steps using different colors in the order of red, green, blue, and cyan as also indicated by the numerical labels on the right. (b) Space-time plot showing the dynamics every 16 time steps ($n\!=\!16n^*$). For these parameters, the plotted lattice values are always in the first band. The lattice has been iterated for over $10^6$ time steps.}
\label{fig:tent-map-lattice1}
\end{figure}

\begin{figure}[h!]
\begin{center}
\includegraphics[width=1.6in]{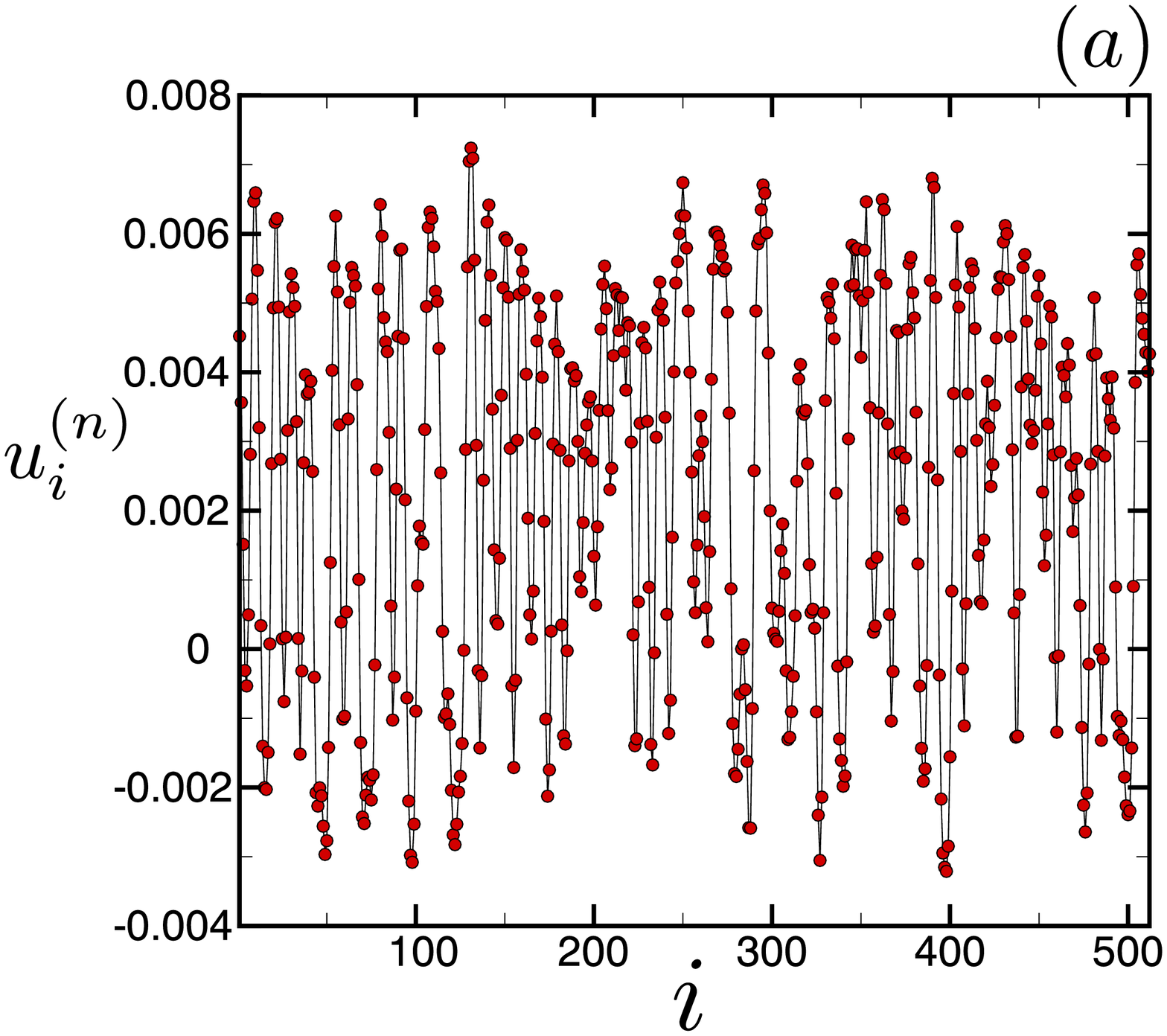} \hspace{-0.1cm}
\includegraphics[width=1.75in]{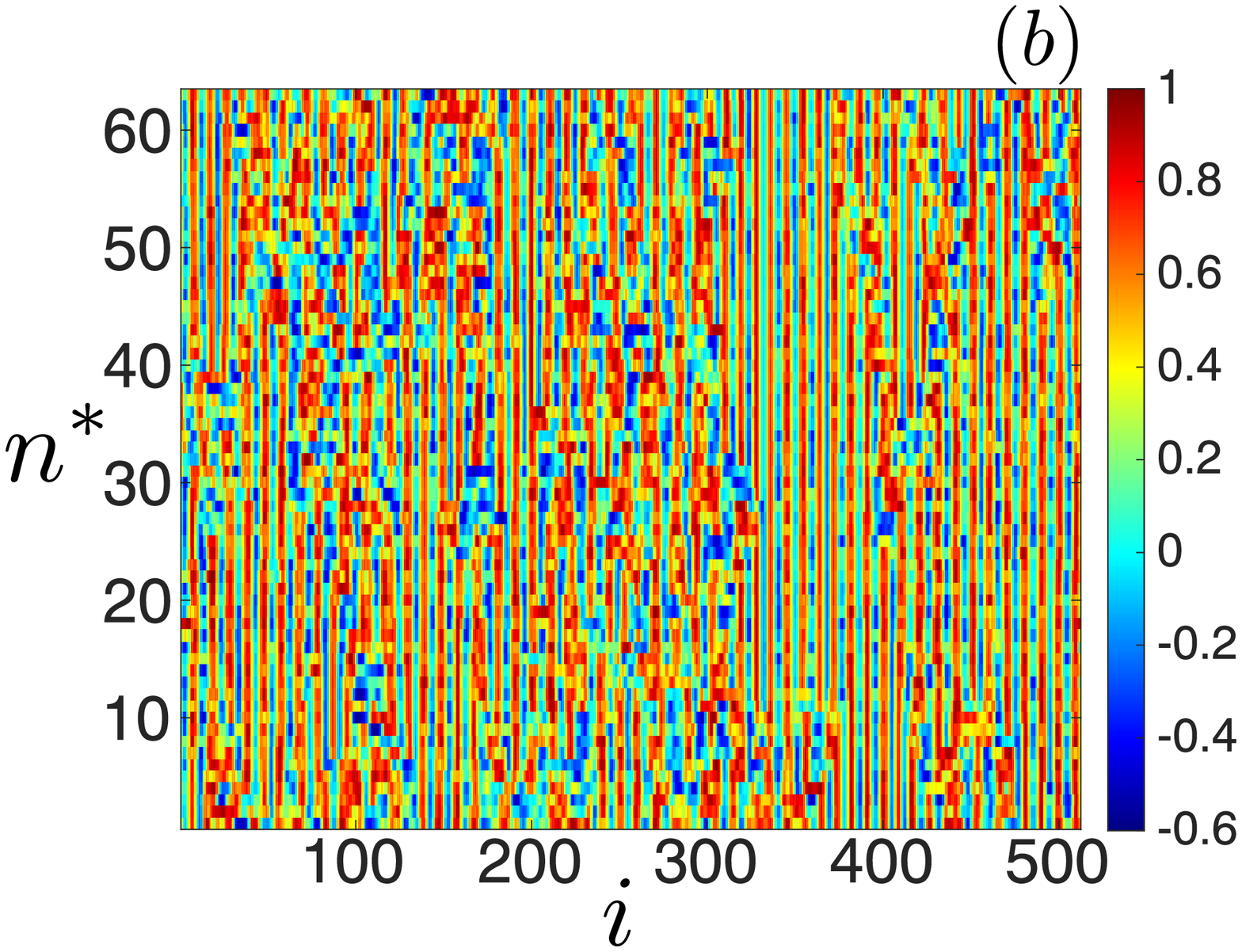}
\end{center}
\caption{The dynamics of diffusively coupled maps in the homogeneous chaos regime without a conservation law ($N\!=\!512$, $\epsilon\!=\!0.65$, $\beta\!=\!0$, $a\!=\!1.6$). (a)~The lattice values are more homogeneously distributed than what is shown in Fig.\ref{fig:tent-map-lattice1}(a). The values of $u_i^{(n)}$  are shown at one instant of time. The lattice was initiated from the same initial condition used in Fig.~\ref{fig:tent-map-lattice1} and the dynamics have been iterated for over $10^6$ time steps. (b) A space-time plot of the dynamics using $n\!=\!16n^*$.}
\label{fig:tent-map-lattice2}
\end{figure}

We first explore lattices of diffusively coupled tent-maps in the absence of a conservation law $(\beta\!=\!0)$. The dynamics of lattices with $N\!=\!512$ diffusively coupled maps are shown in Figs.~\ref{fig:tent-map-lattice1}-\ref{fig:tent-map-lattice2}.  In all of our calculations we use a diffusion coefficient of $\epsilon \!=\! 0.65$.  This allows us to connect our findings with the study by Takeuchi~\emph{et al.} who explored a smaller lattice ($N\!=\!256$) without a conservation law.  It would be interesting to explore the variation of the dynamics as a function of the diffusion but we have not explored this further here.

The four-band chaos case, $a\!=\!1.1$, is presented in Fig.~\ref{fig:tent-map-lattice1}(a) which shows the states of  lattice sites, $u_i^{(n)}$ for $i\!=\!1, \ldots, N$, at four consecutive time steps. Each time step is illustrated using a different color and the sequential order is indicated by the numerical labels on the right.  The entire lattice executes four-band chaos in lock-step. However, the values of the lattice sites vary within each band resulting in the chaotic dynamics. The lattice continues this four-band sequence for the duration of the simulation.

A space-time plot of the lattice dynamics is shown in Fig.~\ref{fig:tent-map-lattice1}(b).  The dynamics are plotted using $n^*$ where $n\!=\!16 n^*$ to more clearly show the variation in the dynamics. For the case shown, this results in the space-time plot illustrating the dynamics in the first band from Fig.~\ref{fig:tent-map-lattice1}(a). Plots illustrating the dynamics by strobing the results to visualize the lattice values in the different bands would yield similar results and are not shown.

Figure~\ref{fig:tent-map-lattice2} illustrates the dynamics of the lattice for the homogeneous chaos case where $a\!=\!1.6$. The band structure is no longer present and only a single instant of time is shown in Fig.~\ref{fig:tent-map-lattice2}(a). There is a spatially varying structure to the lattice values with a wavelength of approximately 10 lattice sites. However, the spatial variation is quite complicated with defect type structures that vary significantly with time which is evident in the space-time plot as indicated by Fig.~\ref{fig:tent-map-lattice2}(b).

\begin{figure}[h!]
\begin{center}
\includegraphics[width=1.7in]{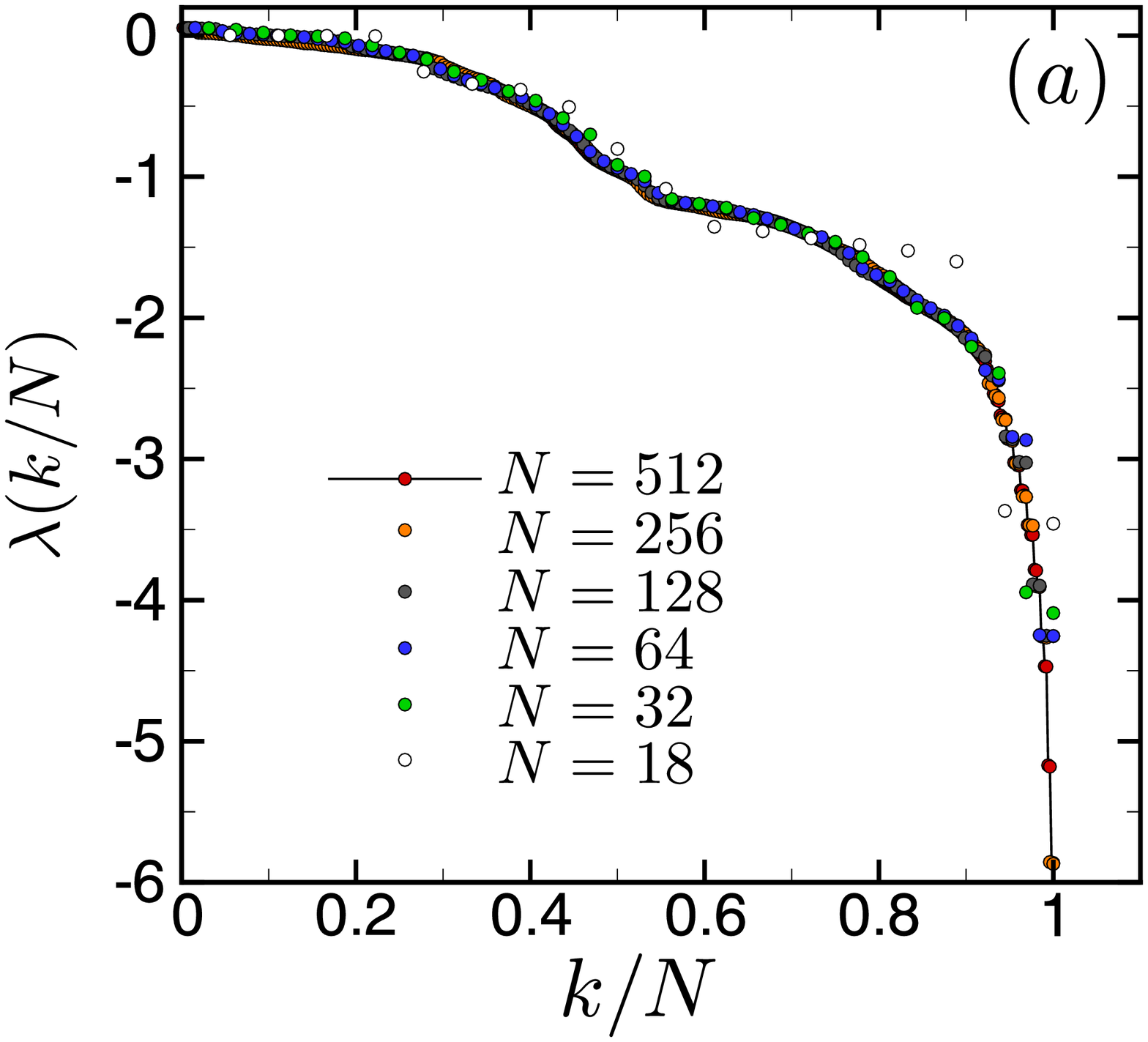} \hspace{-0.6cm}
\includegraphics[width=1.7in]{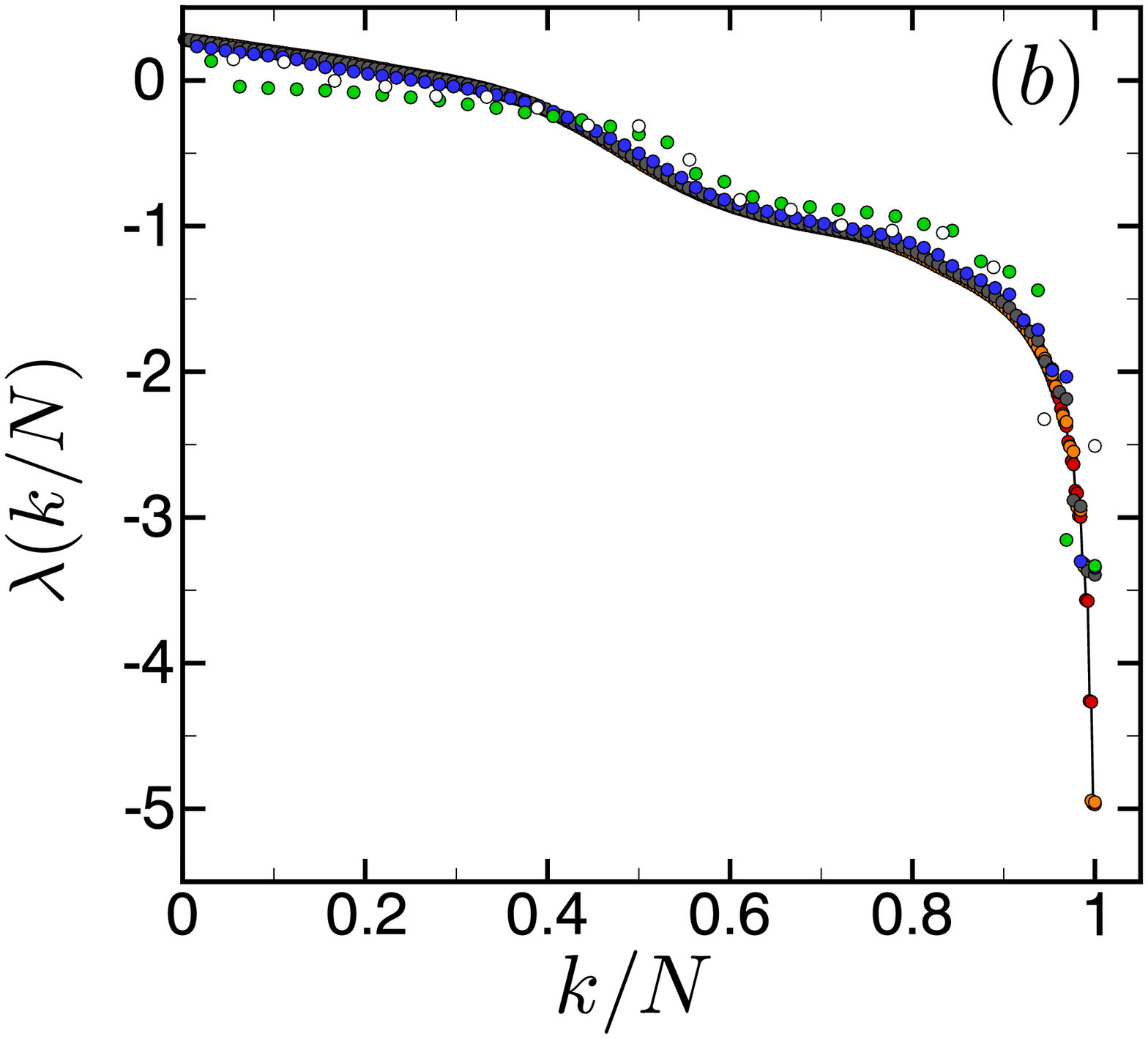}  
\end{center}
\caption{The Lyapunov exponent spectra for diffusively coupled maps without a conservation law ($\beta\!=\!0$). The Lyapunov exponents $\lambda(k/N)$ plotted with the normalized index $k/N$. For increasing $N$, the results collapse to a single curve indicating extensive chaos.  (a)~$a \!=\! 1.1$, the dynamics are extensive for $N \!\gtrsim\! 32$. (b)~$a\!=\!1.6$, the dynamics are extensive for $N \!\gtrsim\! 64$.} 
\label{fig:extensive-chaos}
\end{figure}

The spectra of the Lyapunov exponents are shown in Fig.~\ref{fig:extensive-chaos}(a)-(b) for lattices of different sizes $N$. The spectra are plotted using the scaled index $k/N$ which collapse to a single curve when the chaotic dynamics are in the extensive regime.
\begin{figure}[h!]
\begin{center}
\includegraphics[width=3.5in]{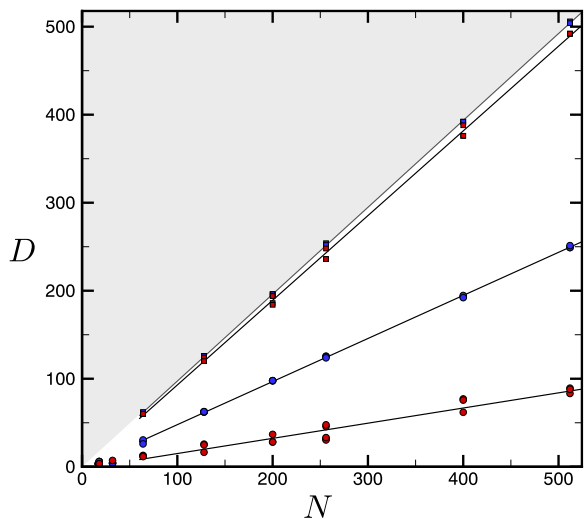}
\end{center}
\caption{The variation of the dimension of the dynamics with system size $N$ for diffusively coupled maps without a conservation law ($\beta\!=\!0$). Fractal dimension $D_\lambda$ (circles), $a\!=\!1.1$ (red, lower) and $a\!=\!1.6$ (blue, upper). Physical dimension $D_{\text{ph}}$ (squares), $a\!=\!1.1$ (red, lower), and $a\!=\!1.6$ (blue, upper). Lines show $D \!\propto\! N$ indicating extensive chaos. The dimension can not be in the gray region. For each $N$, results using at least three different initial conditions are shown.  For $N \!\lesssim\! 18$ the type of dynamics depends upon the initial conditions.} 
\label{fig:dimension-beta0}
\end{figure}

Figure~\ref{fig:extensive-chaos}(a) shows the Lyapunov exponent spectra for the case of four-band chaos, $a\!=\!1.1$. The leading Lyapunov exponent, $\lambda_1 \!=\! 0.052$, is greater than zero indicating chaos. The leading Lyapunov exponent $\lambda_1$ is less than the value of a single map where $\lambda \!=\! 0.095$. This reduction in $\lambda_1$, when compared with the Lyapunov exponent for a single map, is a result of the diffusive coupling. For lattices where $N \!\gtrsim\! 32$ the Lyapunov spectra collapse to a single curve and the chaotic dynamics are extensive.

Figure~\ref{fig:extensive-chaos}(b) shows the Lyapunov spectra for homogeneous chaos ($a\!=\!1.6$) where similar trends are found. In this case, $\lambda_1 \!=\! 0.28$ whereas for a single map  $\lambda \!=\! 0.47$.  The chaotic dynamics become extensive for $N \! \gtrsim  \! 64$. In our study we will focus on lattices with $N\!=\!512$ sites which is in the extensively chaotic regime.

Figure~\ref{fig:dimension-beta0} shows the variation of the fractal dimension $D_\lambda$ (circles) with system size $N$. $D_\lambda$ is computed using the Kaplan-Yorke formula~\cite{kaplan:1979} which only requires knowledge of the spectrum of Lyapunov exponents. Roughly speaking, the fractal dimension is the number of Lyapunov exponents which must be added together to yield a value of zero. This  provides an estimate of the number of degrees of freedom required to describe the chaotic dynamics, on average. This can be expressed as
\begin{equation}
D_\lambda = K + \frac{S_K}{|\lambda_{K+1}|}
\end{equation}
where the summation of the Lyapunov exponents $S_j$ is given by
\begin{equation}
S_j = \sum_{k=1}^j \lambda_k
\end{equation}
and $j \!=\! 1, \ldots, N_\lambda$~\cite{cross:1993}. The integer $K$ is the largest index $j$ for $S_j$ to yield a positive value $S_j \!>\! 0$.

In Fig.~\ref{fig:dimension-beta0}, the upper line through the circles (blue) is for homogeneous chaos, $a\!=\!1.6$, and the lower line through the circles (red) is for four-band chaos, $a\!=\!1.1$. For each $N$, and for each $a$, results are shown for at least three different random initial conditions. $D_\lambda$ for the four-band chaos case  exhibits variations with the choice of initial conditions whereas the variation for the homogeneous chaos case is much smaller. In addition, $D_\lambda$ for homogeneous chaos is significantly larger than that of the four-band chaos as expected because of the larger value of the control parameter $a$. For the same system size, $D_\lambda$ for homogeneous chaos is approximately three times larger than $D_\lambda$ for four-band chaos over the range shown.

For extensive chaos it is expected that $D_\lambda \! \propto \! N$.  This is indicated in Fig.~\ref{fig:dimension-beta0} by the two solid black lines through the circles where $D_\lambda \!=\! 0.17 N \!-\! 2.2$ for $a\!=\!1.1$ and $D_\lambda \!=\! 0.49 N \!-\! 1.1$ for $a\!=\!1.6$. The maximum possible dimension is equal to the number of degrees of freedom, which is the number of lattice sites, and dimension values larger than this are not accessible which is indicated by the gray shaded region.

\begin{figure}[h!]
\begin{center}
\includegraphics[width=1.65in]{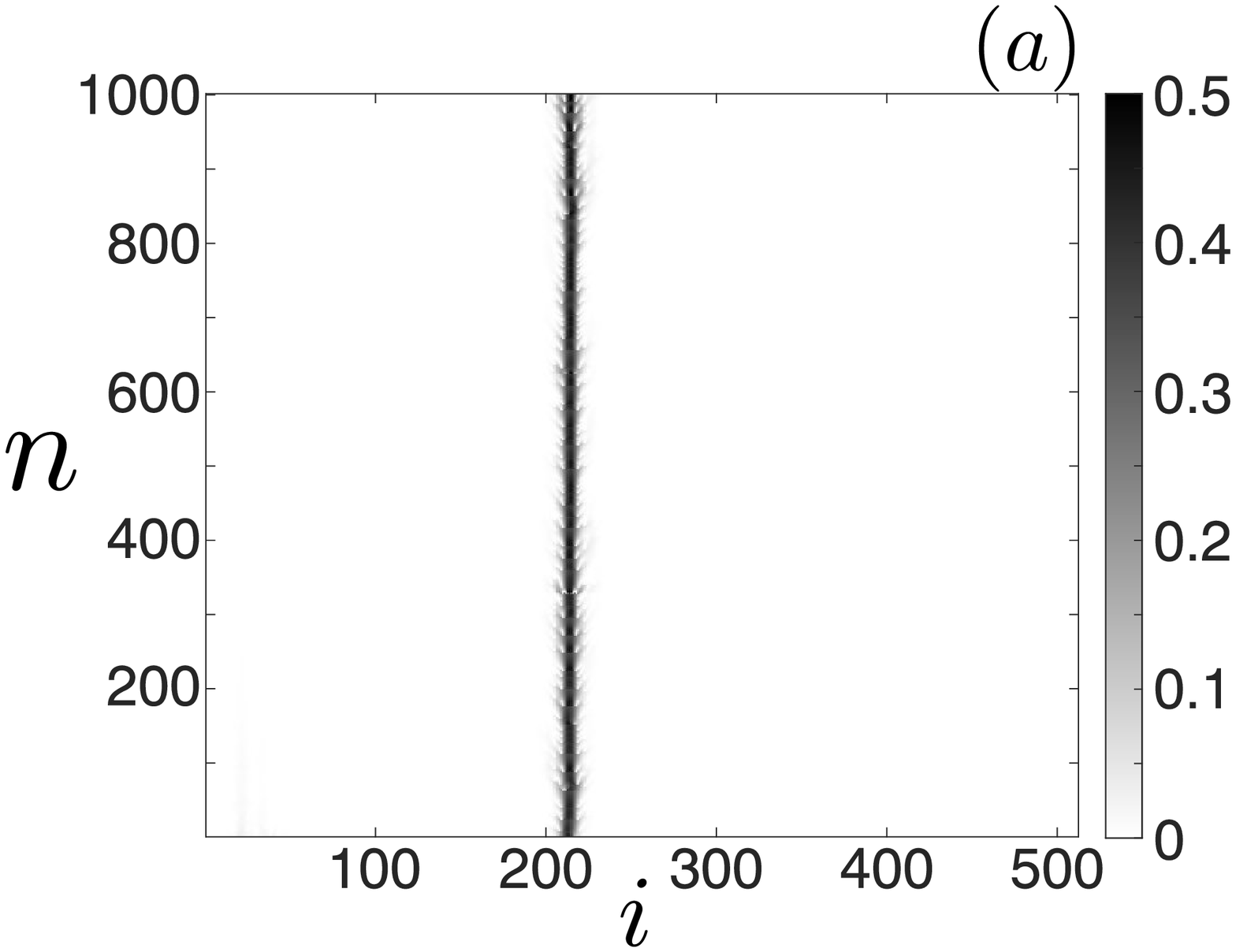}
\includegraphics[width=1.65in]{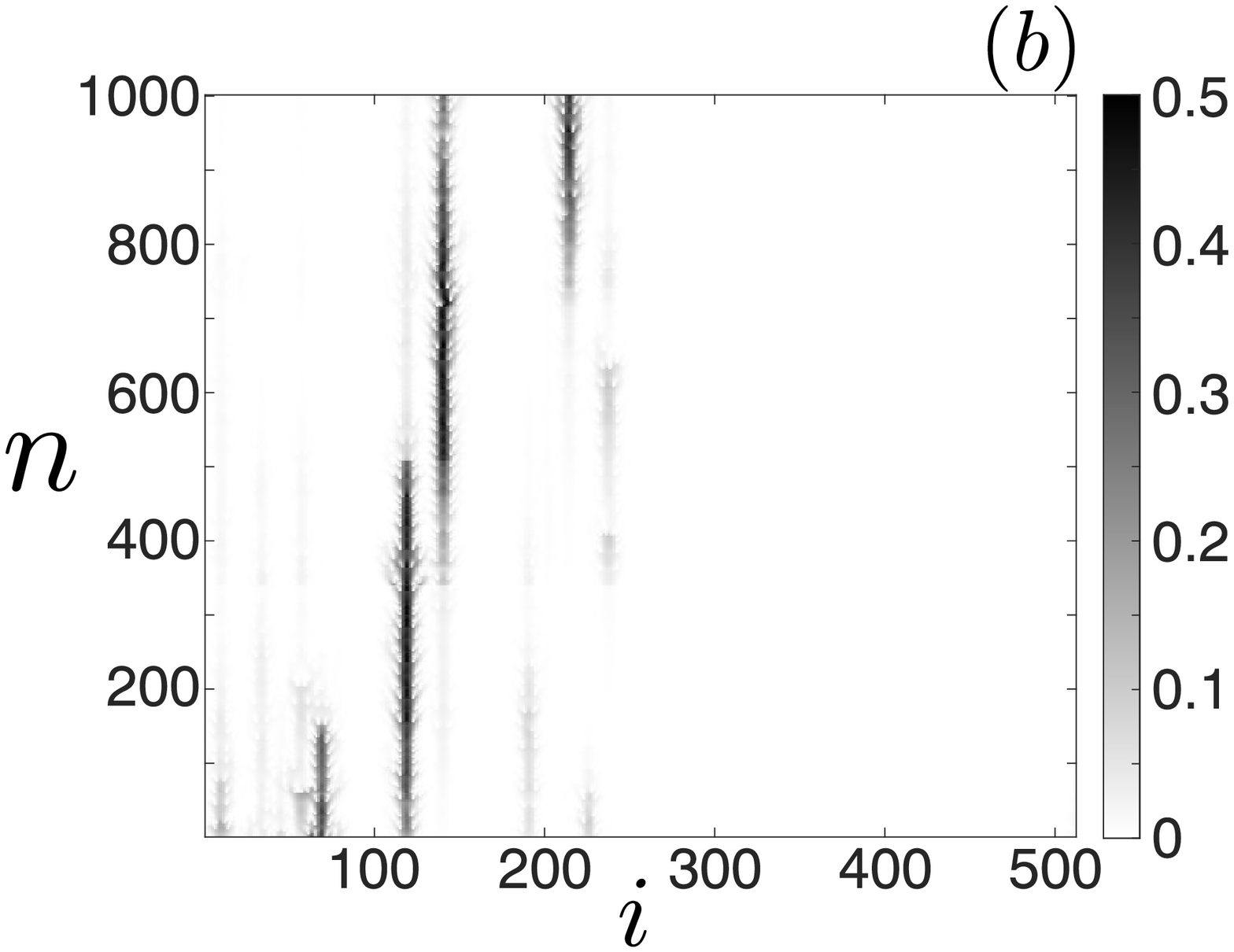} \\
\includegraphics[width=1.65in]{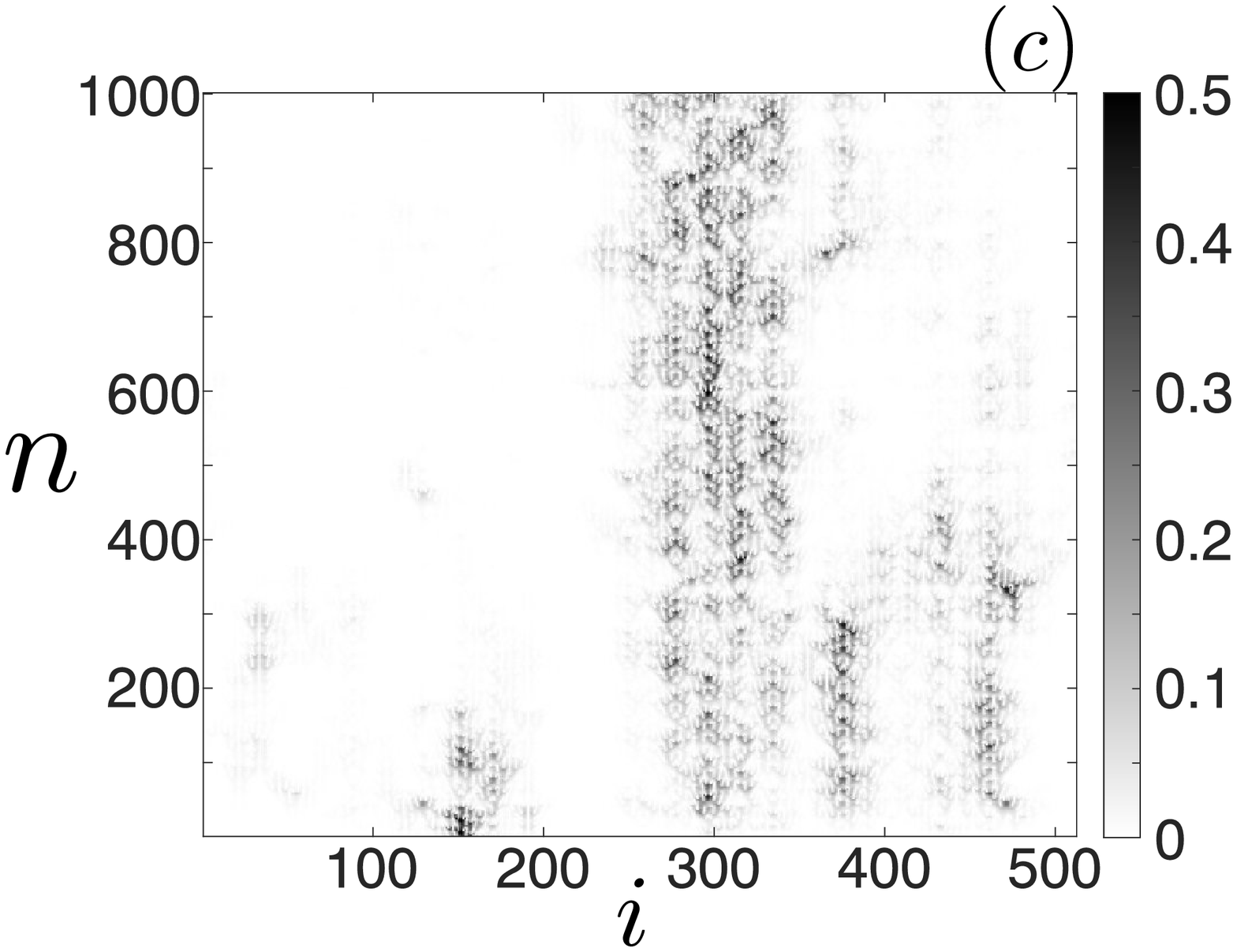}
\includegraphics[width=1.65in]{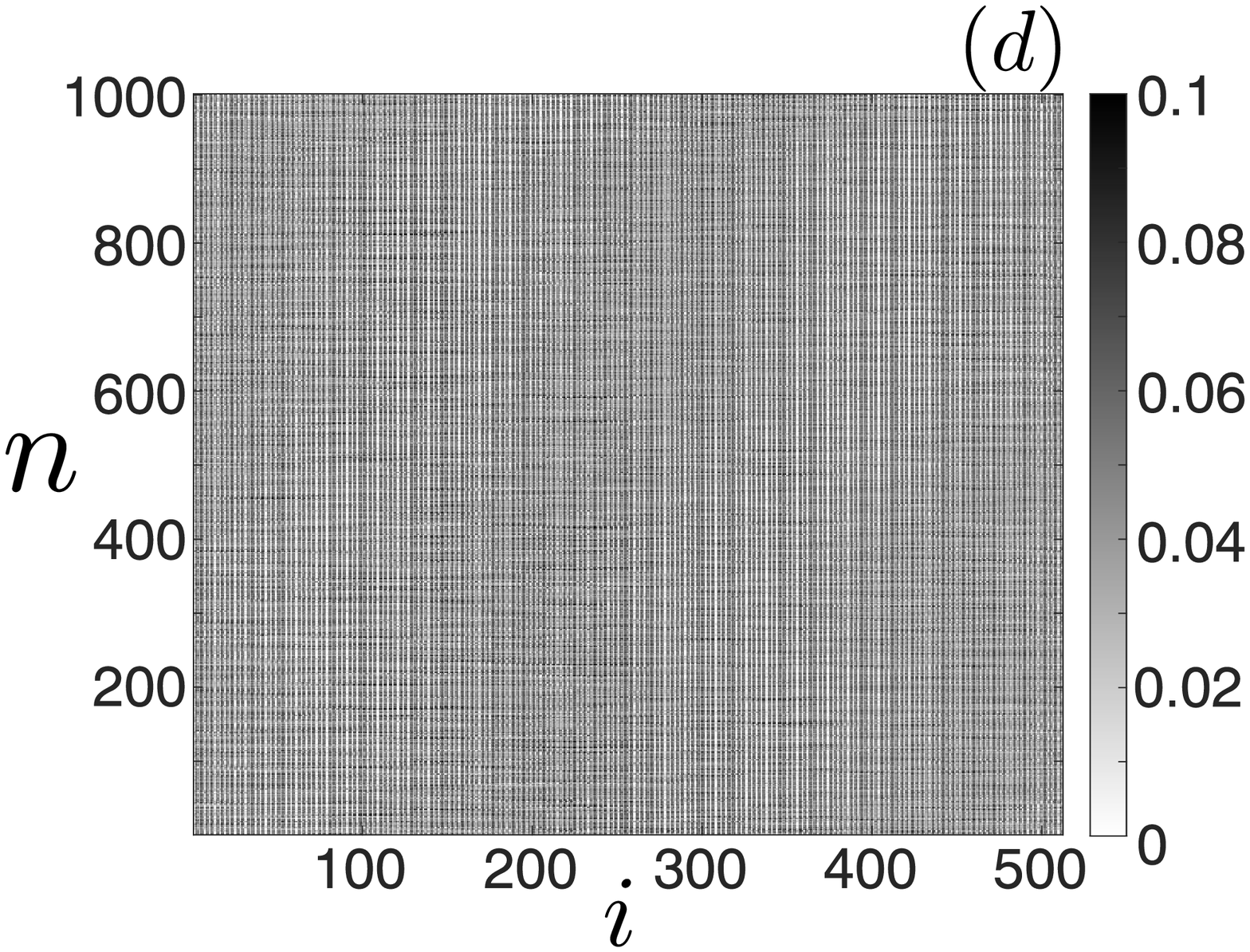}
\end{center}
\caption{Space-time plots of the CLVs of four-band chaos without a conservation law ($N \!=\!512$, $\beta\!=\!0$, $a\!=\!1.1$). The magnitude of $|\vec{v}_k^{\,(n)}|$ is shown where $k$ is the Lyapunov index and $i$ is the index over the components of the CLV. (a)~$k\!=\!1$, (b)~$k\!=\!10$, (c)~$k\!=\!100$, (d)~$k\!=\!500$. For (d) the color scale range has been reduced to account for the strongly delocalized nature of the Lyapunov vector.} 
\label{fig:clvs-a1p1-beta0}
\end{figure}

Figure~\ref{fig:clvs-a1p1-beta0} shows space-time plots of the magnitude of the CLVs, $|\vec{v}_k^{\,(n)}|$, for four values of $k$ for four-band chaos without a conservation law. The dynamics in physical space for these results are shown in Fig.~\ref{fig:tent-map-lattice1}. The four CLVs, as indicated by the CLV index $k$, are selected to indicate the variation in the CLV dynamics over the entire range of the spectrum. In each panel, $i$ is the index representing the $i$th component of that individual CLV and the ordinate axis is time. Dark regions indicate a large value and the light regions indicate a small value which are quantitatively specified by the color bars.

The magnitude of the leading CLV, $k\!=\!1$, is shown in Fig.~\ref{fig:clvs-a1p1-beta0}(a). The leading CLV is very localized in space.  A similar localized structure is found for the leading CLV when using different random initial conditions where the localized structure occurs at different locations in terms of the index $i$. The width of the band of significant CLV magnitude, as indicated by the dark region, encompasses a region of approximately 10 lattice sites. This indicates that small perturbations at these lattice sites grow much faster than elsewhere on the lattice. Such a highly localized CLV indicates a direction in the tangent space where perturbation growth is largest.

A close comparison of the CLV magnitude with the lattice dynamics in physical space, as shown in Fig.~\ref{fig:tent-map-lattice1}, yields that these lattice sites also exhibit significant variations in their dynamics. However, this significant variation in the physical space dynamics is not unique to this region. For example, the dynamics of the lattice in Fig.~\ref{fig:tent-map-lattice1} are also quite complicated in the region $1 \! \lesssim \! i \lesssim \! 90$ whereas the leading CLV magnitude for these index values are very small in Fig.~\ref{fig:clvs-a1p1-beta0}(a).

It is not clear how to identify this location where the CLV magnitude is large and localized beforehand using only knowledge of $u_i^{(n)}$. Similar results have been obtained for Rayleigh-B\'enard convection where the leading Lyapunov vector was compared closely with the flow field dynamics and it was not possible to define flow field features that uniquely locate regions of large Lyapunov vector magnitude~\cite{egolf:2000,scheel:2006,paul:2007,xu:2018,levanger:2019}.

Figure~\ref{fig:clvs-a1p1-beta0}(b)-(d) illustrates the variation of the CLV magnitude with increasing values of the Lyapunov index $k$. For the 10th Lyapunov vector, shown in Fig.~\ref{fig:clvs-a1p1-beta0}(b), the CLV has become less localized overall. For example, at any time $n$ there will be several bands of $i$ where the magnitude is significant, yet it still consists of localized structures. This delocalization increases with increasing Lyapunov index $k$ as indicated by the 100th CLV shown in Fig.\ref{fig:clvs-a1p1-beta0}(c). The 500th CLV is shown in Fig.~\ref{fig:clvs-a1p1-beta0}(d) which is very delocalized and uses a different color bar scale than the other panels in order to discern the variations.

The CLVs are not orthogonal vectors and their direction in the tangent-space is a physically meaningful quantity. Of immediate interest are the relative directions of the different CLVs. If the CLVs have near tangencies their dynamics are coupled with one another since small variations in one of the CLVs can affect the other and vice versa.  If the CLVs do not have near tangencies they are hyperbolically isolated in the tangent space and their dynamics are not coupled with one another. These ideas have been proposed by Yang \emph{et al.}~\cite{yang:2009} to split the tangent space into physical modes, for CLVs with near tangencies, and transient modes, for isolated CLVs.  This decomposition of the tangent space provides important insights into the dynamics.

One way to quantify the relative directions of large numbers of CLVs for long times is to compute the violations of the dominance of Oseledets splitting (DOS)~\cite{oseledec:1968,pikovsky:2016}.  The spectrum of the infinite time Lyapunov exponents are guaranteed to be in descending order $\lambda_1 > \lambda_2 \ldots$, where we have assumed distinct exponents for simplicity of the discussion, due to the isolation of the subspaces described by the different expansion rates in the tangent space~\cite{oseledec:1968}.

However, the finite time covariant Lyapunov exponents (CLEs) may not follow this strict ordering for all time and there will be periods of time where this ordering is violated. It is important to highlight that these violations can only be computed using the finite time CLEs and not the finite time Lyapunov exponents computed using the Gram-Schmidt vectors despite the fact that they converge to the same values in the infinite time limit.  It has been shown that the presence of violations between two finite time CLEs indicate near tangencies between the two corresponding CLVs~\cite{pugh:2003,bochi:2005}.

Following the typical convention~\cite{takeuchi:2011}, we use $\tilde{\lambda}_k^\tau(t)$ to indicate the kth finite time CLE at time $t$ whose value is determined using the time interval from $t$ to $t \!+\!\tau$ where $\tau$ is a constant.  In our calculations we have used $\tau \!=\! 5$. The difference between all possible pairs of $\tilde{\lambda}_k^\tau(t)$ can be expressed as
\begin{equation}
    \Delta \lambda_{k_1,k_2}^\tau(t) = \tilde{\lambda}_{k_1}^\tau(t) - \tilde{\lambda}_{k_2}^\tau (t)
    \label{eq:deltalambda}
\end{equation}
where $k_1$ and $k_2$ are Lyapunov vector indices with each index ranging from 1 to $N_\lambda$. The amount of violation is then computed as
\begin{equation}
    \nu_{k_1,k_2}^\tau = \langle 1 - \mathcal{H} [\Delta \lambda_{k_1,k_2}^\tau(t)] \rangle
    \label{eq:vDOS}
\end{equation}
where a unit step function is indicated by $\mathcal{H}$ and the angle brackets represent a time average. Therefore, $\nu_{k_1,k_2}^\tau$ represents the fraction of the time a violation occurs between the two CLVs given by the $k_1$ and $k_2$ indices. The violation measure is within the range $0 \!\le\! \nu_{k_1,k_2}^\tau \!\le\! 1$ where $\nu_{k_1,k_2}^\tau\!=\!1$ indicates pure violation for all time and $\nu_{k_1,k_2}^\tau\!=\!0$ indicates the absence of any violations.

\begin{figure}[h!]
\begin{center}
\includegraphics[width=2.25in]{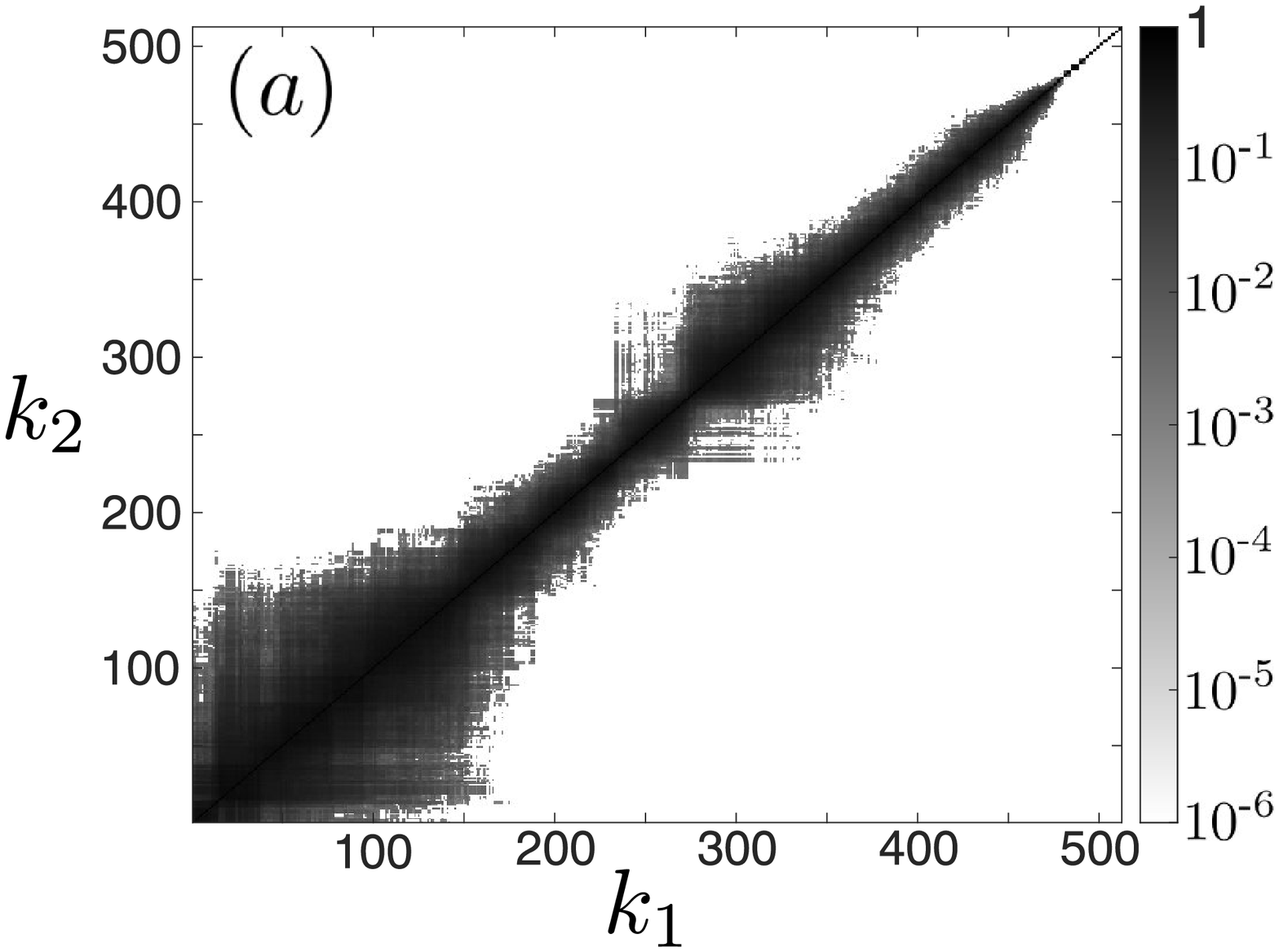} \\ 
\includegraphics[width=2.25in]{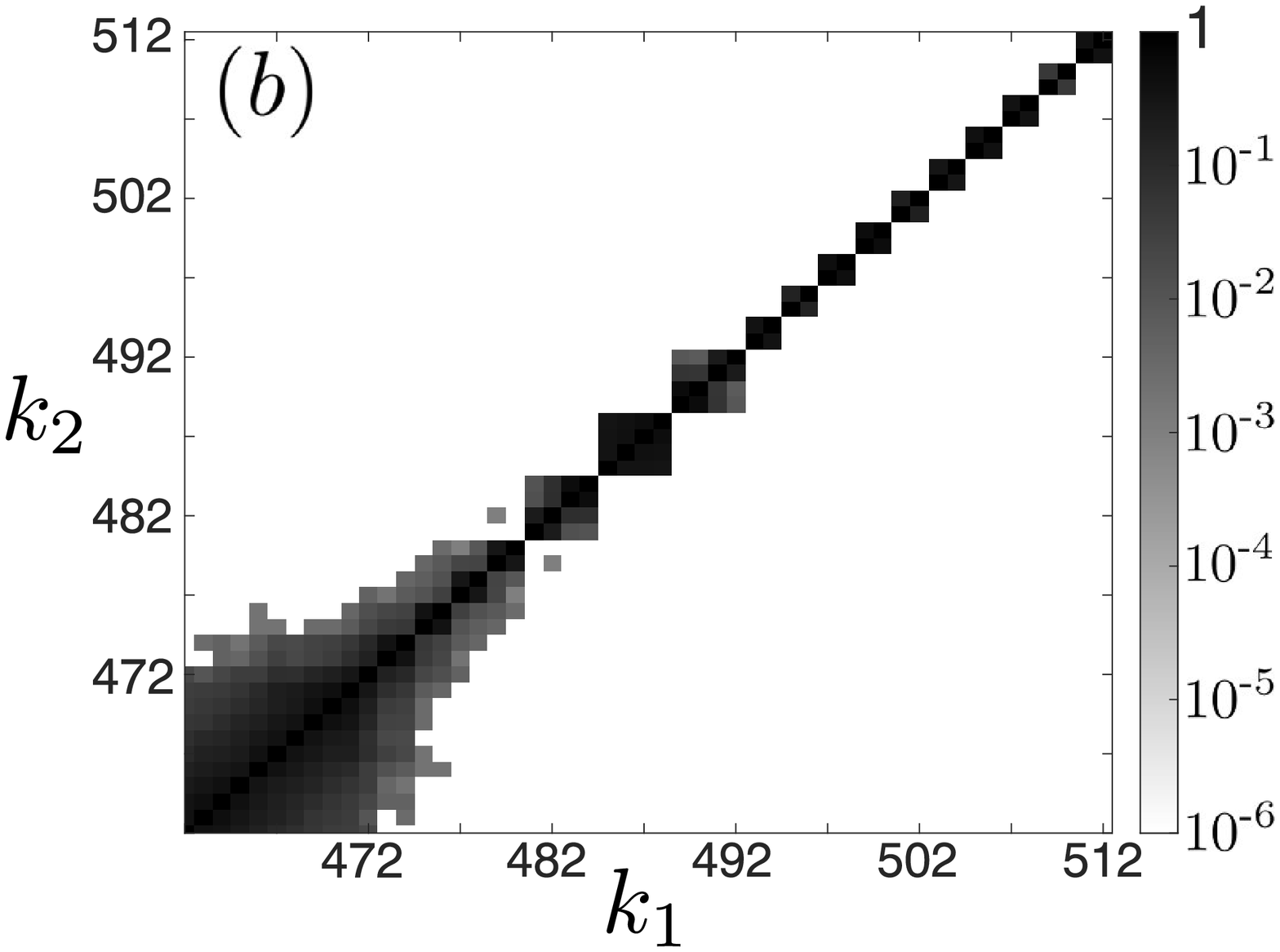}
\end{center}
\caption{Violations of the DOS for four-band chaos without a conservation law ($N \!=\!512$, $\beta\!=\!0$, $a\!=\!1.1$) indicating 480 physical modes ($D_\text{ph} \!=\! 480$) and 32 transient modes. Contours of the violations $\nu_{k_1,k_2}^\tau$, defined in Eq.~(\ref{eq:vDOS}), are plotted on a log scale where dark regions represent violations and light regions represent the absence of violations. $k_1$ and $k_2$ are CLV indices. (a)~The entire lattice. (b)~A close-up of the violations for the last 50 CLVs.} 
\label{fig:vdos-a1p1-beta0}
\end{figure}

The violations of the DOS for the lattice of diffusively coupled tent maps in the absence of a conservation law is shown in Fig.~\ref{fig:vdos-a1p1-beta0}. The results are shown using a log scale where dark regions indicate violations and light regions indicate the absence of violations as indicated by the color bars. The diagonal from the lower left to the upper right is black and represents that each CLV, when compared with itself, yields pure violation. The figure is symmetric about the diagonal because $\nu_{k_1,k_2}^\tau$ represents the pairwise comparison between two CLVs and is independent of the order of the indices. The region above the diagonal is obtained by reversing the order of indices in Eqs.~(\ref{eq:deltalambda})-(\ref{eq:vDOS}).

The violations of the DOS for the entire lattice are shown in Fig.~\ref{fig:vdos-a1p1-beta0}(a). The dark regions represent significant violations which indicate that the two CLVs being compared,  given by $k_1$ and $k_2$, have frequent near tangencies and can be thought of as entangled or coupled CLVs. The large band of entangled CLVs, due to their near tangencies, have been called the physical modes~\cite{yang:2009,takeuchi:2011}.

A close-up of the violations among the final 50 CLVs is shown in Fig.~\ref{fig:vdos-a1p1-beta0}(b). This reveals a transition from the physical modes to isolated modes that occur in groups of 4 and 2 modes. These isolated modes are referred to as the transient modes. In Fig.~\ref{fig:vdos-a1p1-beta0} there are 480 physical modes and 32 isolated modes. It has been suggested~\cite{yang:2009} that the number of physical modes yields a measure of the dimension of the dynamics, called the physical dimension $D_\text{ph}$, which may provide an estimate for the dimension of the inertial manifold.

For this lattice of diffusively coupled tent maps without a conservation law, our results indicate $D_\text{ph} \!=\! 480$. Values of $D_\text{ph}$ for lattices of varying size without a conservation law are shown in Fig.~\ref{fig:dimension-beta0} using the square symbols. $D_\text{ph}$ is over five times larger than the fractal dimension, $D_\lambda \!=\! 83.2$. It is clear that the physical dimension, for these results, is nearly equal to the system size $N$ which represents the maximum possible value for the dimension of these one-dimensional lattices as represented by the gray region in Fig.~\ref{fig:dimension-beta0}.

\begin{figure}[h!]
\begin{center}
\includegraphics[width=1.65in]{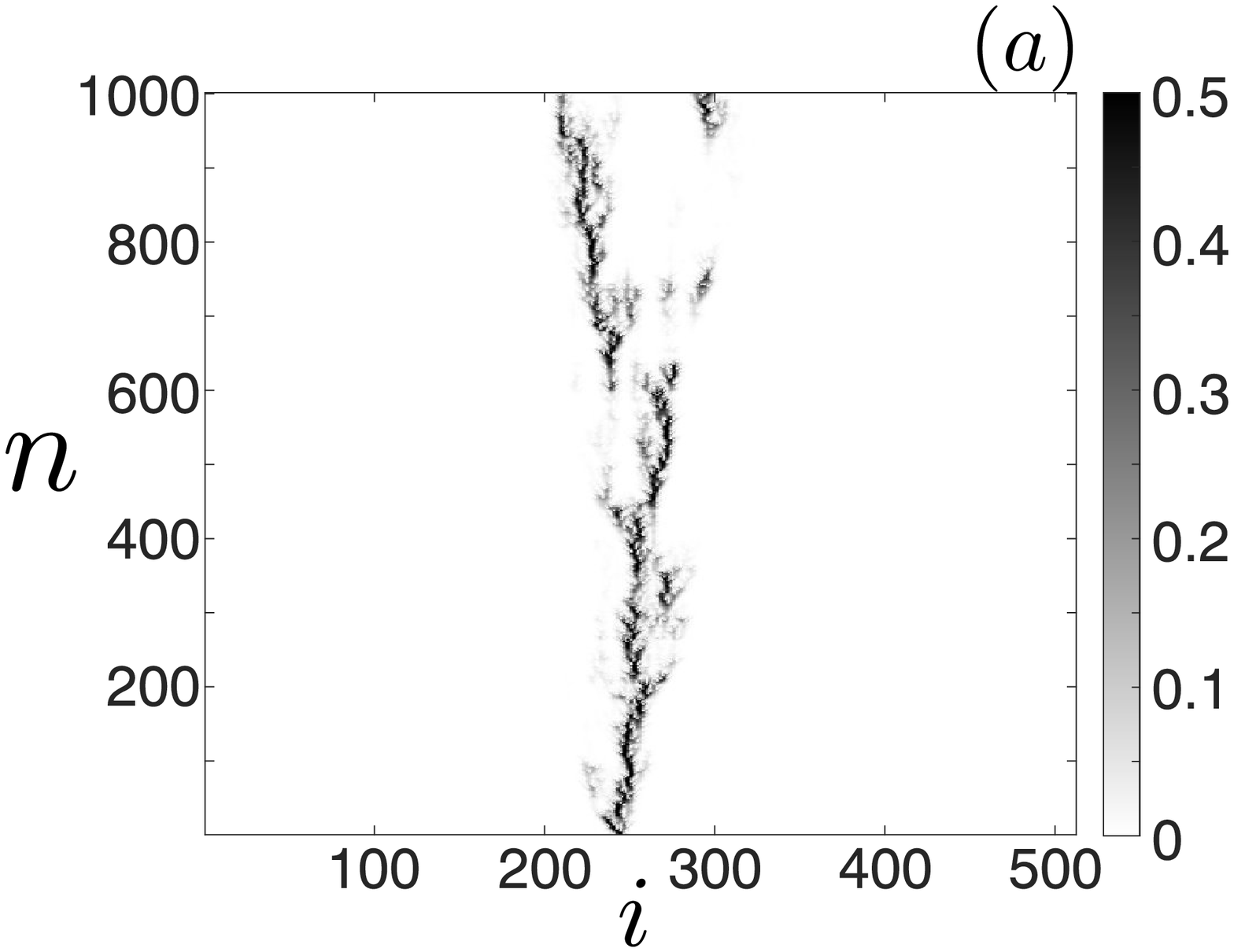}
\includegraphics[width=1.65in]{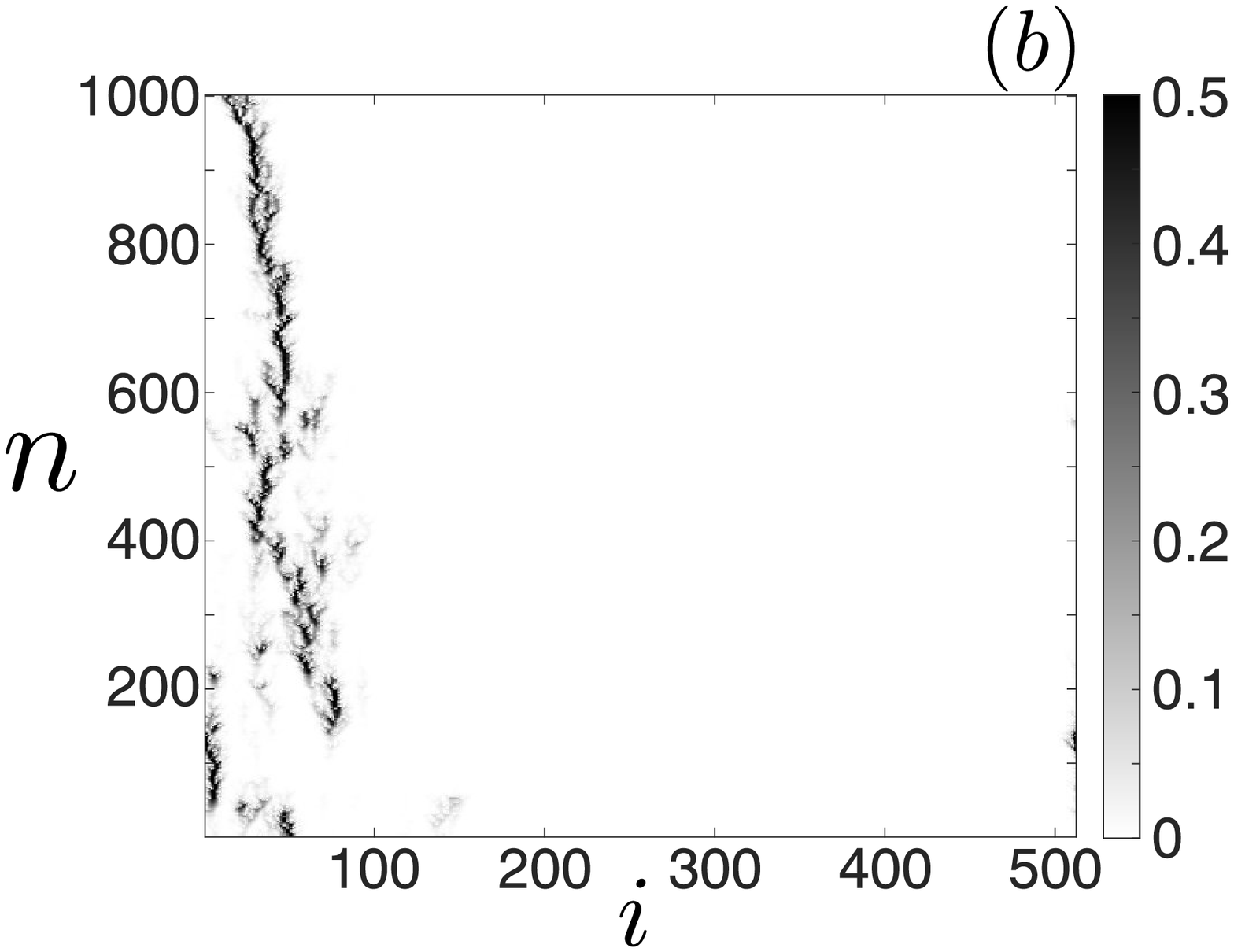} \\ 
\includegraphics[width=1.65in]{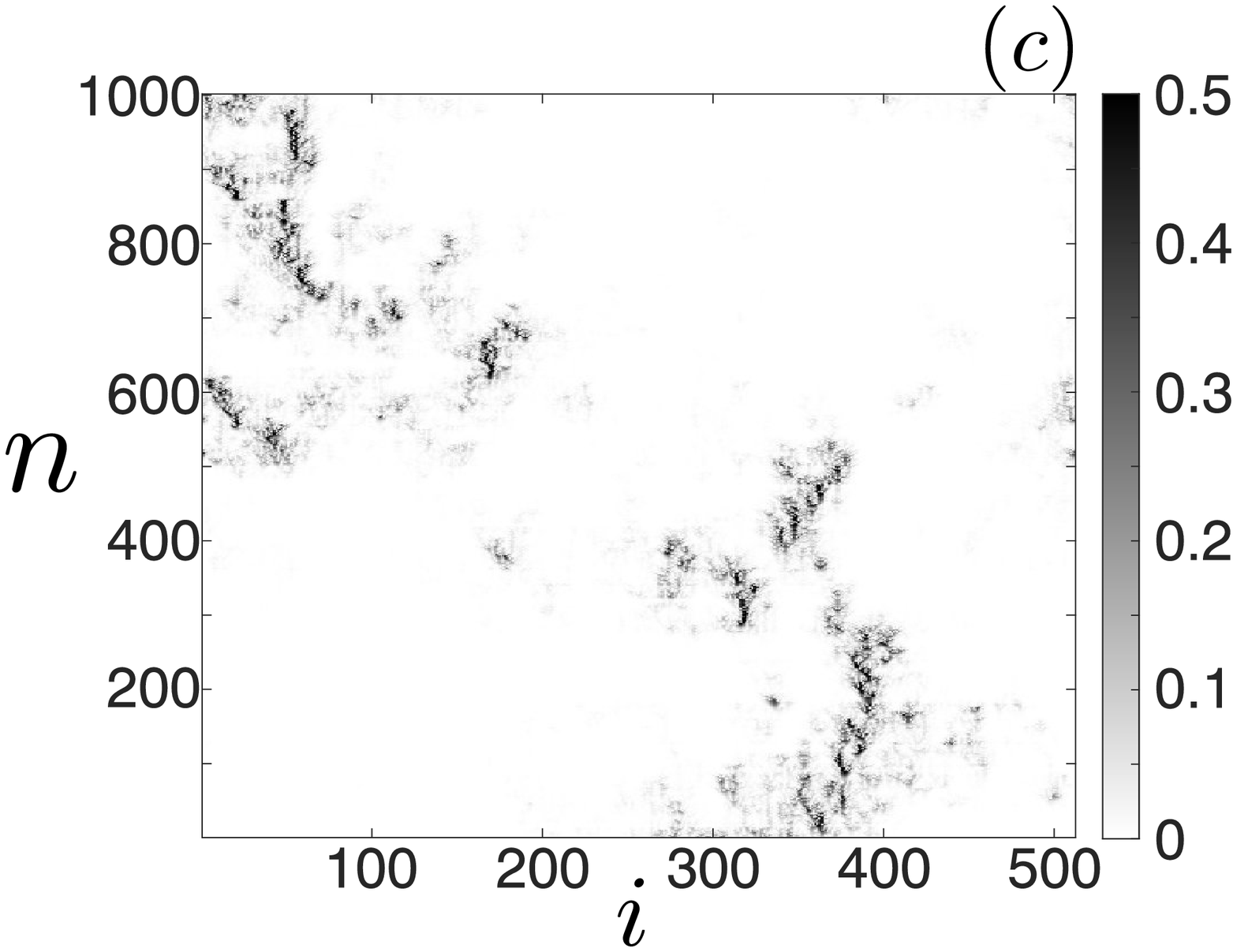} 
\includegraphics[width=1.65in]{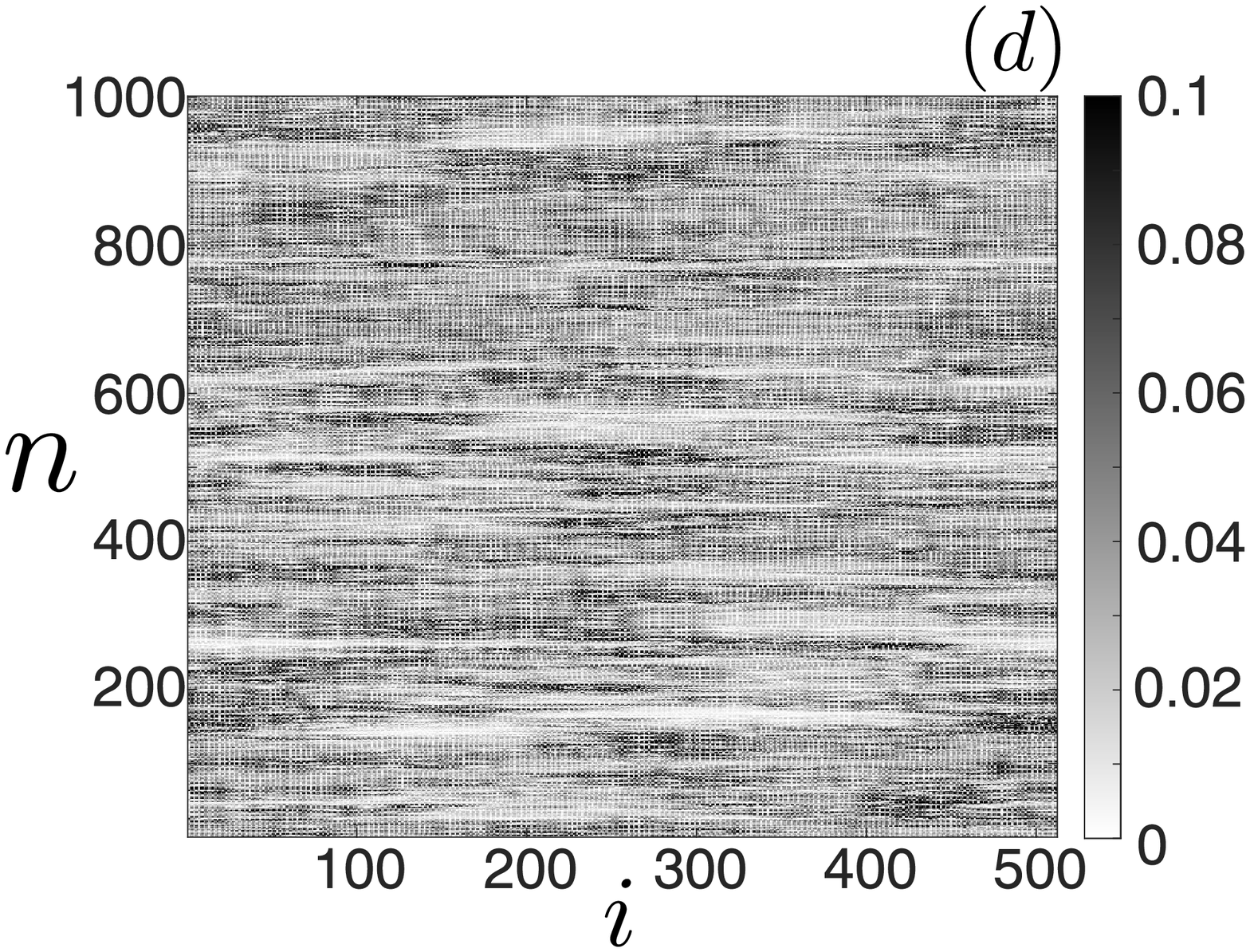}
\end{center}
\caption{Space-time plots of the CLVs of homogeneous chaos without a conservation law ($N \!=\!512$, $\beta\!=\!0$, $a\!=\!1.6$).  $|\vec{v}_k^{\,(n)}|$ is shown for: (a)~$k\!=\!1$, (b)~$k\!=\!10$, (c)~$k\!=\!100$, (d)~$k\!=\!500$. The gray scale range for~(d) has been reduced to visualize the delocalized spatial variation.} 
\label{fig:clvs-a1p6-beta0}
\end{figure}

We now explore the dynamics of a lattice where each individual map has the control parameter set to $a\!=\!1.6$. In this case, this is a lattice of diffusively coupled maps without a conservation law where each individual maps would exhibit homogeneous chaos if isolated. The dynamics of this lattice are shown in Fig.~\ref{fig:tent-map-lattice2}.  Space-time plots of the CLVs are shown in Fig.~\ref{fig:clvs-a1p6-beta0} for (a)~$k\!=\!1$, (b)~ $k\!=\!10$, (c)~$k\!=\!100$, and (d)~$k\!=\!500$ using the same conventions of Fig.~\ref{fig:clvs-a1p1-beta0}.

The spatiotemporal dynamics of the leading CLV are shown in Fig.~\ref{fig:clvs-a1p6-beta0}(a). The leading CLV is highly localized yet the spatial variation is larger than what was found in Fig.~\ref{fig:clvs-a1p1-beta0}(a).  Despite its localized nature, the leading CLV shown in Fig.~\ref{fig:clvs-a1p6-beta0}(a), is fractured with the presence of branching structures. As the CLV index increases in Fig.~\ref{fig:clvs-a1p6-beta0}(b)-(d), the CLVs again delocalize with a very homogeneous spatial variation for the 500th CLV shown in panel~(d).

\begin{figure}[h!]
\begin{center}
\includegraphics[width=2.25in]{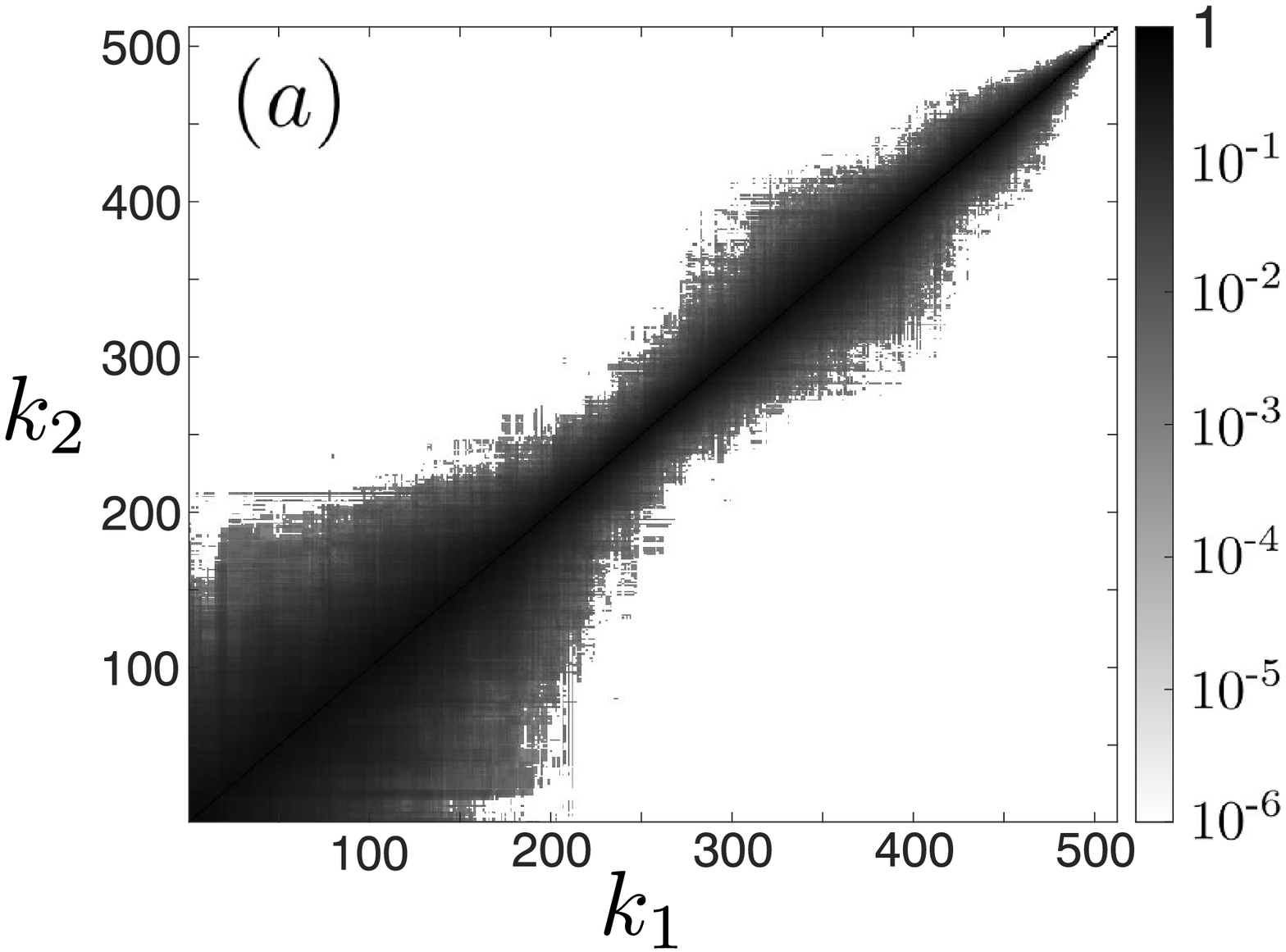} \\ 
\includegraphics[width=2.25in]{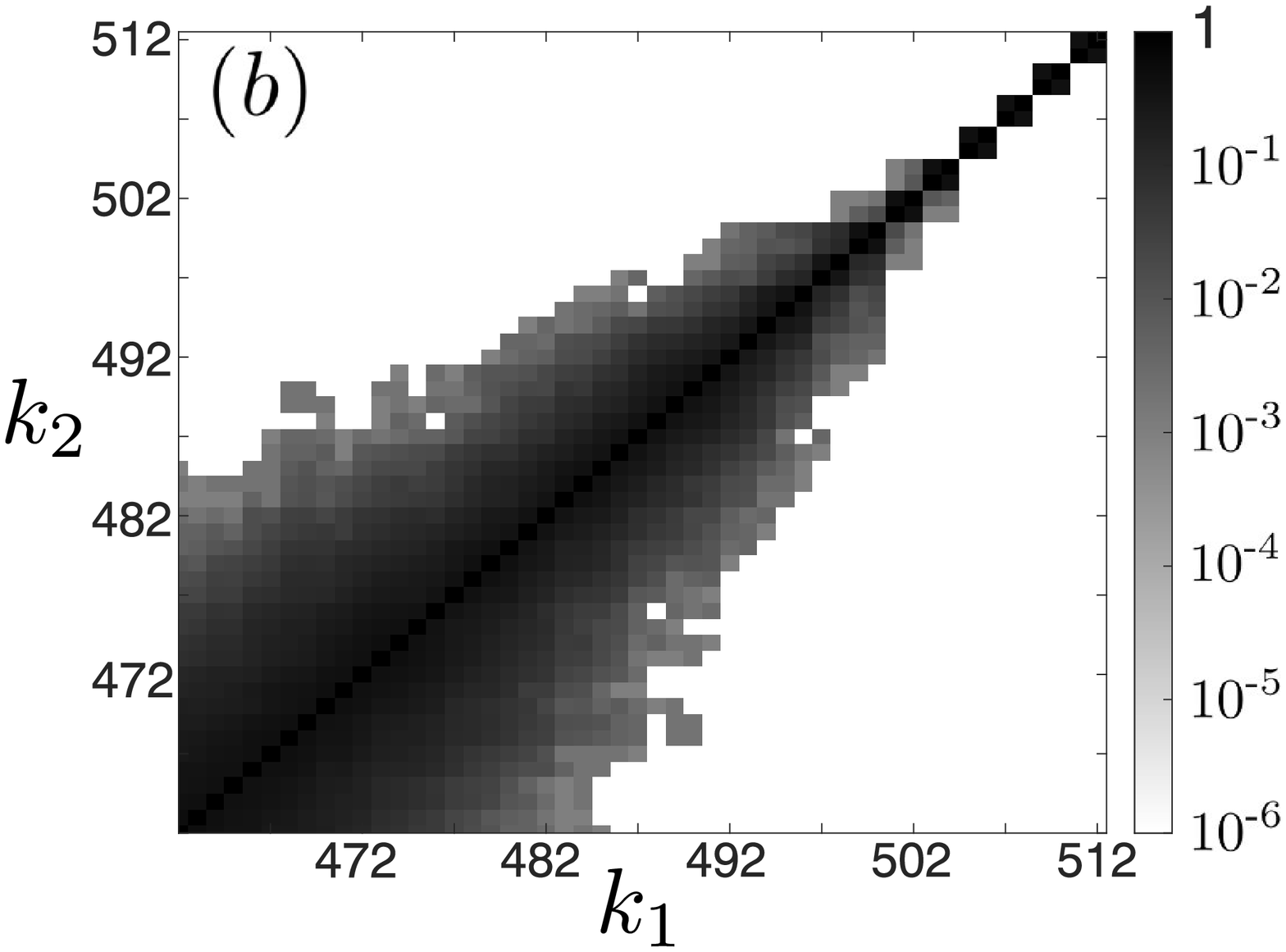}
\end{center}
\caption{Violations of the dominance of Oseledets splitting for homogeneous chaos without a conservation law ($N \!=\!512$, $\beta\!=\!0$, $a\!=\!1.6$) indicating $D_{\text{ph}}=504$ and 8 transient modes. (a)~The entire lattice. (b) A close-up of the last 50 CLVs.} 
\label{fig:vdos-a1p6-beta0}
\end{figure}

The violations of the DOS for this lattice are shown in Fig.~\ref{fig:vdos-a1p6-beta0}. Again there is a large band of entangled physical modes which eventually transitions to transient modes at large values of the index. The transition between the physical and transient modes is evident in the close-up of the violations for the final 50 CLVs shown in Fig.~\ref{fig:vdos-a1p6-beta0}(b). A close inspection reveals 504 physical modes and 8 transient modes which all occur in pairs. As expected, the increased value of the control parameter $a$ has resulted in an increase in the number of physical modes when compared with Fig.~\ref{fig:vdos-a1p1-beta0}. Even for this increased value of $a$ with more complex dynamics, the splitting of the tangent space into physical and transient modes remains.

\subsection{Chaotic Dynamics with a Conservation Law}
\label{section:with-conservation-law}

We explore the dynamics of diffusively coupled maps with a conservation law by setting $\beta \!=\!1$ in Eq.~(\ref{eq:cml-full}). We investigate how the conservation law affects the dynamics, the splitting of the tangent space, and the spatiotemporal features of the CLVs. 

The synchronous conservation law reduces the number of degrees of freedom of the system by one. As a result, the total number of CLVs comprising the spectrum is $N\!-\!1$. Therefore, when the conservation law is present we use $N_\lambda \!=\! N \!-\! 1$ when computing the full spectrum of CLVs. 

\begin{figure}[h!]
\begin{center}
\includegraphics[width=1.6in]{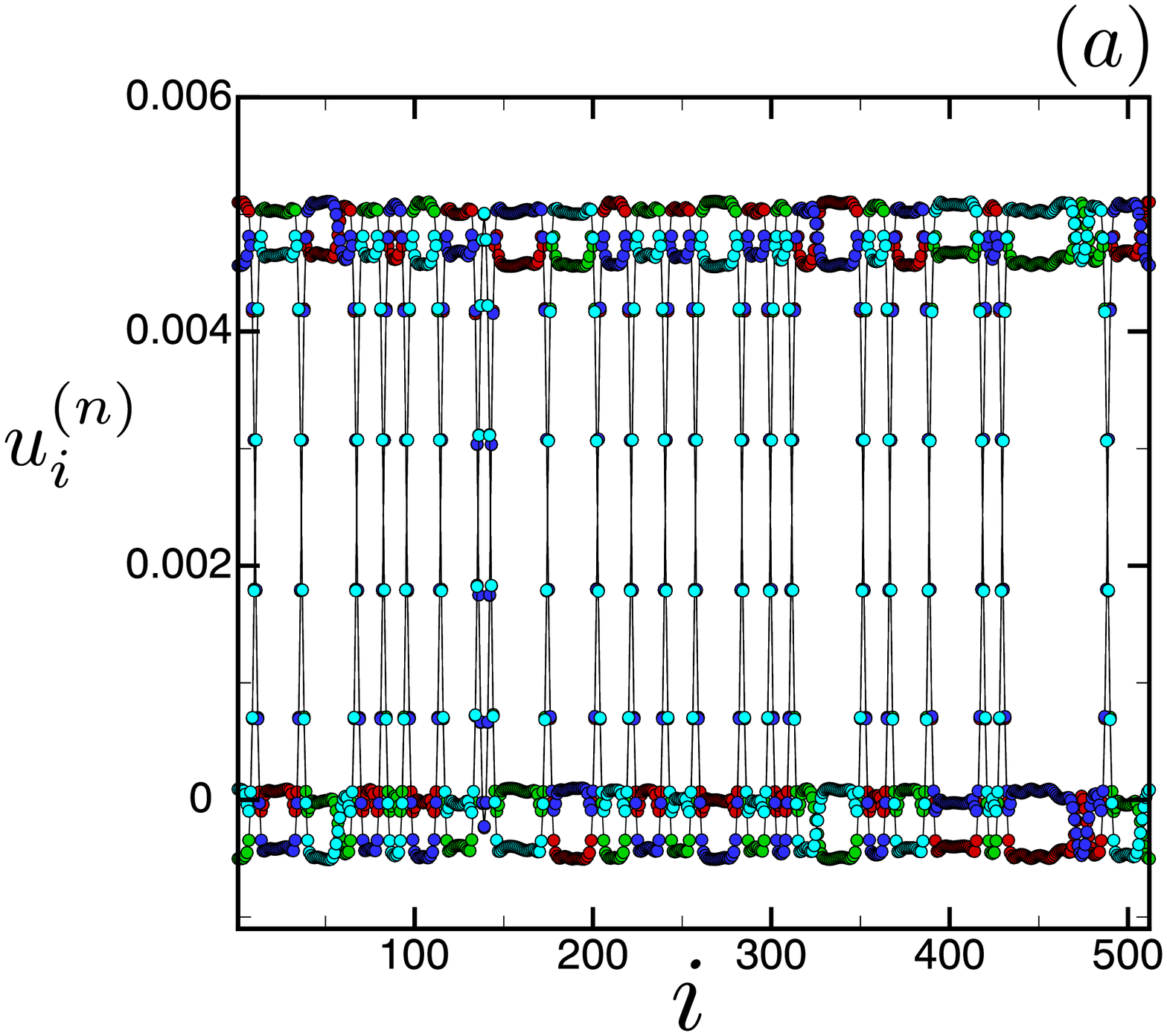} \hspace{-0.1cm}
\includegraphics[width=1.75in]{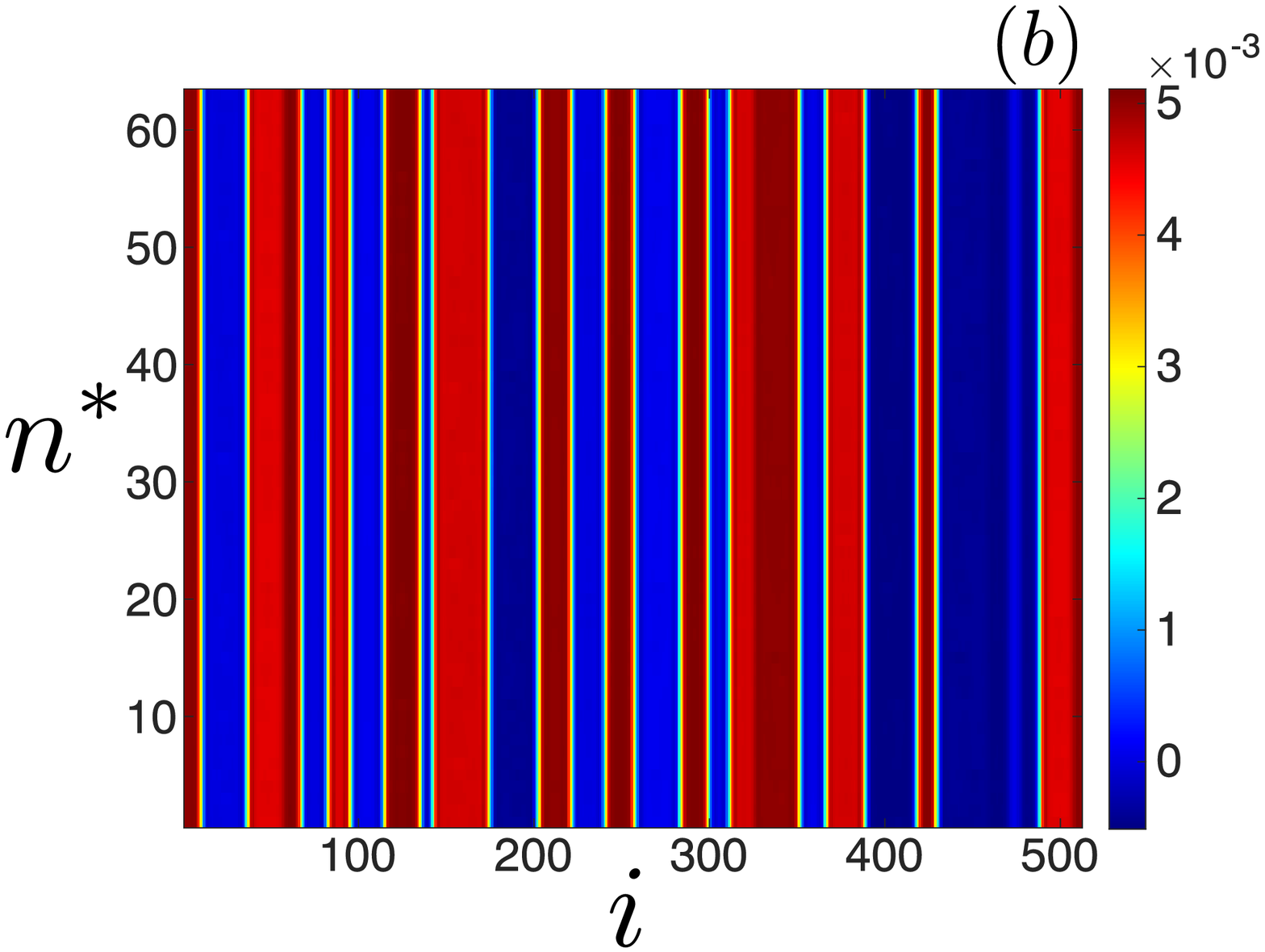}
\end{center}
\caption{The dynamics of diffusively coupled maps in the four-band chaos regime with a conservation law ($N \!=\!512$, $\beta\!=\!1$, $a\!=\!1.1$). (a) The values of the lattice sites at four consecutive time steps which are color coded in the order red, blue, green, and cyan. (b)~A space-time plot of the lattice for every 16 time steps ($n\!=\!16 n^*$). By showing every $16$th time step the regions of different magnitude (red and blue) and the immobile barriers between them (green) are evident.}
\label{fig:tent-map-lattice3}
\end{figure}

The dynamics of a diffusively coupled lattice in the four-band chaos regime with a conservation law is shown in Fig.~\ref{fig:tent-map-lattice3}. In Fig.~\ref{fig:tent-map-lattice3}(a) the state of the maps are shown at four consecutive time steps which have been color coded in the sequence: red, blue, green, and cyan. The four band structure remains yet there are several new interesting features. The four bands are now much closer together as indicated by the values of the maps which are now within the range $-5 \!\times\! 10^{-4} \!\lesssim\! u_i \!\lesssim\! 5 \!\times\! 10^{-2}$ as opposed to $-0.1 \!\lesssim\! u_i \!\lesssim\! 1$ in the absence of the conservation law. In addition, the lattice values with a conservation law now include kinks, or barriers, between the lattice values within each band. These defect structures were not present in the absence of a conservation law as shown in Fig.~\ref{fig:tent-map-lattice1}. The banded nature of the dynamics is also evident by the striped form of the space-time plot shown in Fig.~\ref{fig:tent-map-lattice3}(b).

\begin{figure}[h!]
\begin{center}
\includegraphics[width=3in]{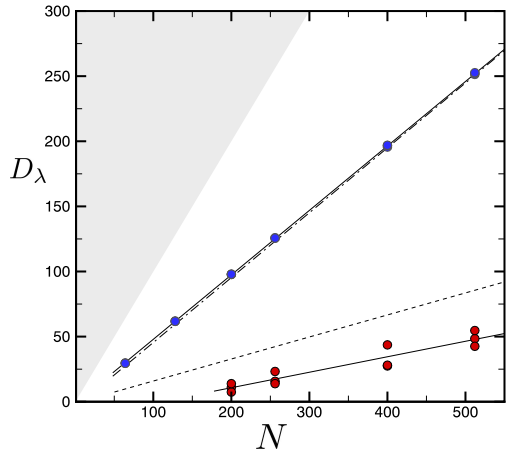}
\end{center}
\caption{The influence of the conservation law on the variation of $D_\lambda$ with $N$. Circles are diffusively coupled maps with a conservation law ($\beta\!=\!1$). Lower circles (red) are for four band chaos ($a\!=\!1.1$). Dashed line is average result without conservation law. For four-band chaos the conservation law significantly reduces $D_\lambda$. Upper circles (blue) are for homogeneneous chaos ($a\!=\!1.6$), dash-dotted line is average result without a conservation law. $D_\lambda$ does not vary significantly with a conservation law for homogeneous chaos. Solid lines are linear fits and the gray region indicates inaccessible values. Results are shown using three different random initial conditions for $a=1.1$ and $a=1.6$ at each $N$.} 
\label{fig:dlambda-comparison-with-beta}
\end{figure}

Insight into the complexity of the dynamics is given by the magnitude of the fractal dimension. The variation of $D_\lambda$ with $\beta$ is shown in Fig.~\ref{fig:dlambda-comparison-with-beta} as a function of system size $N$. The diffusively coupled lattice with a conservation law is shown by the circles (red, lower). Chaotic dynamics were only found for $N \! \ge \! 200$ in our simulations. We did not attempt to quantify precisely the system size where this transition occurs. The solid black line through the red circles (lower) is a linear fit through the data points indicating extensive chaos. The dashed line is the variation of $D_\lambda$ for the four band chaos case without a conservation law from Fig.~\ref{fig:dlambda-comparison-with-beta} for reference. The fractal dimension has decreased by nearly a factor of two due to the conservation law.

\begin{figure}[h!]
\begin{center}
\includegraphics[width=3in]{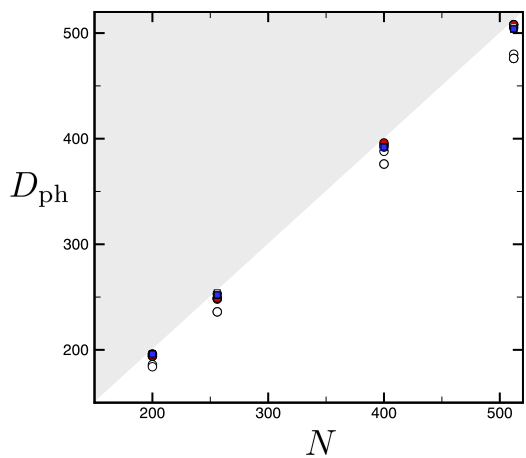}
\end{center}
\caption{The influence of the conservation law on the variation of $D_\text{ph}$ with $N$: circles (red) $a \! = \! 1.1$, $\beta \! = \! 1$; squares (blue) $a \! = \! 1.6$, $\beta \! = \! 1$; circles (open) $a \! = \! 1.1$, $\beta \! = \! 0$; squares (open) $a \! = \! 1.6$, $\beta \! = \! 0$. For each case, results for three different random initial conditions are shown.}
\label{fig:dphysical-comparison-with-beta}
\end{figure}

The variation of $D_\text{ph}$ with $N$ is shown in Fig.~\ref{fig:dphysical-comparison-with-beta}. The results for the four band chaos case with the conservation law are shown as red circles and, for reference, $D_\text{ph}$  for $\beta\!=\!0$ is shown as open circles.  The physical dimension does not change significantly due to the conservation law where the number of physical modes is consistently near the total number of degrees of freedom. 

\begin{figure}[h!]
\begin{center}
\includegraphics[width=1.65in]{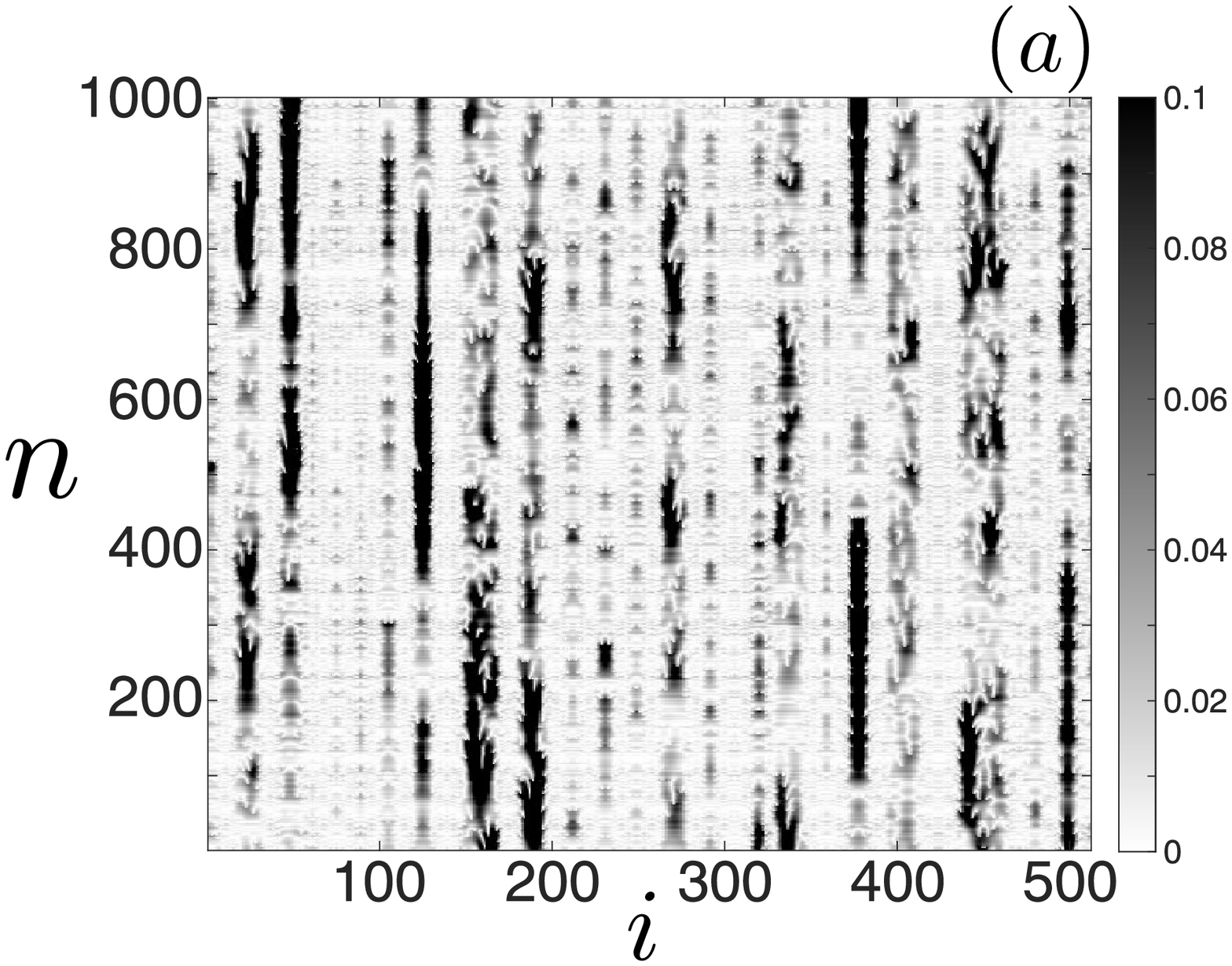}
\includegraphics[width=1.65in]{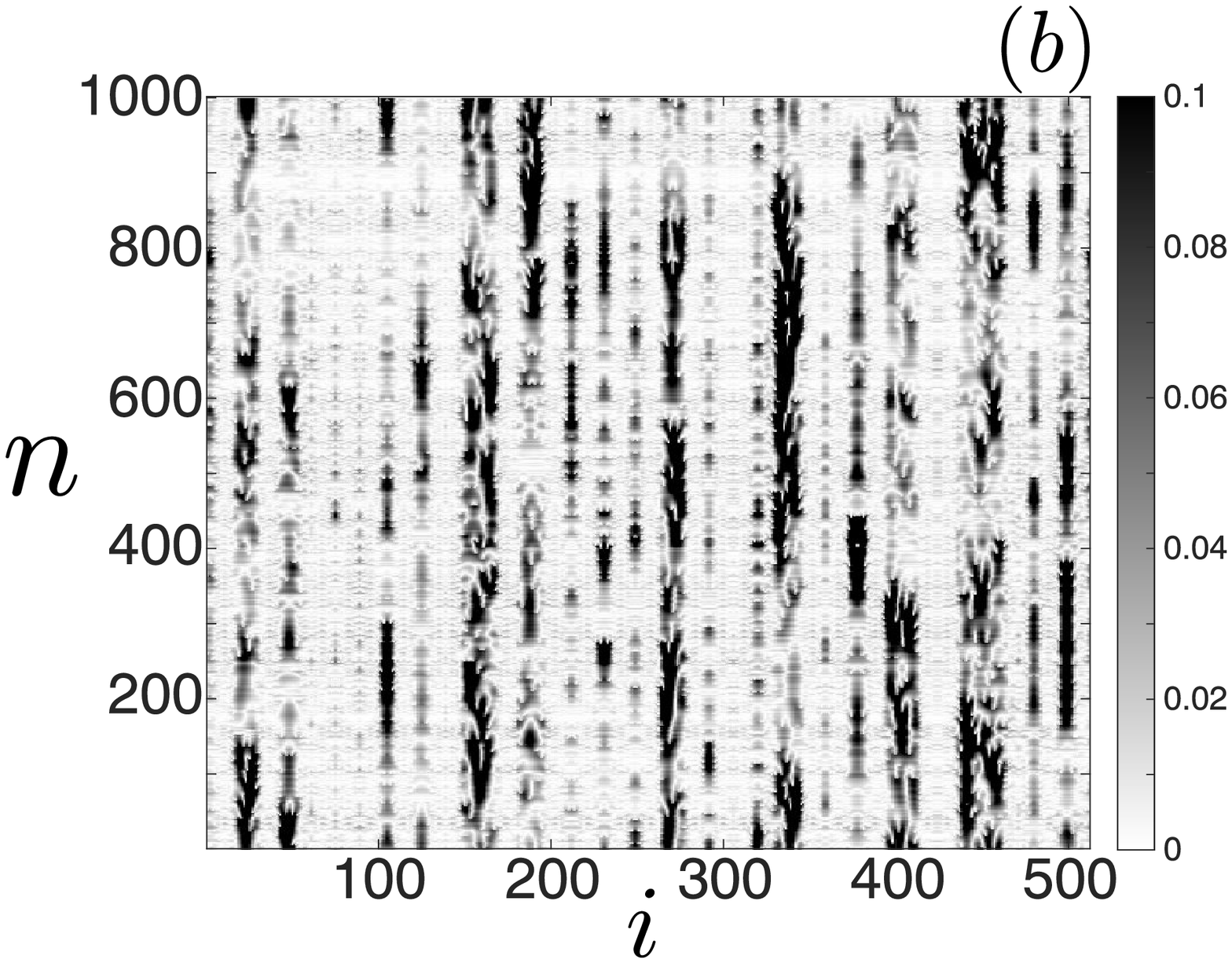} \\
\includegraphics[width=1.65in]{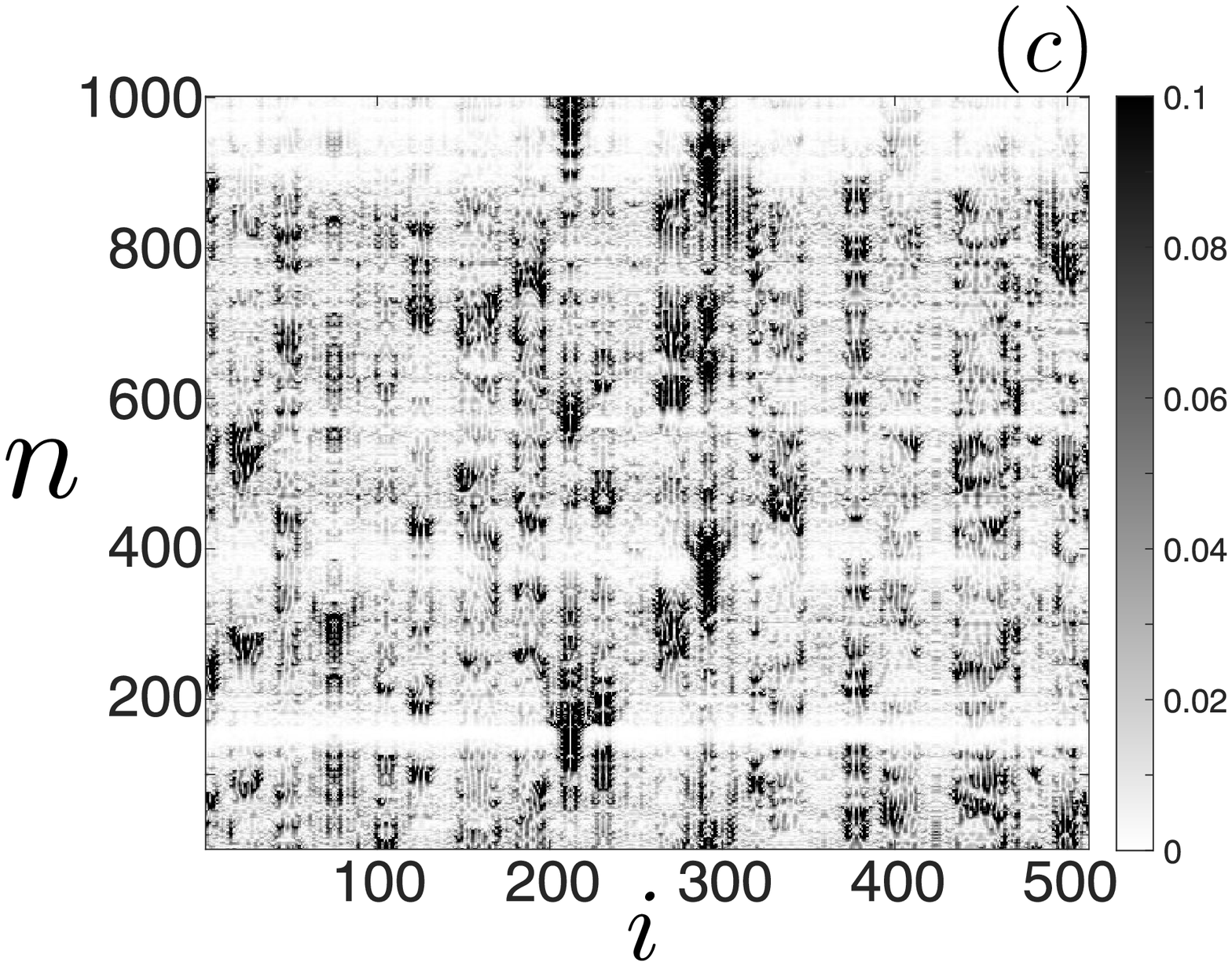} 
\includegraphics[width=1.65in]{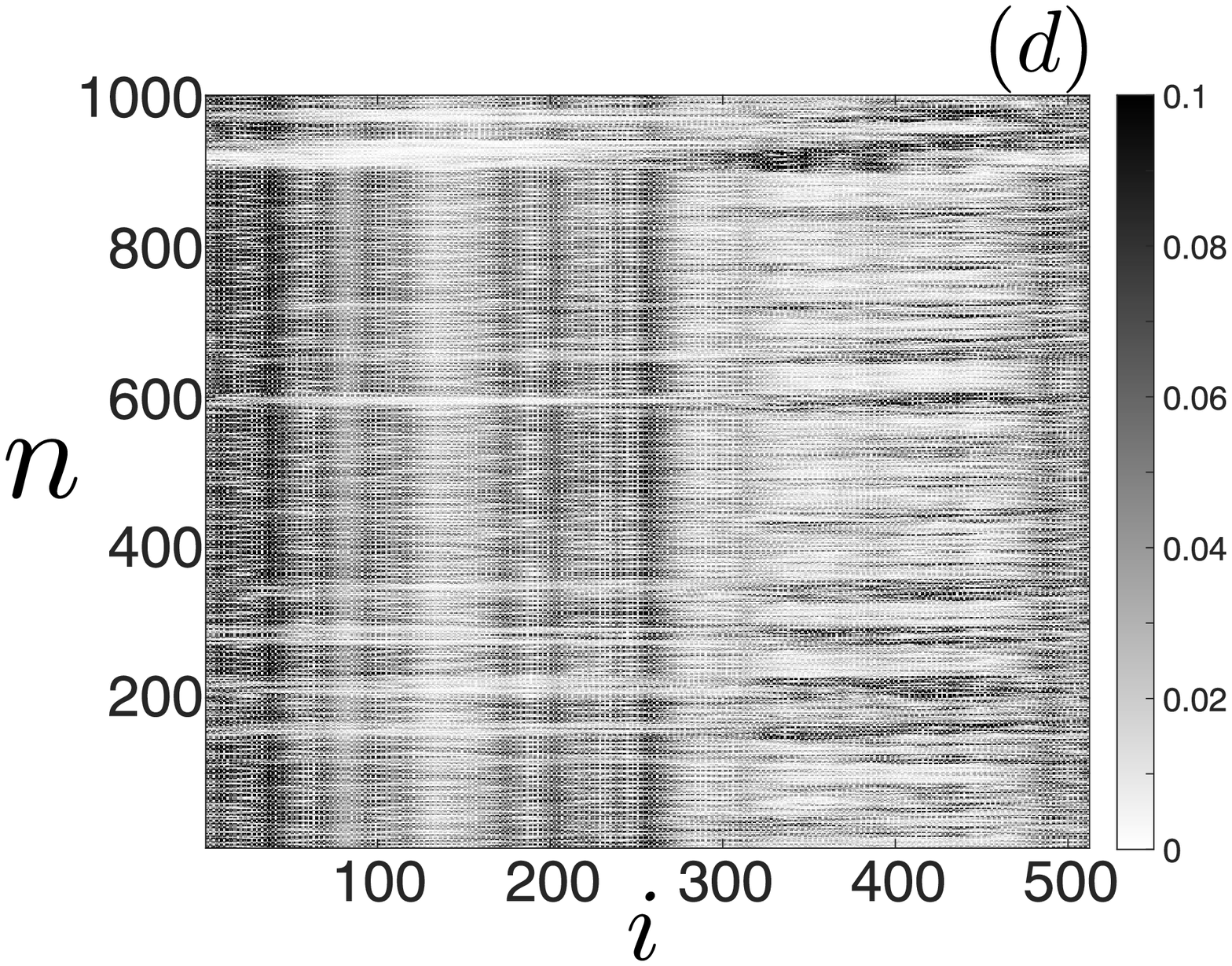}
\end{center}
\caption{Space-time plots of the CLVs of four-band chaos with a conservation law ($N \!=\!512$, $\beta\!=\!1$, $a\!=\!1.1$).  $|\vec{v}_k^{\,(n)}|$ is shown for: (a)~$k\!=\!1$, (b)~$k\!=\!10$, (c)~$k\!=\!100$, (d)~$k\!=\!500$.} 
\label{fig:clvs-a1p1-beta1}
\end{figure}

Space-time plots of the CLVs are shown in Fig.~\ref{fig:clvs-a1p1-beta1} for diffusively coupled maps in the four band chaos regime with a conservation law.  The leading CLV is shown in Fig.~\ref{fig:clvs-a1p1-beta1}(a) which shows a striking difference when compared to the $\beta\!=\!0$ result shown in Fig.~\ref{fig:clvs-a1p1-beta0}(a). The leading CLV exhibits significant delocalization when the conservation law is present. This indicates the presence of many locations along the lattice, at any time $n$, where small perturbations would grow rapidly. The delocalization of the CLVs increases with increasing CLV index as illustrated in Fig.~\ref{fig:clvs-a1p1-beta1}(b)-(d).

\begin{figure}[h!]
\begin{center}
\includegraphics[width=2.25in]{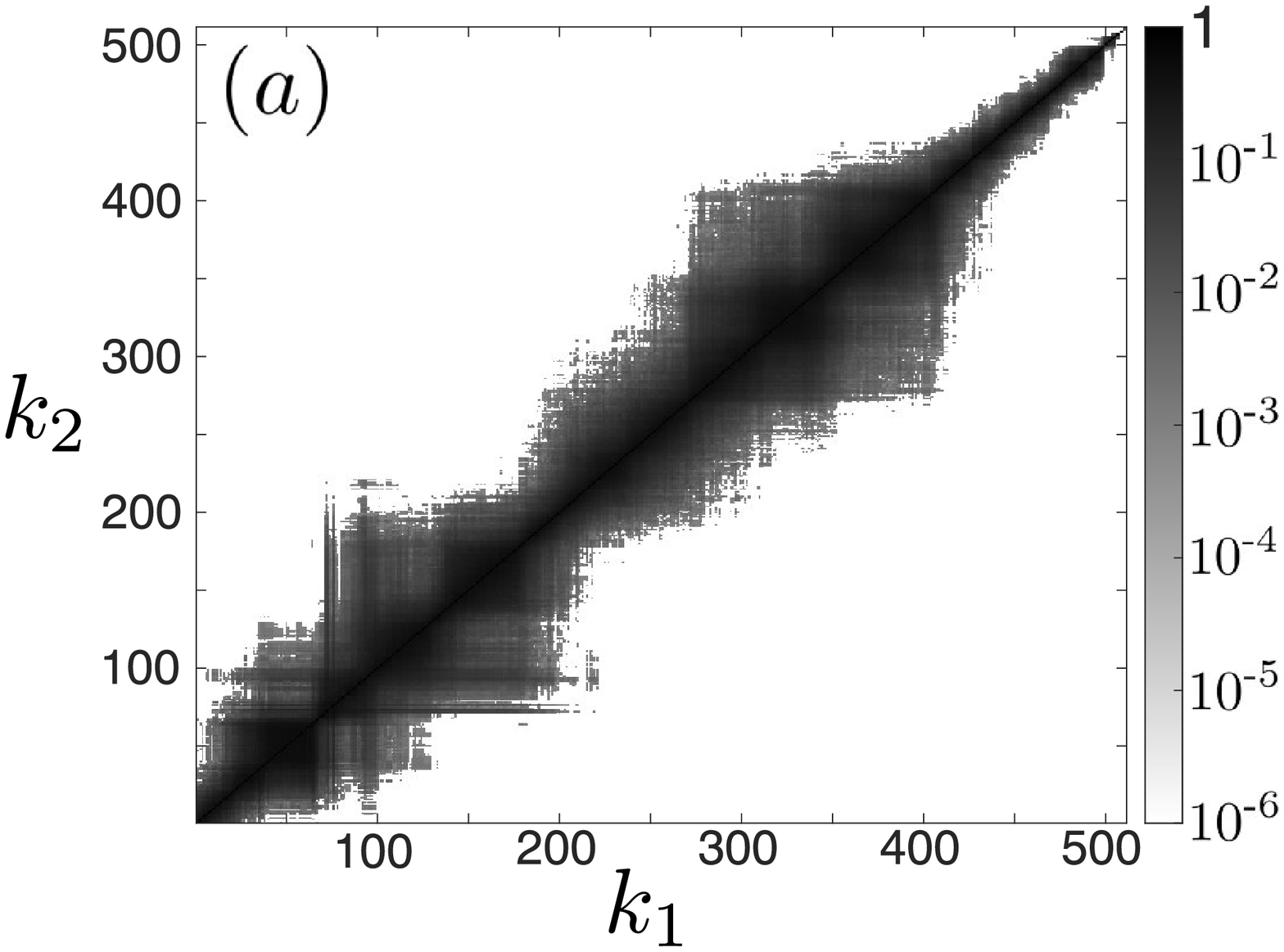}
\includegraphics[width=2.25in]{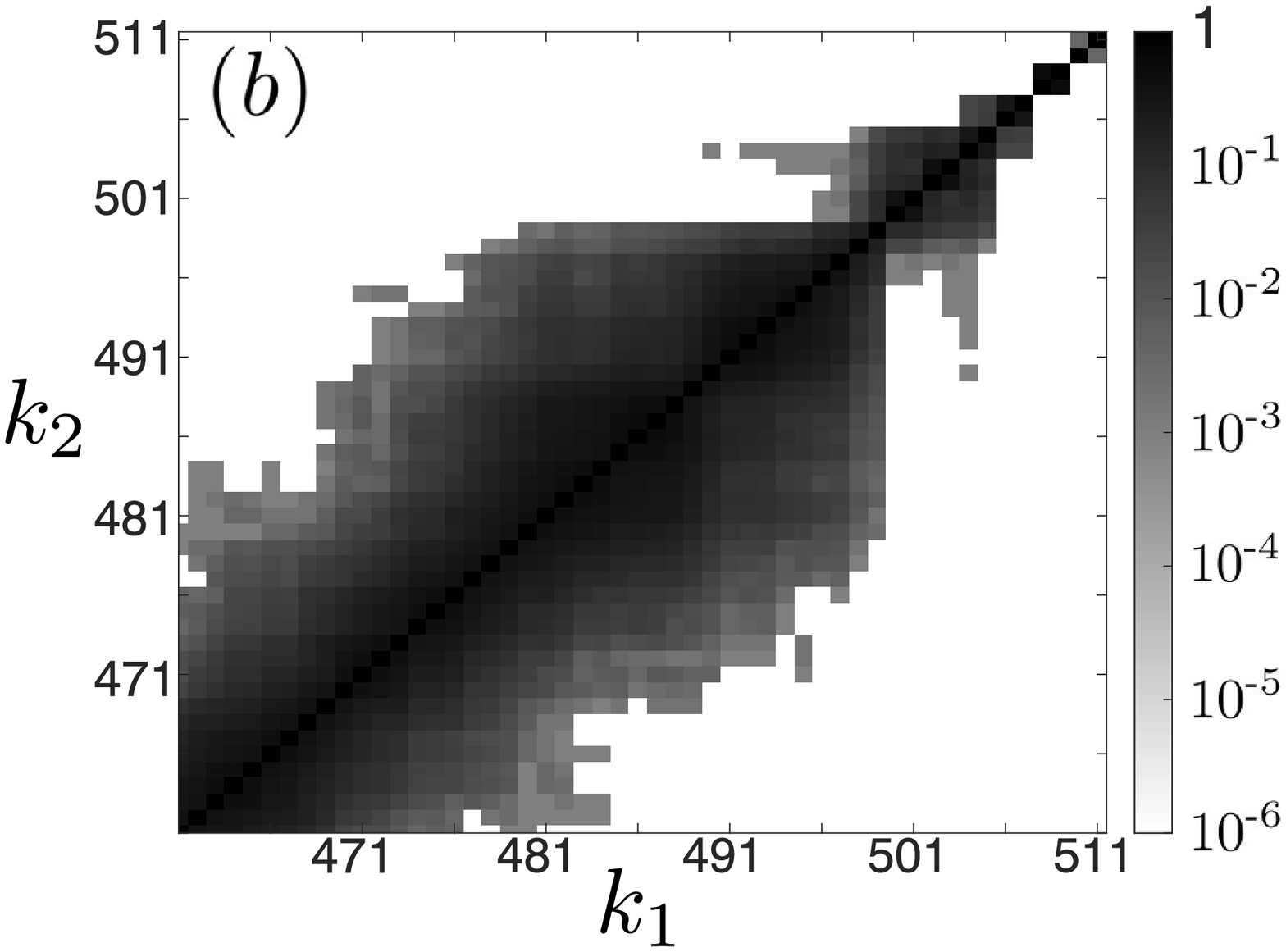}
\end{center}
\caption{Violations of the DOS for four-band chaos with a conservation law ($N \!=\!512$, $\beta\!=\!1$, $a\!=\!1.1$) indicating $D_\text{ph} \!=\! 507$ and 4 transient modes. Dark regions represent violations and light regions represent the absence of violations. (a)~Entire lattice. (b)~Close-up of the last 50 CLVs.} 
\label{fig:vdos-a1p1-beta1}
\end{figure}

The violation of the DOS are shown in Fig.~\ref{fig:vdos-a1p1-beta1} for the four band chaos case with a conservation law. There is a large band of entangled physical modes that are followed by the transient modes. A closer inspection reveals $D_\text{ph} \!=\! 507$ with 4 transient modes.

\begin{figure}[h!]
\begin{center}
\includegraphics[width=1.6in]{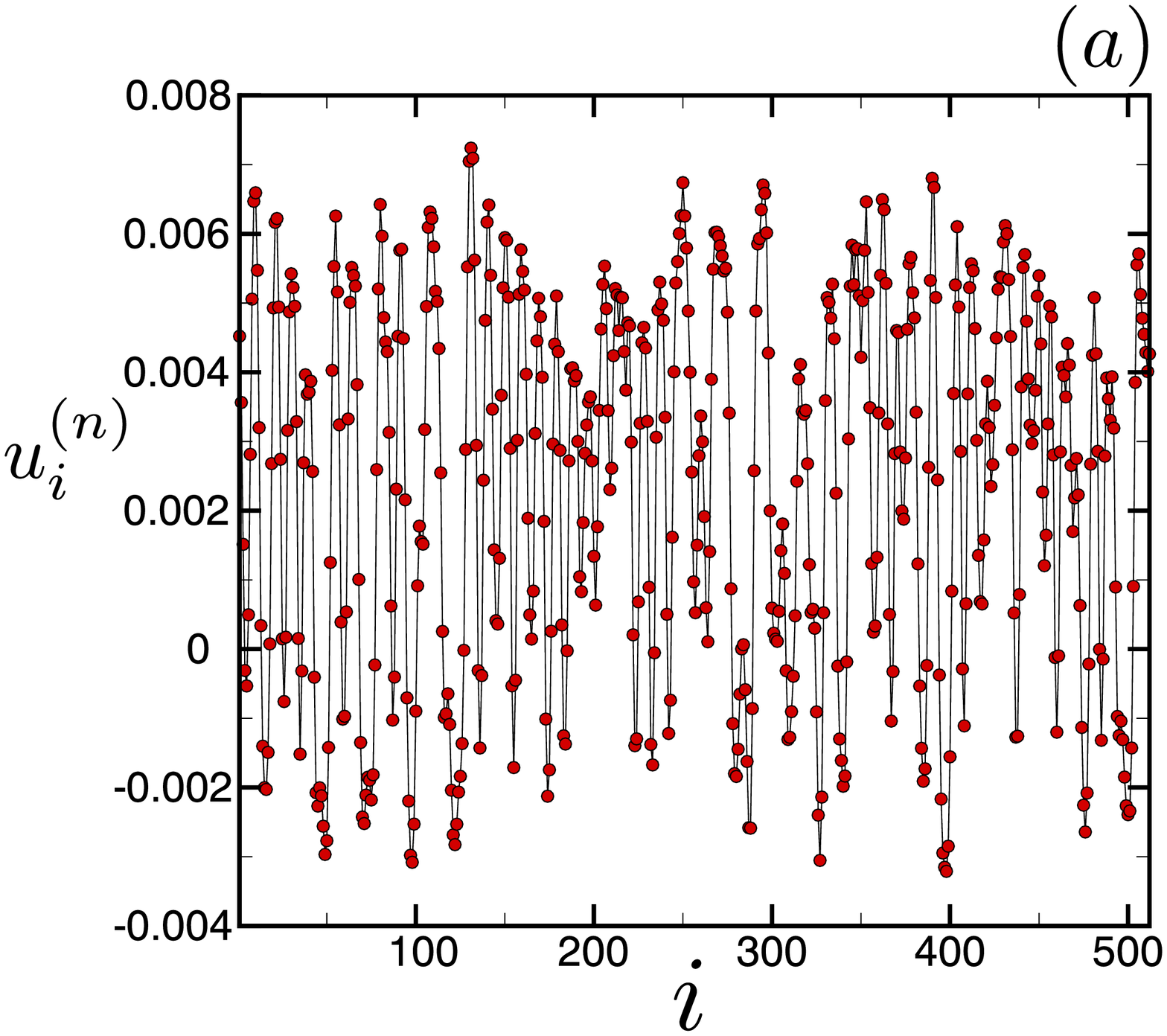} \hspace{-0.1cm}
\includegraphics[width=1.75in]{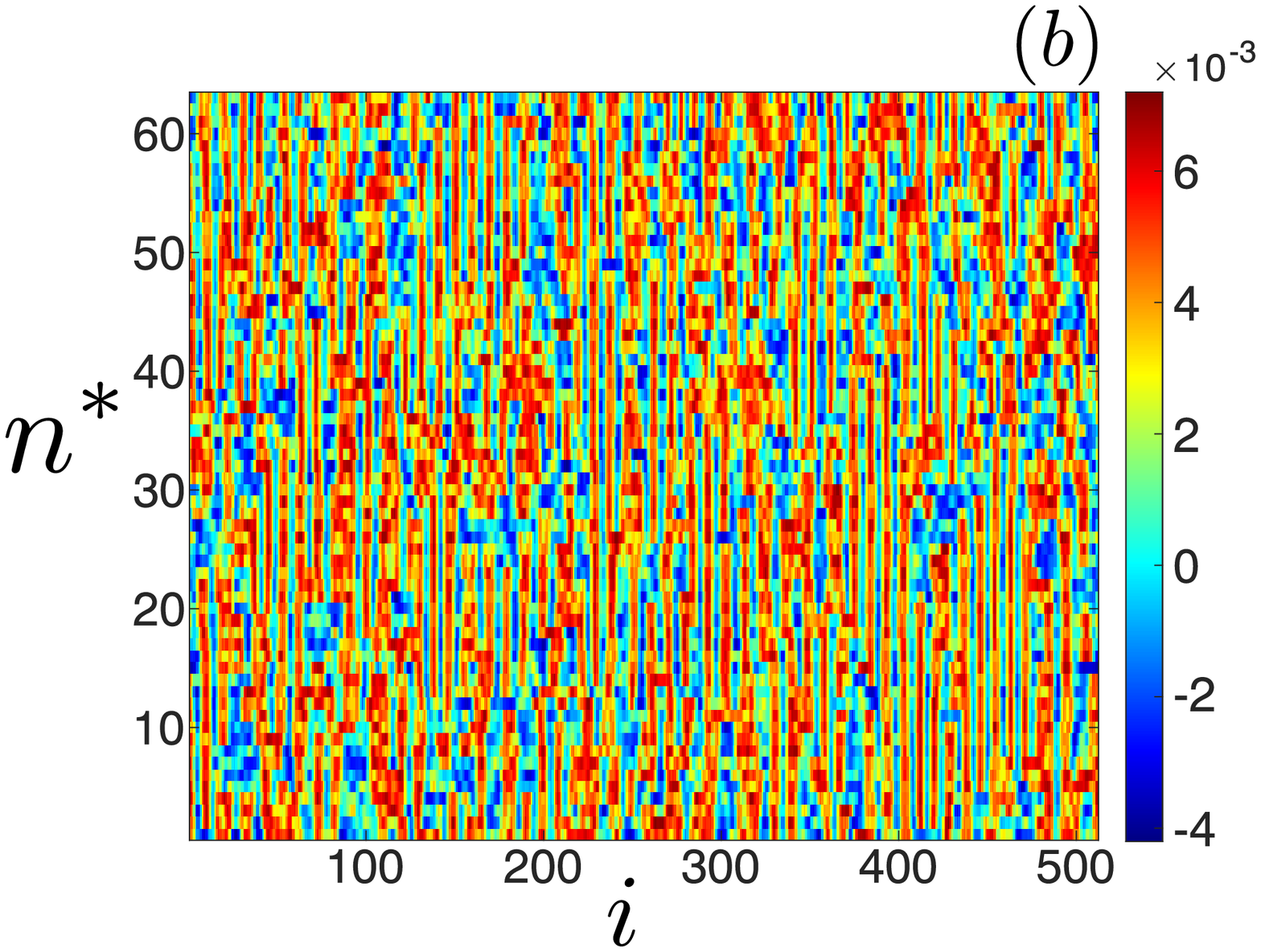}
\end{center}
\caption{The dynamics of diffusively coupled maps in the homogeneous chaos regime with a conservation law ($N\!=\!512$, $\beta\!=\!1$, $a\!=\!1.6$). (a)~Lattice values at one instant of time. (b)~Space-time plot of the dynamics showing every 16th time step ($n\!=\!16n^*$).}
\label{fig:tent-map-lattice4}
\end{figure}

We now investigate how the dynamics and CLVs in the homogeneous chaos regime are affected by a conservation law. The dynamics of the lattice are illustrated in Fig.~\ref{fig:tent-map-lattice4}. Again the lattice values are now tightly constrained to within a small band as shown in Fig.~\ref{fig:tent-map-lattice4}(a).  In general, the lattice values remain homogeneously distributed within this range and there is not a four-band structure nor the presence of any kinks or defect structures. The complexity of the dynamics is illustrated further by the significant variations shown in the space-time plot of Fig.~~\ref{fig:tent-map-lattice4}(b).

The variation of the fractal dimension with system size for these dynamics is shown in Fig.~\ref{fig:dlambda-comparison-with-beta}. The diffusively coupled lattice in the homogeneous chaos regime with a conservation law is illustrated by the blue circles (upper) where chaotic dynamics were found for $N \!\ge\! 64$. The variation of $D_\lambda$ with $N$ for these parameters with $\beta\!=\!0$ is indicated by the dash-dotted line which is included for reference. A comparison of these results suggests that the fractal dimension for the homogeneous chaos case is not significantly affected by the conservation law.

The variation of the physical dimension with system size for the homogeneous chaos case with a conservation law is shown in Fig.~\ref{fig:dphysical-comparison-with-beta} as the blue squares.  The corresponding result where $\beta=0$ is shown as the open squares. Again the physical dimension is not significantly affected by the presence of a conservation law and its value remains close to the total number of degrees of freedom of the system.

\begin{figure}[h!]
\begin{center}
\includegraphics[width=1.65in]{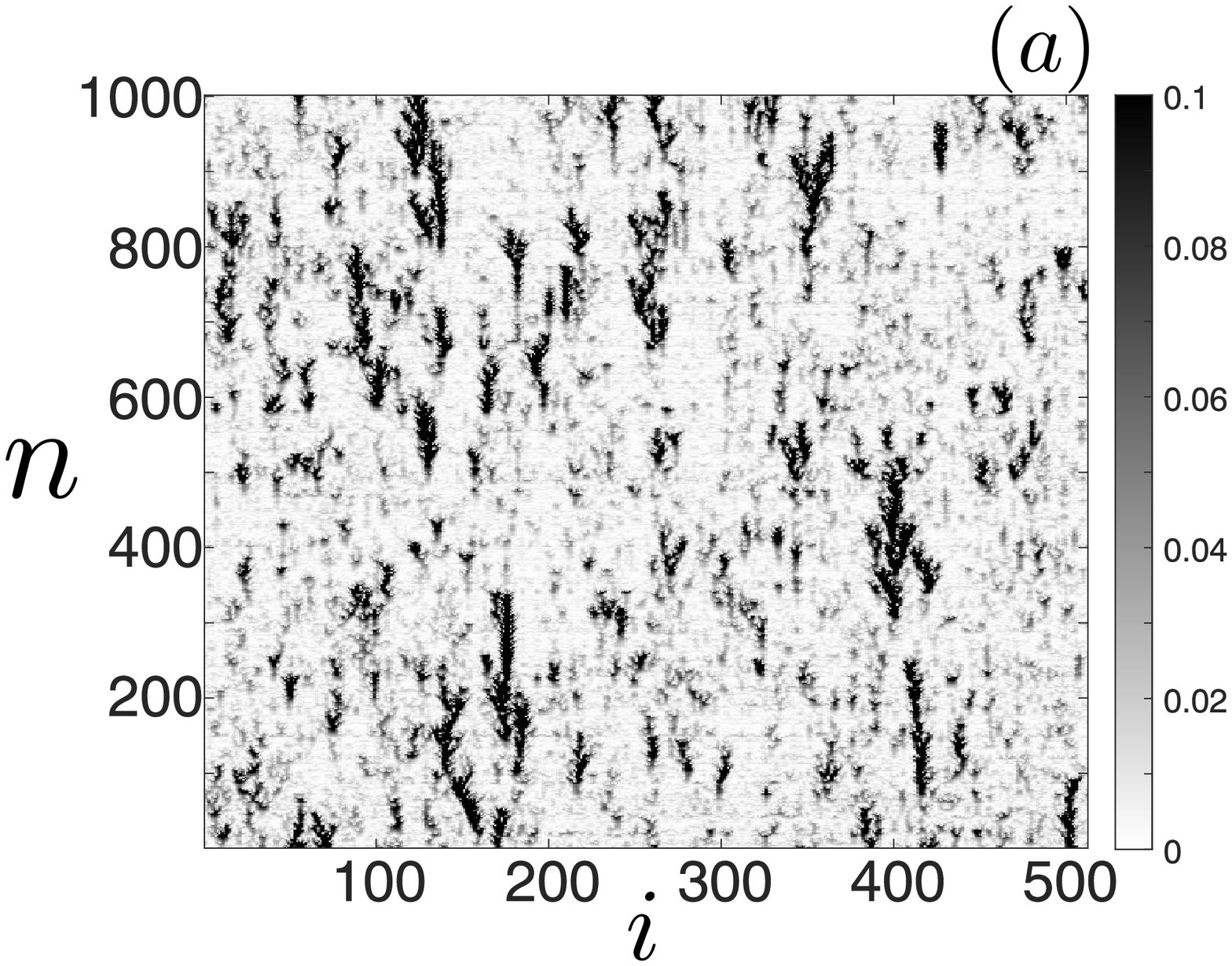}
\includegraphics[width=1.65in]{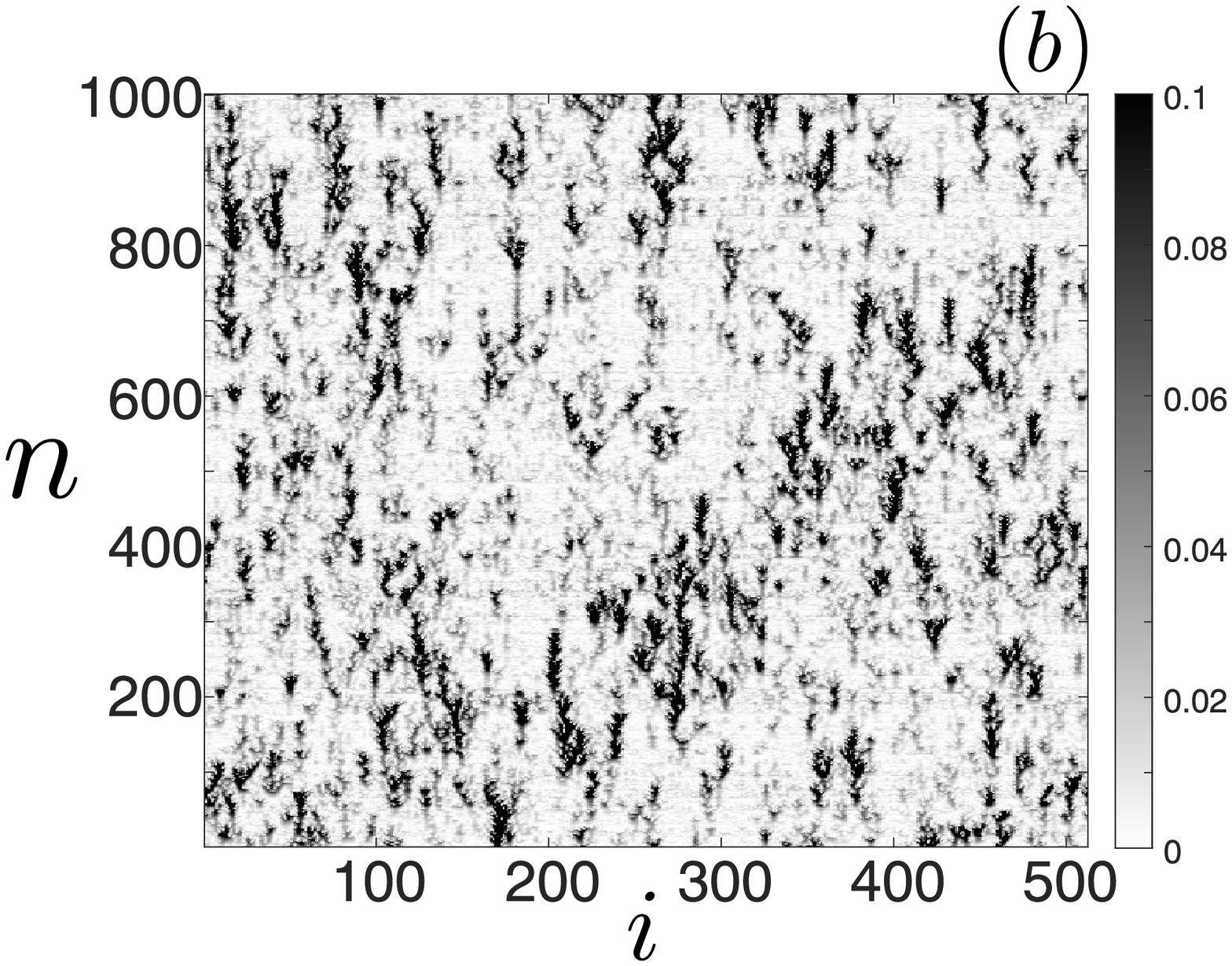} \\
\includegraphics[width=1.65in]{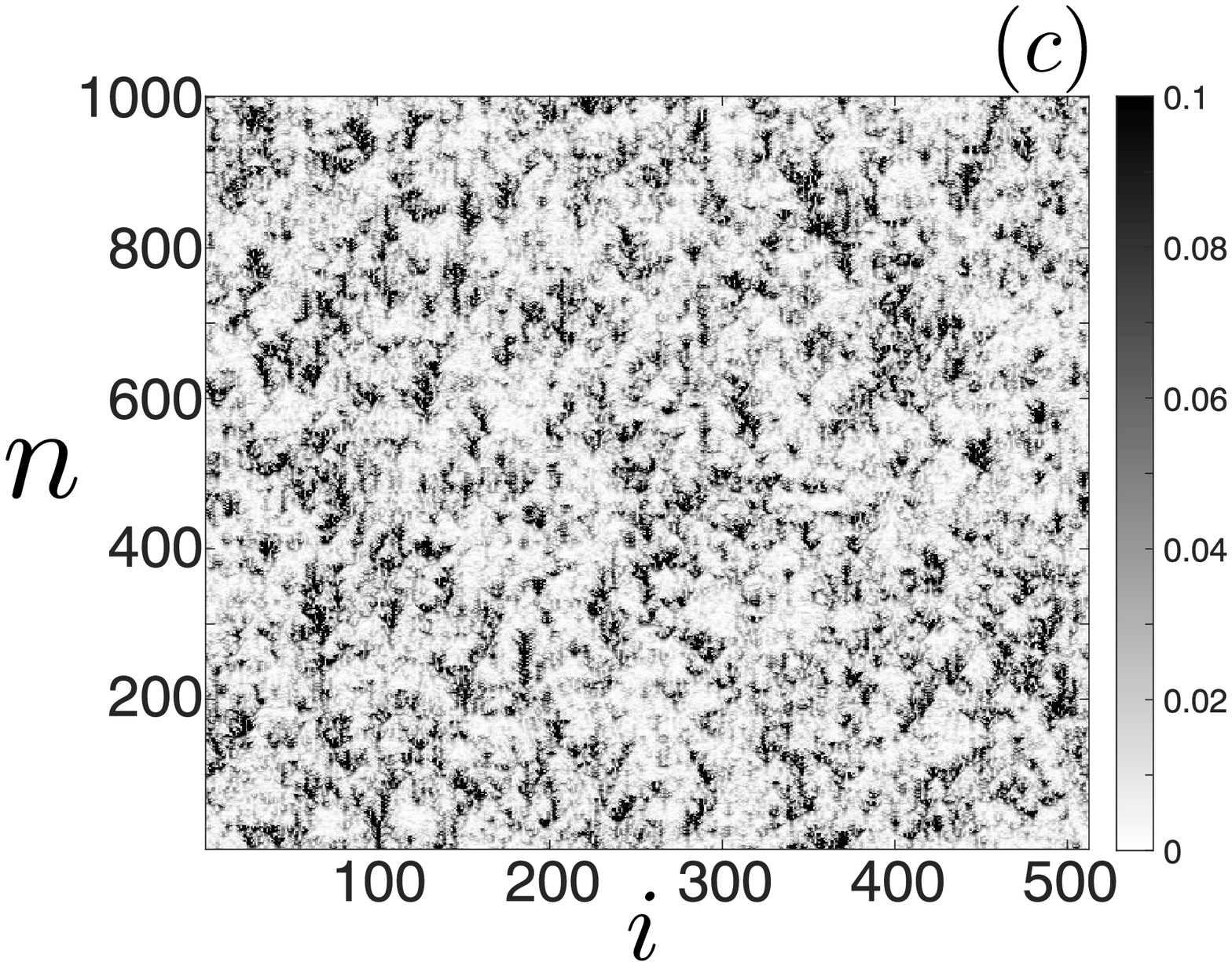} 
\includegraphics[width=1.65in]{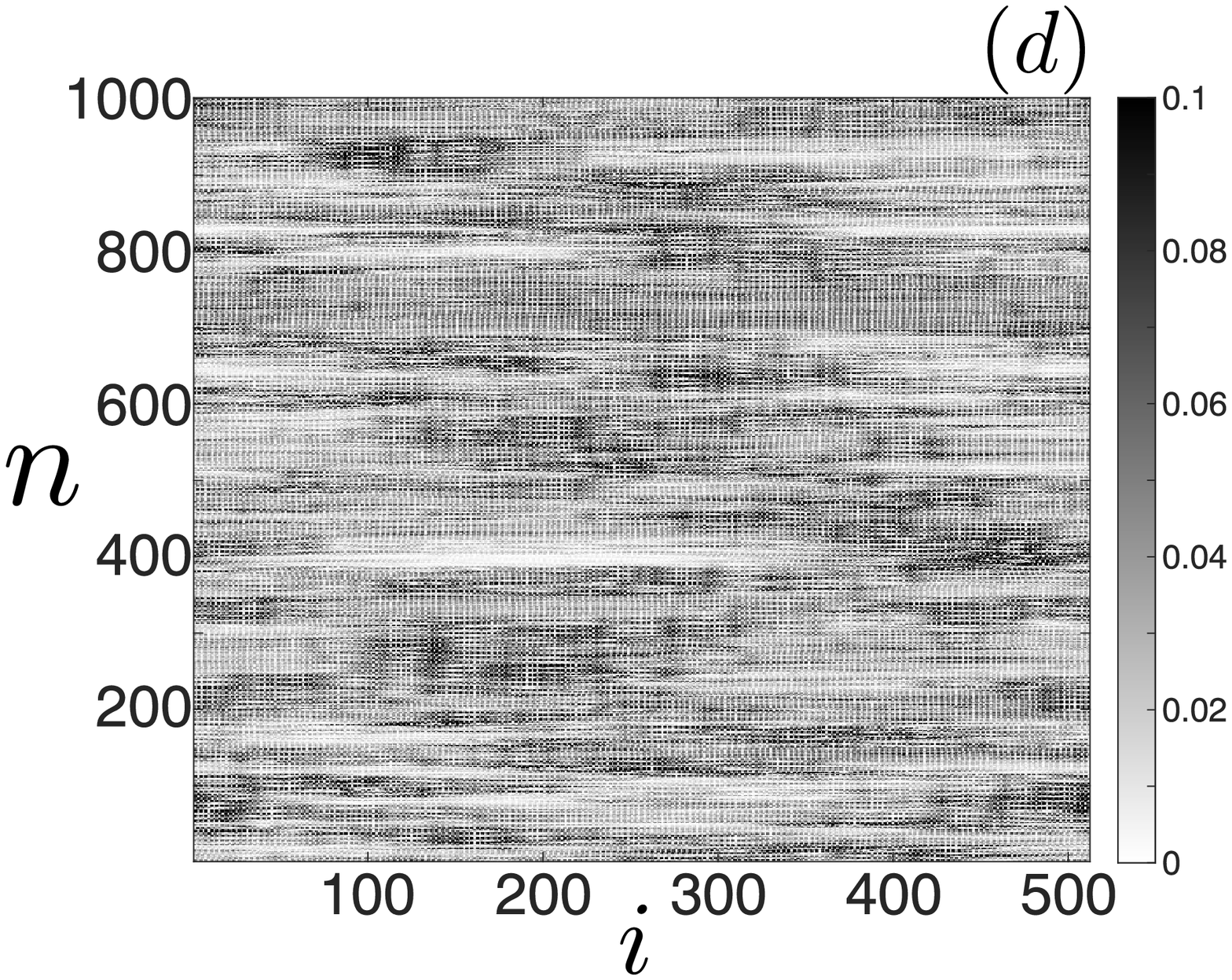}
\end{center}
\caption{Space-time plots of the CLVs of homogeneous chaos with a conservation law ($N \!=\!512$, $\beta\!=\!1$, $a\!=\!1.6$).  $|\vec{v}_k^{\,(n)}|$ is shown for: (a)~$k\!=\!1$, (b)~$k\!=\!10$, (c)~$k\!=\!100$, (d)~$k\!=\!500$.} 
\label{fig:clvs-a1p6-beta1}
\end{figure}

The spatiotemporal features of the CLVs for the homogeneous chaos case with a conservation law is shown in Fig.~\ref{fig:clvs-a1p6-beta1}. Again the CLVs are now highly delocalized due to the presence of the conservation law. This is apparent by comparing the delocalized leading CLV shown in Fig.~\ref{fig:clvs-a1p6-beta1}(a) for $\beta\!=\!1$ with the much more spatially localized structure of the leading CLV shown in Fig.~\ref{fig:clvs-a1p1-beta1}(a) for $\beta\!=\!0$. In Fig.~\ref{fig:clvs-a1p6-beta1} the delocalization increases as the index of the CLV increases.

\begin{figure}[h!]
\begin{center}
\includegraphics[width=2.25in]{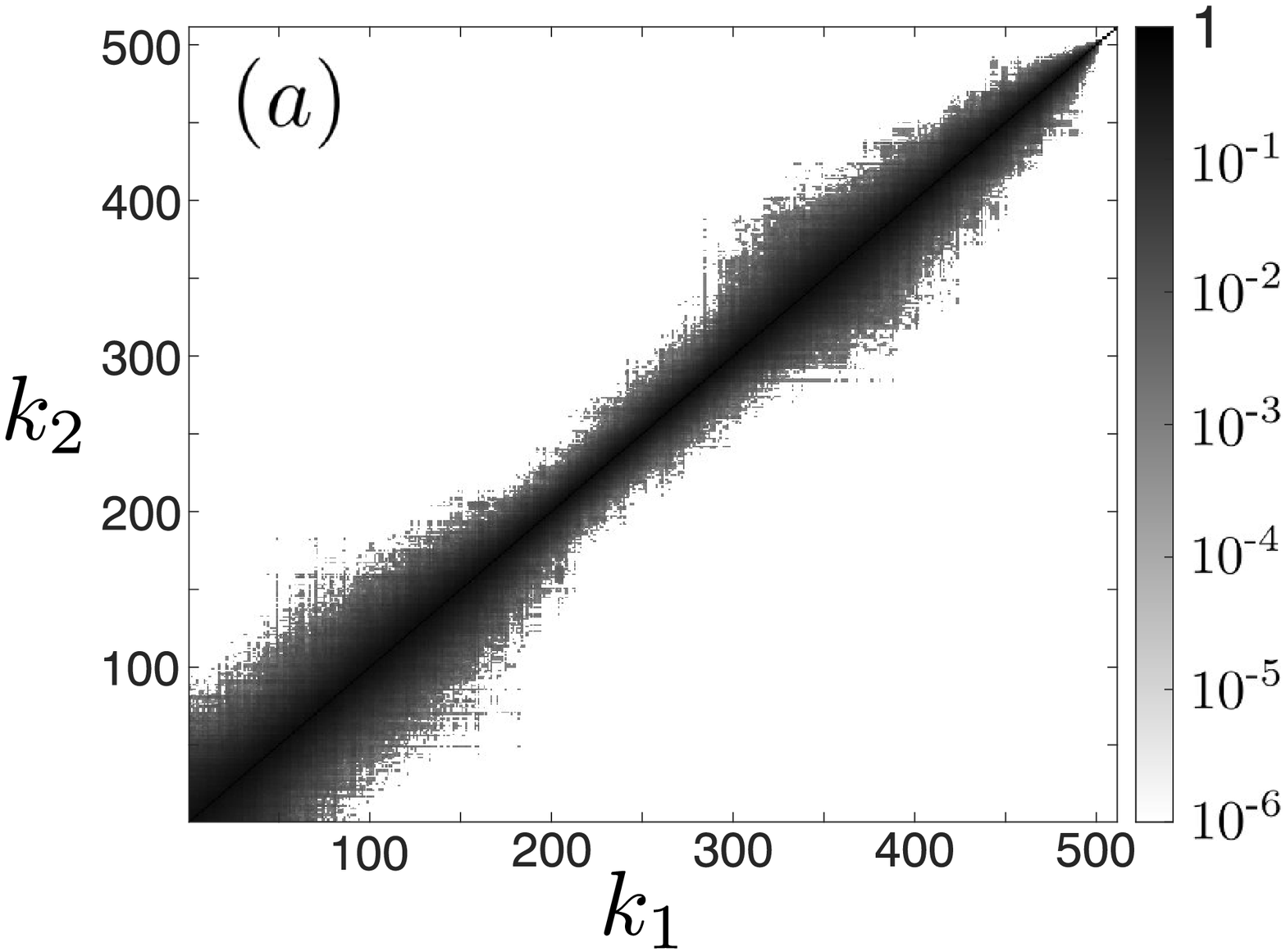} \\ 
\includegraphics[width=2.25in]{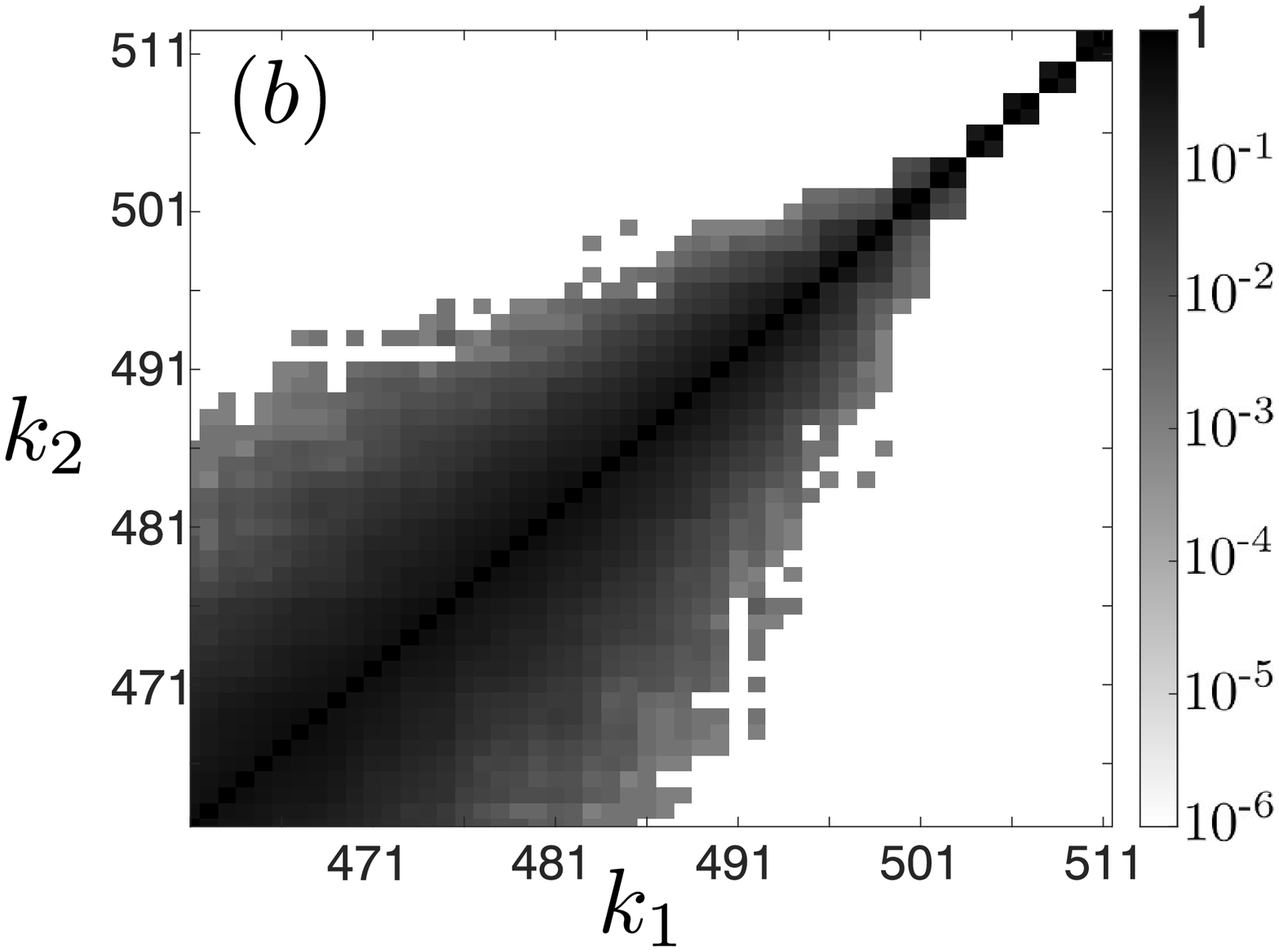}
\end{center}
\caption{Violations of the DOS for homogeneous chaos with a conservation law ($N \!=\!512$, $\epsilon\!=\!0.65$, $\beta\!=\!1$, $a\!=\!1.6$) indicating $D_{\text{ph}}\!=\!503$ and 8 transient modes. (a)~Entire lattice. (b) Close-up of the last 50 CLVs.} 
\label{fig:vdos-a1p6-beta1}
\end{figure}

The violation of the DOS for the lattice in the homogeneous chaos regime with a conservation law is shown in Fig.~\ref{fig:vdos-a1p6-beta1}. These results indicate a large number of entangled physical modes that are followed by  transient modes that appear in pairs. For these parameters, $D_\text{ph} \!=\! 503$ and there are 8 transient modes.

A comparison of the violations of the DOS with, and without a conservation law, suggests that the degree of entanglement of the leading physical modes decreases with the inclusion of the conservation law. The degree of entanglement refers to the number of adjacent CLVs that are tangled with one another as indicated by a significant violation of the DOS. For example, in Fig.~\ref{fig:vdos-a1p1-beta0}(a) ($\beta\!=\!0$, $a\!=\!1.1$) nearly all of the CLVs with indices $1 \!\le\! k_1 \lesssim \!150$ and $1\! \le\! k_2\! \lesssim\! 150$ exhibit significant violations. However, when the conservation law is included, as shown in Fig.~\ref{fig:vdos-a1p1-beta1}(a), far fewer of the leading CLVs are entangled.

As a measure of this entanglement we compute the fraction of the CLV pairs, $\xi$, that exhibit a significant amount of violation of the DOS. In our calculations, we have set the threshold indicating a significant amount of violation as $\nu_{k_1,k_2}^\tau \!\ge\! 1 \times 10^{-3}$. We only consider distinct pairs of CLVs, $k_1\!\ne\!k_2$, since all CLVs are always in violation when compared with themselves. This threshold value of $\nu_{k_1,k_2}^\tau$ corresponds to a pair of CLVs that exhibit a violation of the DOS for 0.1\% of the time or more.

Our conclusions do not change significantly with small changes in the value of this threshold. We also point out that this threshold value is in the middle of the color scale used when plotting the violations of the DOS in Figs.~\ref{fig:vdos-a1p1-beta0}, \ref{fig:vdos-a1p6-beta0}, \ref{fig:vdos-a1p1-beta1}, and~\ref{fig:vdos-a1p6-beta1}.  A CLV pair that exhibits a significant value of violation is given a value of unity, $\alpha_{k_1,k_2}\!=\!1$, and a CLV pair that does not exhibit a significant violation is given a value of zero, $\alpha_{k_1,k_2}\!=\!0$. The final reported value of $\xi$ is the average value of these $\alpha$ over all of the CLV pairs considered, $\xi = \bar{\alpha}$. Therefore, $0 \! \le \! \xi \! \le \! 1$ where $\xi\!=\!1$ when all distinct pairs of CLVs have a significant amount of violation and a value of $\xi\!=\!0$ when there is the absence of any violation.

We first consider the case of four-band chaos with a conservation law. A closer inspection of Fig.~\ref{fig:vdos-a1p1-beta0}(a) indicates the presence of violations for nearly all pairs of CLVs with indices $1 \! \le \! k \! \lesssim \! 150$. The value for the amount of entanglement for the first 150 CLVs yields $\xi \!=\! 0.96$.  This indicates that 96\% of the distinct pairs of CLVs, with indices of 150 or less, exhibit a significant amount of entanglement. This is in contrast to Fig.~\ref{fig:vdos-a1p1-beta1} which indicates much less entanglement occurring for $k \lesssim 150$ when the conservation law is imposed. The amount of entanglement for this case yields $\xi \!=\! 0.71$. These results indicate that the amount of entanglement of the leading CLVs decreases in the presence of a conservation law.  In addition, the fractal dimension of the dynamics, $D_\lambda$, decreases in the presence of a conservation law as indicated by Fig.~\ref{fig:dlambda-comparison-with-beta}.

A similar conclusion can be drawn regarding the amount of entanglement for the case of homogeneous chaos. Figure~\ref{fig:vdos-a1p6-beta0}(a) indicates the presence of entanglement for the first 200 CLVs in the absence of a conservation law. For these CLVs the amount of entanglement is $\xi \!=\! 0.96$. There is a decrease in this entanglement when a conservation law is present as shown in Fig.~\ref{fig:vdos-a1p6-beta1}(a). Using only the first 200 CLVs yields a decrease in the amount of entanglement to $\xi = 0.57$. For this case, the degree of entanglement decreases significantly in the presence of a conservation law while $D_\lambda$ remains nearly the same as shown in Fig.~\ref{fig:dlambda-comparison-with-beta}.

\subsection{Chaotic Dynamics with a Broken Conservation Law}
\label{section:broken-conservation-law}

We now explore the variation in the chaotic dynamics as a function of degree of enforcement of the conservation law. In the previous sections we explored the two cases of no conservation law ($\beta\!=\!0$) and a fully imposed conservation law ($\beta\!=\!1$). We now investigate the dynamics over the entire range of values of $0 \!\le\! \beta \!\le\! 1$.
\begin{figure}[h!]
\begin{center}
\includegraphics[width=3.0in]{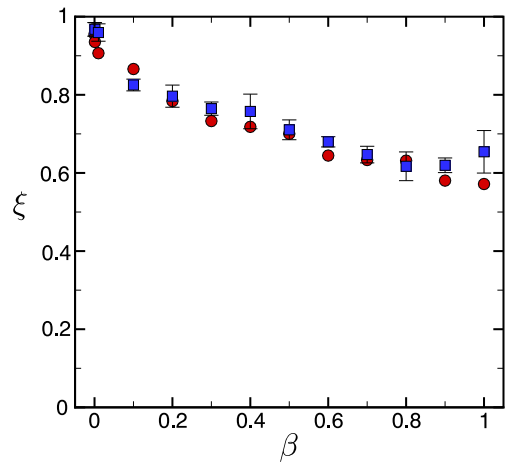}  
\end{center}
\caption{The variation of the entanglement of the leading CLVs, $\xi$, with the strength of the conservation law as indicated by $\beta$: squares (blue) $a\!=\!1.1$, circles (red) $a\!=\!1.6$. For $a=1.1$, $\xi$ is computed using the first 150 CLVs.  Symbols are mean values of $\xi$ using 3 different random initial conditions at each $\beta$, error bars are the standard deviation. For $a\!=\!1.6$, $\xi$ is computed using the fist 200 CLVs. The variation of $\xi$ with initial conditions for $a\!=\!1.6$ is much smaller and results are shown for one initial condition. Values shown for $\beta$ from 0 to 1 in increments of 0.1 in addition to $\beta \!=\! 0.001$ and  $0.01$.} 
\label{fig:xi-a11}
\end{figure}

The variation of the amount of entanglement of the CLVs, $\xi$, with $\beta$ is shown in Fig.~\ref{fig:xi-a11} for both four-band chaos (circles, red) and homogeneous chaos (squares, blue). For four-band chaos, the leading 150 CLVs are used and for the homogeneous chaos case the leading 200 CLVs are used which aligns with the discussion in \S\ref{section:with-conservation-law}.  For both four-band chaos and homogeneous chaos, it is clear that the amount of entanglement is largest for $\beta\!=\!0$ and that it is significantly reduced for $\beta\!=\!1$.

For the four-band chaos results, the amount of entanglement decreases with increasing values of $\beta$. In this case, there is some variation in the dynamics as a function of the initial conditions that are used. As a result, we report results at each value of $\beta$ for simulations that were started from three different random initial conditions. The square symbols indicate the average value of $\xi$, and the error bars are the standard deviation of $\xi$, over these initial conditions.

A similar trend in the variation of $\xi$ with $\beta$ is shown for the case of homogeneous chaos as indicated by the circles in Fig.~\ref{fig:xi-a11}. For this case, less than one percent variation was found in the value of $\xi$ with different random initial conditions and the results shown in Fig.~\ref{fig:xi-a11} at each $\beta$ are for a single random initial condition. 

In \S\ref{section:with-conservation-law} it was shown that the spatial localization of the magnitude of the CLVs varied significantly with and without a conservation law. Our investigation indicated that a fully imposed conservation law resulted in a significant amount of delocalization.  This is evident by comparing Fig.~\ref{fig:clvs-a1p6-beta0}(a) and~\ref{fig:clvs-a1p6-beta1}(a) for the cases of $\beta\!=\!0$ and $1$, respectively. We have also explored how these findings vary as a function of $\beta$. In the following we show results only for the homogeneous chaos case although similar trends are also found for the case of four-band chaos.
\begin{figure}[h!]
\begin{center}
\includegraphics[width=1.68in]{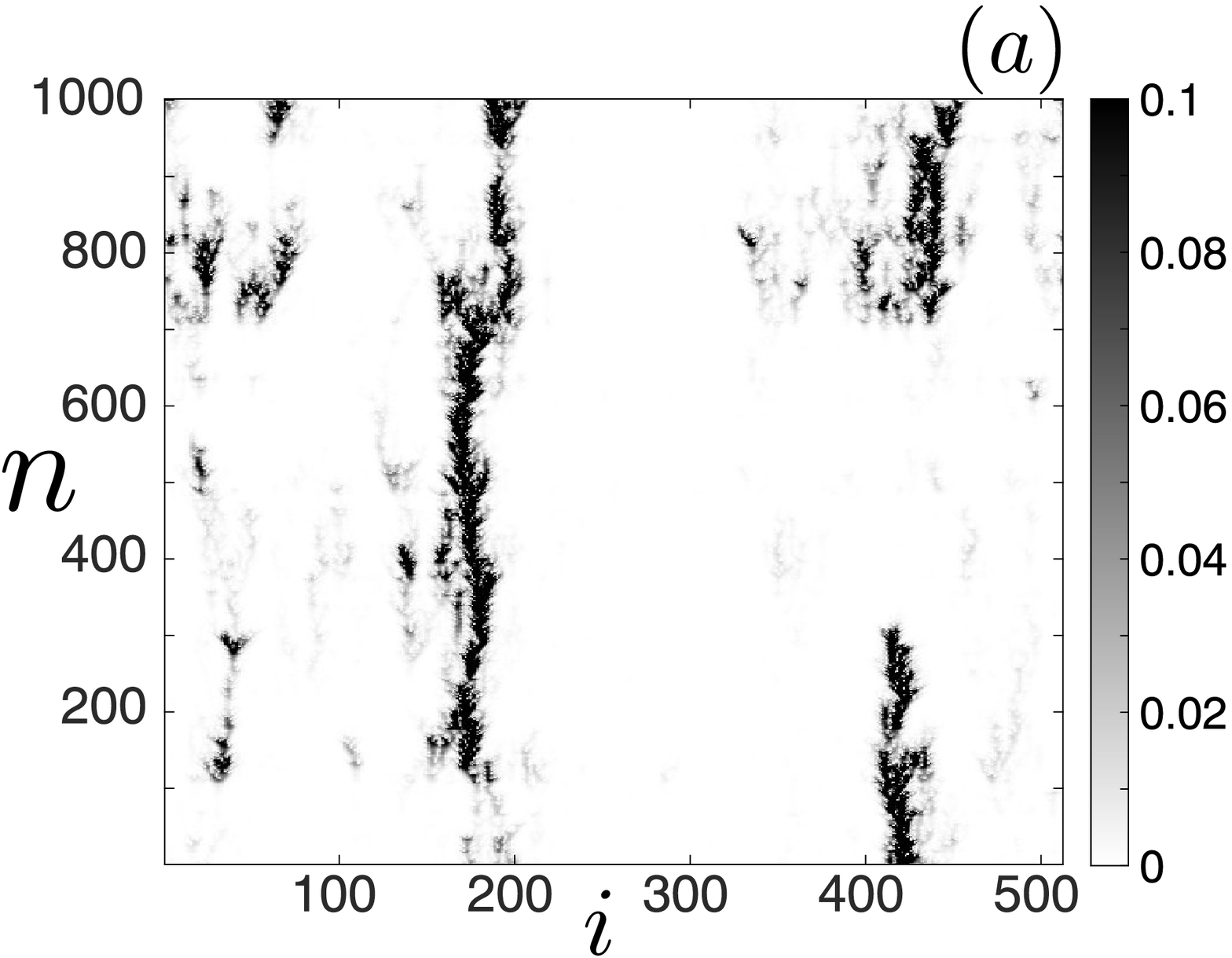}  
\includegraphics[width=1.68in]{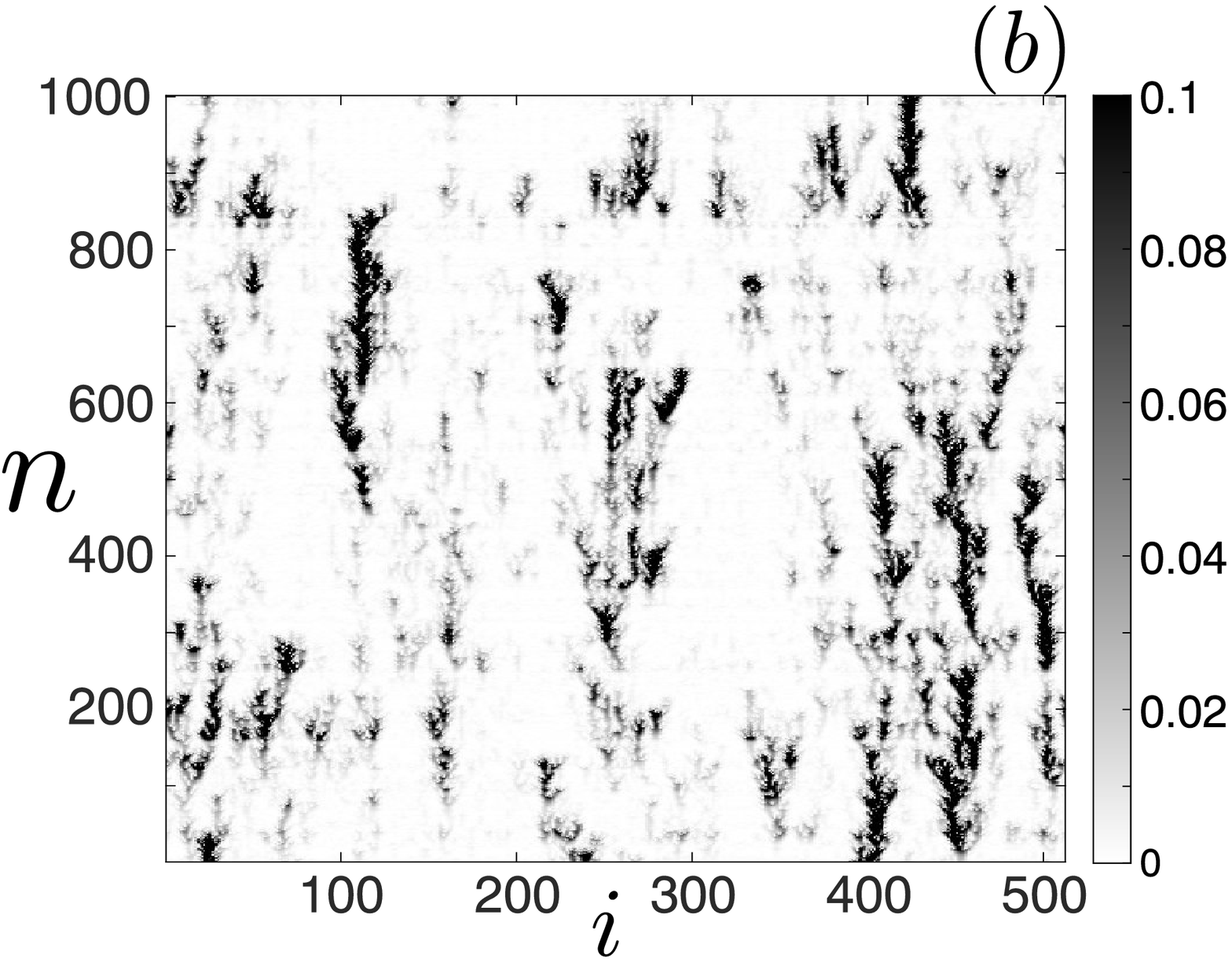}  \\
\includegraphics[width=1.68in]{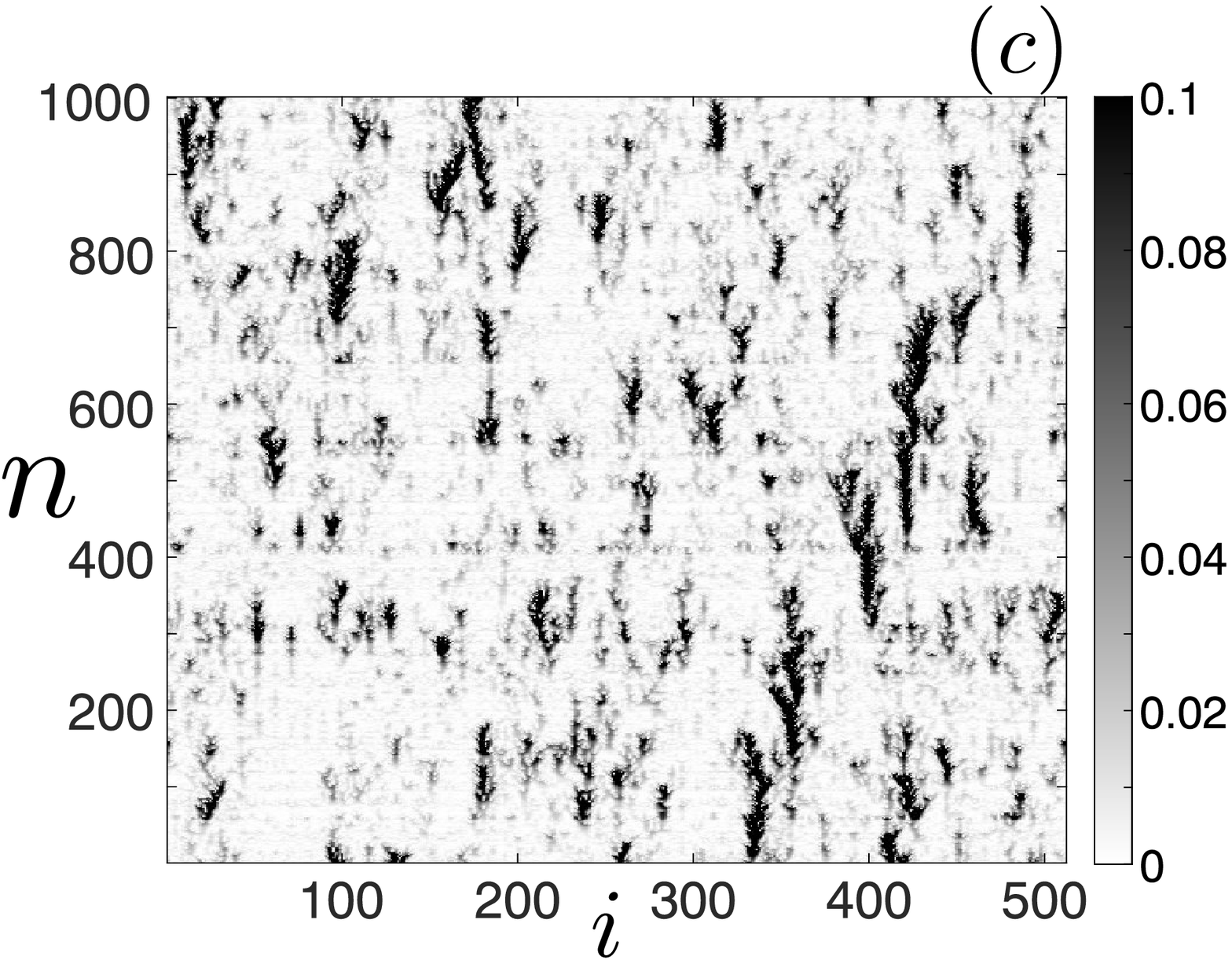}  
\includegraphics[width=1.68in]{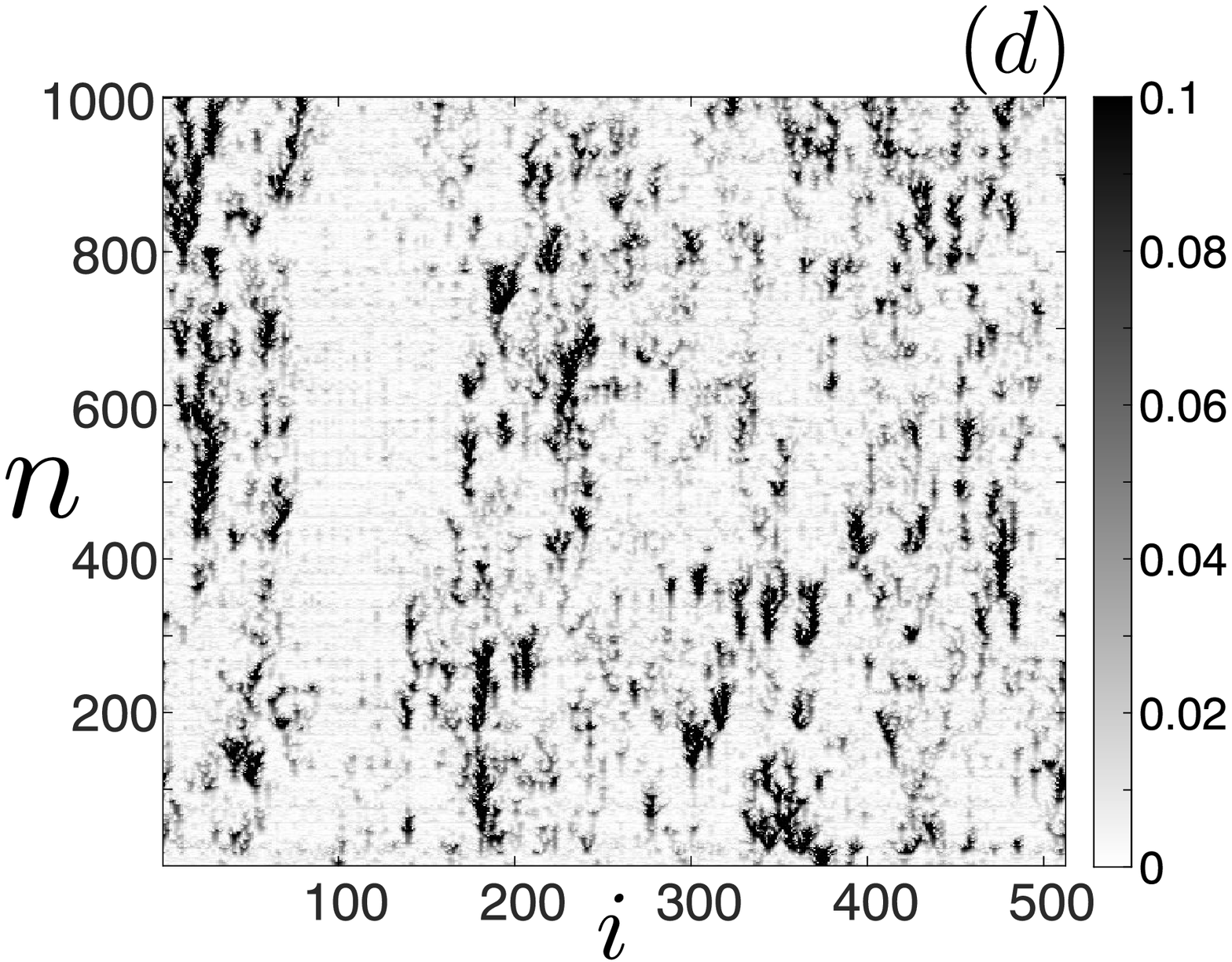}
\end{center}
\caption{Space-time plots of the magnitude of the leading CLV, $|\vec{v}_1^{\,(n)}|$, as a function of $\beta$ for homogeneous chaos ($a \!=\! 1.6$). (a)~$\beta \!=\! 0.01$, (b)~$\beta \!=\! 0.1$, (c)~$\beta \!=\! 0.5$, (d)~$\beta \!=\! 0.9$. For $\beta\!=\!0$ see Fig.~\ref{fig:clvs-a1p6-beta0}(a), for $\beta\!=\!1$ see Fig.~\ref{fig:clvs-a1p6-beta1}(a).} 
\label{fig:clv-with-beta}
\end{figure}

Figure~\ref{fig:clv-with-beta} illustrates space-time plots of the magnitude of the leading CLV, $|\vec{v}_1^{\,(n)}|$, for four different values of $\beta$. Figure~\ref{fig:clv-with-beta}(a) shows results for a very weakly imposed conservation law where $\beta \!=\! 0.01$. A comparison of Fig.~\ref{fig:clv-with-beta}(a) with Fig.~\ref{fig:clvs-a1p6-beta0}(a) indicates the significant delocalization that has occurred even for this small amount of global coupling. In Fig.~\ref{fig:clv-with-beta}(a) the regions of large magnitude of the leading CLV are contained by several stripe structures which also contain small scale structures including branching type features.

Figure~\ref{fig:clv-with-beta}(b) shows the space-time plot for the leading CLV for $\beta\!=\!0.1$. It is clear that the regions containing significant CLV magnitude has increased and that the spatial localization has decreased. This trend of decreasing localization continues with increasing values of $\beta$. Figure~\ref{fig:clv-with-beta}(c) shows the space-time plot for $\beta\!=\!0.5$ which is now highly delocalized. Figure~\ref{fig:clv-with-beta}(d) shows the highly delocalized leading CLV for $\beta\!=\!0.9$. In general, the trend is that the spatial localization decreases with increasing values of $\beta$. We emphasize that even a very weakly imposed conservation law leads to a significant amount of delocalization of the leading CLV. This suggests that a conservation law, even a weakly imposed one, may result in dynamics that are more sensitive to perturbations that occur over a wider range of spatial locations. 

\section{Conclusion}
\label{section:conclusion}

We have used CMLs to quantitatively explore high-dimensional spatiotemporal chaos. This has made it possible to explore chaotic dynamics with physical dimensions on the order of 500 for a wide range of conditions. This required the computation of over 500 CLVs for long times which remains a very difficult calculation using the partial differential equations that govern many laboratory systems. By exploring two values of the control parameter, and for lattices with a conservation law of varying strength, we were able to investigate difficult fundamental questions.

Our results suggest that a conservation law, even when imposed weakly, strongly delocalizes the spatial dependence of the CLVs for the conditions we explore. This was found to be true for the four band chaos regime as well as the homogeneous chaos regime.  In addition, the entanglement of the leading CLVs, composing the leading part of the physical modes in our study, is significantly affected by the presence of a conservation law. As the strength of the conservation law is increased, the entanglement of the leading CLVs with their neighbors decreases. We anticipate that these findings will provide valuable insights to guide future investigations using CLVs aimed at building a better physical understanding of larger and more complex laboratory scale systems with conservation laws.

~\vspace{0.2cm}

\noindent Acknowledgments: MRP acknowledges support from NSF fund number CMMI-2138055. Portions of the computations were performed using the Advanced Research Computing (ARC) center at Virginia Tech. 

\bibliographystyle{unsrt}

\begin{thebibliography}{10}

\bibitem{cross:1993}
M.~C. Cross and P.~C. Hohenberg.
\newblock Pattern formation outside of equilibrium.
\newblock {\em Rev. Mod. Phys.}, 65(3 II):851--1112, 1993.

\bibitem{guckenheimer:1983}
J.~Guckenheimer and P.~Holmes.
\newblock {\em Nonlinear Oscillations, Dynamical Systems, and Bifurcations of
  Vector Fields}.
\newblock Springer, 1983.

\bibitem{abarbanel:1996}
H.~D.~I. Abarbanel.
\newblock {\em Analysis of Observed Chaotic Data}.
\newblock Springer, 1996.

\bibitem{cross:2009}
M.~C. Cross and H.~S. Greenside.
\newblock {\em Pattern formation and dynamics in nonequilibrium systems}.
\newblock Cambridge University Press, 2009.

\bibitem{rowley:2009}
C.~W. Rowley, I.~Mezi\'c, S.~Bagheri, P.~Schlatter, and D.~S. Henningson.
\newblock Spectral analysis of nonlinear flows.
\newblock {\em J. Fluid Mech.}, 641:115--127, 2009.

\bibitem{mezic:2013}
I.~Mezi\'c.
\newblock Analysis of fluid flows via spectral properties of the {K}oopman
  operator.
\newblock {\em Annu. Rev. Fluid Mech. Rev.}, 45:357--378, 2013.

\bibitem{schmid:2010}
P.~J. Schmid.
\newblock Dynamic mode decomposition of numerical and experimental data.
\newblock {\em J. Fluid Mech.}, 656:5--28, 2010.

\bibitem{tu:2014}
J.~Tu, C.~W. Rowley, D.~M. Luchtenburg, S.~L. Brunton, and J.~N. Kutz.
\newblock On dynamic mode decomposition: theory and applications.
\newblock {\em J. Comput. Dyn.}, 1(2):391--421, 2014.

\bibitem{brunton:2016}
S.~L. Brunton, J.~L. Proctor, and N.~Kutz.
\newblock Discovering governing equations from data by sparse identification of
  nonlinear dynamical systems.
\newblock {\em Proc. Nat. Acad. Sci. USA}, 113(15):3932--3937, 2016.

\bibitem{pathak:2018}
J.~Pathak, B.~Hunt, M.~Girvan, Z.~Lu, and E.~Ott.
\newblock Model-free prediction of large spatiotemporally chaotic systems from
  data: a reservoir computing approach.
\newblock {\em Phys. Rev. Lett.}, 120:024102, 2018.

\bibitem{waleffe:1997}
F.~Waleffe.
\newblock On a self-sustaining process in shear flows.
\newblock {\em Phys. Fluids}, 9:883--900, 1997.

\bibitem{kerswell:2005}
R.~R. Kerswell.
\newblock Recent progress in understanding the transition to turbulence in a
  pipe.
\newblock {\em Nonlinearity}, 18(6):R17--R44, 2005.

\bibitem{kawahara:2012}
G.~Kawahara, M.~Uhlmann, and L.~van {V}een.
\newblock The significance of simple invariant solutions in turbulent flows.
\newblock {\em Annu. Rev. Fluid Mech.}, 44:203--225, 2012.

\bibitem{eckmann:1985}
J.~P. Eckmann and D.~Ruelle.
\newblock Ergodic theory of chaos and strange attractors.
\newblock {\em Rev. Mod. Phys.}, 57(3):617--656, 1985.

\bibitem{pikovsky:2016}
A.~Pikovsky and A.~Politi.
\newblock {\em Lyapunov exponents: a tool to explore complex dynamics}.
\newblock Cambridge University Press, 2016.

\bibitem{kaneko:1993}
K.~Kaneko, editor.
\newblock {\em Theory and applications of coupled map lattices}.
\newblock John Wiley \& Sons, Ltd., 1993.

\bibitem{oono:1987}
Y.~Oono and S.~Puri.
\newblock Computational efficient modeling of ordering of quenched phases.
\newblock {\em Phys. Rev. Lett.}, 58(8):836--839, 1987.

\bibitem{kaneko:1989}
K.~Kaneko.
\newblock Spatiotemporal chaos in one- and two-dimensional coupled map
  lattices.
\newblock {\em Physica D}, 37:60--82, 1989.

\bibitem{takeuchi:2011}
K.~A. Takeuchi, H.~Yang, F.~Ginelli, G.~Radons, and H.~Chat\`{e}.
\newblock Hyperbolic decoupling of tangent space and effective dimension of
  dissipative systems.
\newblock {\em Phys. Rev. E}, 84:046214, 2011.

\bibitem{ginelli:2007}
F.~Ginelli, P.~Poggi, A.~Turchi, H.~Chat\`{e}, R.~Livi, and A.~Politi.
\newblock Characterizing dynamics with covariant {L}yapunov vectors.
\newblock {\em Phys. Rev. Lett.}, 99:130601, 2007.

\bibitem{wolfe:2007}
C.~L. Wolfe and R.~M. Samelson.
\newblock An efficient method for recovering {L}yapunov vectors from singular
  vectors.
\newblock {\em Tellus}, 59A:355--366, 2007.

\bibitem{kuptsov:2012}
P.~V. Kuptsov and U.~Parlitz.
\newblock Theory and computation of covariant {L}yapunov vectors.
\newblock {\em J. Nonlinear Sci.}, 22(5):727, 2012.

\bibitem{kaplan:1979}
J.~L. Kaplan and J.~A. Yorke.
\newblock Lecture notes in math.
\newblock {\em 730}, pages 204--227, 1979.

\bibitem{egolf:2000}
D.~A. Egolf, I.~V. Melnikov, W.~Pesch, and R.~E. Ecke.
\newblock Mechanisms of extensive spatiotemporal chaos in
  {R}ayleigh-{B}\'{e}nard convection.
\newblock {\em Nature}, 404:733--736, 2000.

\bibitem{scheel:2006}
J.~D. Scheel and M.~C. Cross.
\newblock Lyapunov exponents for small aspect ratio {R}ayleigh-{B}\'{e}nard
  convection.
\newblock {\em Phys. Rev. E}, 74:066301, 2006.

\bibitem{paul:2007}
M.~R. Paul, M.~I. Einarsson, P.~F. Fischer, and M.~C. Cross.
\newblock Extensive chaos in {R}ayleigh-{B}\'{e}nard convection.
\newblock {\em Phys. Rev. E}, 75:045203, 2007.

\bibitem{xu:2018}
M.~Xu and M.~R. Paul.
\newblock Spatiotemporal dynamics of the covariant {L}yapunov vectors of
  chaotic convection.
\newblock {\em Phys. Rev. E}, 97:032216, 2018.

\bibitem{levanger:2019}
R.~Levanger, M.~Xu, J.~Cyranka, M.~F. Schatz, K.~Mischaikow, and M.~R. Paul.
\newblock Correlations between the leading {L}yapunov vector and pattern
  defects for chaotic {R}ayleigh-{B}\'enard convection.
\newblock {\em Chaos}, 29:053103, 2019.

\bibitem{yang:2009}
H.~Yang, K.~A. Takeuchi, F.~Ginelli, H.~Chat\`{e}, and G.~Radons.
\newblock Hyperbolicity and the effective dimension of spatially extended
  dissipative systems.
\newblock {\em Phys. Rev. Lett.}, 102:074102, 2009.

\bibitem{oseledec:1968}
V.~I. Oseledec.
\newblock A multiplicative ergodic theorem. {L}yapunov characteristic numbers
  for dynamical systems.
\newblock {\em Trans. Moscow Math. Soc.}, 19:197 -- 231, 1968.

\bibitem{pugh:2003}
C.~Pugh, M.~Shub, and A.~Starkov.
\newblock Stable ergodicity.
\newblock {\em Bull. Am. Math. Soc.}, 41(1):1--41, 2003.

\bibitem{bochi:2005}
J.~Bochi and M.~Viani.
\newblock The {L}yapunov exponents of generic volume-preserving and symplectic
  maps.
\newblock {\em Ann. Math.}, 161:1423--1485, 2005.

\end{thebibliography}

\end{document}